\documentclass[12pt,a4paper]{article}
\pdfoutput=1
\usepackage{jcappub}
\usepackage{natbib}
\usepackage{graphicx}
\usepackage[english]{babel}
\usepackage{color}
\usepackage{multirow}
\usepackage{ulem}

\def\bea{\begin{eqnarray}}
\def\eea{\end{eqnarray}}
\def\be{\begin{equation}}
\def\ee{\end{equation}}
\newcommand{\Dpp}{D_{pp}}
\newcommand{\Dxx}{D_{xx}}
\newcommand{\ddp}{{\partial\over\partial p}}

\definecolor{orange}{RGB}{255,127,0}
\definecolor{maroon}{RGB}{128, 0, 0}
\definecolor{brown}{RGB}{150, 75, 0}

\title{The isotropic radio background revisited}

\author[a,b]{Nicolao Fornengo}
\author[c]{Roberto A. Lineros}
\author[a,b]{Marco Regis}
\author[d]{Marco Taoso}

\affiliation[a]{Dipartimento di Fisica Teorica, Universit\`{a} di Torino, via P. Giuria 1, I--10125 Torino, Italy}
\affiliation[b]{Istituto Nazionale di Fisica Nucleare, Sezione di Torino, via P. Giuria 1, I--10125 Torino, Italy}
\affiliation[c]{Instituto de F\'{\i}sica Corpuscular -- CSIC/U. Valencia, Parc Cient\'{\i}fic, calle Catedr\'{a}tico Jos\'{e} Beltr\'{a}n, 2, E-46980 Paterna, Spain}
\affiliation[d]{Institut de Physique Th$\acute{\mbox{e}}$orique, CEA/Saclay, F-91191 Gif-sur-Yvette C$\acute{\mbox{e}}$dex, France}

\emailAdd{fornengo@to.infn.it}
\emailAdd{rlineros@ific.uv.es}
\emailAdd{regis@to.infn.it}
\emailAdd{taoso@cea.fr }

\abstract{
We present an extensive analysis on the determination of the isotropic radio background.
We consider six different radio maps, ranging from 22 MHz to 2.3 GHz and covering a large fraction of the sky.
The large scale emission is modeled as a linear combination of an isotropic component plus the Galactic synchrotron radiation and thermal bremsstrahlung.
Point-like and extended sources are either masked or accounted for by means of a template.
We find a robust estimate of the isotropic radio background, with limited scatter among different Galactic models.
The level of the isotropic background lies significantly above the contribution obtained by integrating the number counts of observed extragalactic sources.
Since the isotropic component dominates at high latitudes, thus making the profile of the total emission flat, a Galactic origin for such excess appears unlikely.
We conclude that, unless a systematic offset is present in the maps, and provided that our current understanding of the Galactic synchrotron emission is reasonable, extragalactic sources well below the current experimental threshold seem to account for the majority of the brightness of the extragalactic radio sky.
} 

\date{\today}
\begin{document}
\maketitle

\section{Introduction}
\label{sec:Intro}
Two main avenues can be followed to estimate the total brightness of the extragalactic radio sky.
The first and most obvious way is to simply collect all the radio waves arriving from the sky in a detector, which, nowadays, typically is a single dish telescope or a balloon/satellite.
The main problem of this estimate resides in the subtraction of the foreground emission from within our Galaxy.
The second way consists instead in adding up the single contributions from all the extragalactic sources. It can be done by individually observing the sources or by statistically evaluating their contribution from the fluctuations of the intensity below the detection threshold. To this aim, synthesis arrays are the most powerful telescope currently developed. Their interferometric working mechanism prevents, on the other hand, the detection of a truly isotropic emission.

By themselves, both methods cannot provide an ultimate answer to the radio background estimate.
In the first case, a degeneracy with other contributions, like the monopole of the cosmic-ray (CR) emission in the Milky-Way, a zero-level offset in the maps, or local emissions, is unavoidable.
In the second case, since the experimental sensitivity is finite, there is always an open window for a contribution from faint populations of sources.
Our understanding of the total brightness of the extragalactic radio sky can be settled down, only if the estimates from the two techniques match.
Unfortunately, at present, they do not match. 

Indeed, the balloon--borne experiment ARCADE 2 (Absolute Radiometer for Cosmology, Astrophysics and Diffuse Emission)~\cite{Singal:2009xq} recently reported an isotropic radio emission (based on new observations at 3 to 10 GHz and on a re-analysis of old maps at lower frequencies) which significantly exceeds the expected contributions from known extragalactic sources.
The isotropic background has been estimated by subtracting a Galactic component. The latter was computed employing two different methods: using a simple Galactic plane-parallel model with a $cosec(b)$ dependence on the galactic latitude; correlating the radio emission with a map of the CII recombination line. These two techniques give consistent results.

The goal of the paper is to revisit the isotropic background estimate, employing a more sophisticated analysis, to assess the robustness of such excess.
In particular, we will use state-of-the-art models for the Galactic foreground.

The source contributions to counts and to the sky brightness are linked by the Rayleigh-Jeans law:
\be
\frac{dN}{dS}(S)\,S\,dS=\frac{2\,k_B}{c^2}\,\nu_{\rm obs}^2\,dT\;,
\ee
where $dN/dS$ is the number of sources per steradian and per unit flux in the range of flux $(S,S+dS)$, $\nu_{\rm obs}$ is the radio-frequency of observation, $T$ is the brightness temperature, $k_B$ is the Boltzmann constant, and $c$ is the speed of light.
The total brightness of the extragalactic sky is thus given by:
\be
T_E=\int\ dT=\frac{c^2}{2\,k_B\,\nu_{obs}^2}\ \int_{0}^{S_{\rm max}}\ dS \ \frac{dN}{dS}(S)\,S\;,
\label{eq:totT}
\ee
where $S_{\rm max}$ is the flux density of the brightest sources.
However, observationally we can obviously measure source counts only down to a certain threshold $S_{\rm thr}>0$ because of finite sensitivities.
Thus, part of the integral in Eq.~\ref{eq:totT} has to rely on an extrapolation of $dN/dS$ at low fluxes.

The most recent estimate of $T_E$ from number counts is presented in Ref. \cite{Vernstrom:2013vva} from 1.4 GHz data, with the $dN/dS$ derived down to the $\mu$Jy level. The extrapolation to lower fluxes does not affect the isotropic temperature determination if such extrapolation is assumed to follow from standard evolutionary models, like the one in Ref. \cite{Condon:1984}. The reported background temperature in Ref. \cite{Vernstrom:2013vva} is 115 mK.

On the other hand, by subtracting a model for the foreground Galactic emission~\cite{Fixsen:2009xn}, the ARCADE Collaboration isolated, in radio maps from 22 MHz to 10 GHz, an isotropic temperature (on top of the CMB blackbody contribution), which can be fitted by a power law $T_E=T_s \left(\nu/{\rm GHz}\right)^{\alpha}$ with $\alpha=-2.62\pm0.04$ and $T_s=1.19\pm 0.14$ K~\cite{Fixsen:2009xn}.
This leads to about 500 mK at 1.4 GHz, so exceeding the counts-based estimate by a factor of 4-5.

Various astrophysical solutions for the mismatch have been discussed. The contribution from radio supernovae, radio quiet quasars and diffuse emission from intergalactic medium and clusters have been shown to be quite modest \cite{Singal:2009dv} and cannot provide a sufficiently high bump in the sub-$\mu$Jy range of number counts. Star forming galaxies with a non-standard evolutionary model are constrained by bounds from the far IR-radio correlation, gamma-rays, and P(D) analyses \cite{Singal:2009dv,Lacki:2010uz,Vernstrom:2011xt,Ponente:2011se,Condon:2012ug,Vernstrom:2013vva}.
The required level of contribution can be also efficiently probed by anisotropy studies in the case of clustered sources~\cite{Holder:2012nm}.
If the origin of the excess is from a population of extragalactic sources, they have to be very faint (significantly contributing to number counts well below the $\mu$Jy level) and very numerous \cite{Singal:2009dv,Vernstrom:2011xt,Vernstrom:2013vva}.

A possibility in this direction is represented by synchrotron radiation induced by annihilations of dark-matter particles in halos of extragalactic structures, as proposed in Ref. \cite{paper1}.
This has been discussed also in Refs. \cite{Fornengo:2011xk,Hooper:2012jc}. Other exotic explanations include \cite{Yang:2012qi,Lawson:2012zu,Cline:2012hb}.

In this paper, we revisit the determination of the isotropic radio background from radio maps.
An attempt in the same direction has been recently undertaken by Ref.~\cite{Subrahmanyan:2013eqa}.
Here we consider a more realistic and sophisticated Galactic description, we increase the number of maps employed, and adopt a different statistical technique.

In Sect. \ref{sec:Surveys}, we present the maps which are more relevant for the isotropic background study.
Sect. \ref{sec:Models} describes how the Galactic emission is accounted for.
The emission observed in radio skymaps from few MHz to hundreds of GHz is the result of diverse emission mechanisms and sources. 
The Milky Way activity accelerates CR electrons and positrons and amplifies magnetic fields, and a large fraction of the Galactic radio emission is produced via synchrotron radiation. 
In addition, thermal bremsstrahlung coming from warm and hot gas in the interstellar medium can contribute at such frequencies but is more localized nearby the Galactic plane.
Discrete sources also provide an important fraction of the total emission.
Finally, the extragalactic background from unresolved sources can dominate at high-latitudes.

We adopt a 4D modeling (3 spatial dimensions plus momentum) of CR synchrotron radiation, solving the CR transport equation by means of the publicly available code 
GALPROP~\cite{Strong:1998pw}\footnote{We used the GALPROP version 54.1.984, downloaded from Ref. \cite{galpropweb}. The implementation in GALPROP of the magnetic field model employed in this analysis, as well as the theoretical and observational maps shown in this work, are available at \cite{mapsweb} or by writing to taoso@cea.fr or regis@to.infn.it }. 
The propagation models are tuned to fit local CR spectra and gamma-ray data~\cite{Strong:2010pr}.
We employ a detailed description of the magnetic field following Refs.~\cite{Jansson:2012rt,Jansson:2012pc}.
Discrete sources are either masked or described by means of a template as discussed in Sect. \ref{sec:Fit}, where we also describe the fitting procedure.
The resulting estimates of the isotropic radio background is presented in Sect. \ref{sec:Results}.
Caveats and consequences of the analysis are discussed in Sect. \ref{sec:Discussion}, while Sect. \ref{sec:Conclusions} concludes.

\section{Radio surveys}
\label{sec:Surveys}

The requirements we follow to choose the radio maps employed in this work are: a) a good coverage of high latitudes and b) a large fraction of the sky observed.
The first condition is obviously necessary, since we aim at estimating the extragalactic background which mostly affects high latitude data. The second condition is instead needed to have a proper gauging of the Galactic component (in addition, of course, of allowing an increased  statistics).

\begin{figure*}[t]
\begin{center}
 \includegraphics[width=0.32\textwidth]{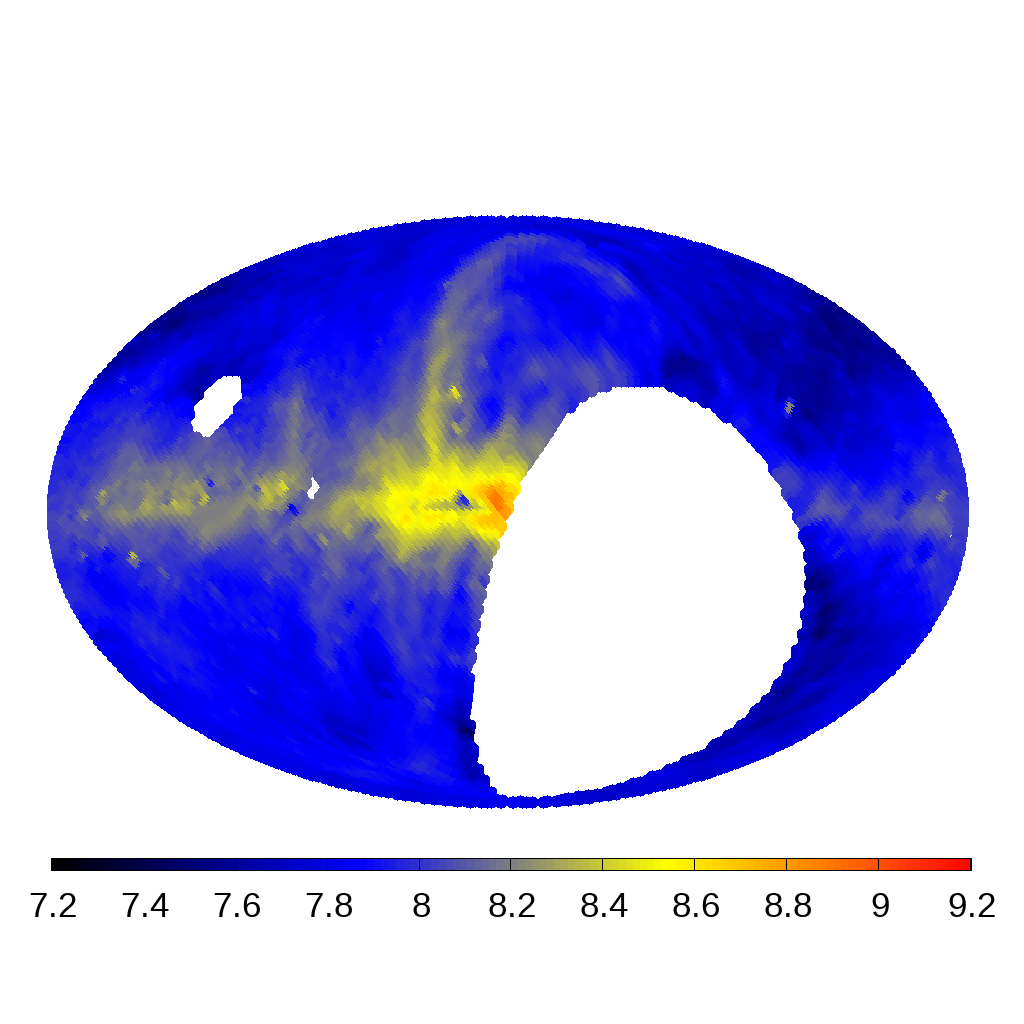}
 \includegraphics[width=0.32\textwidth]{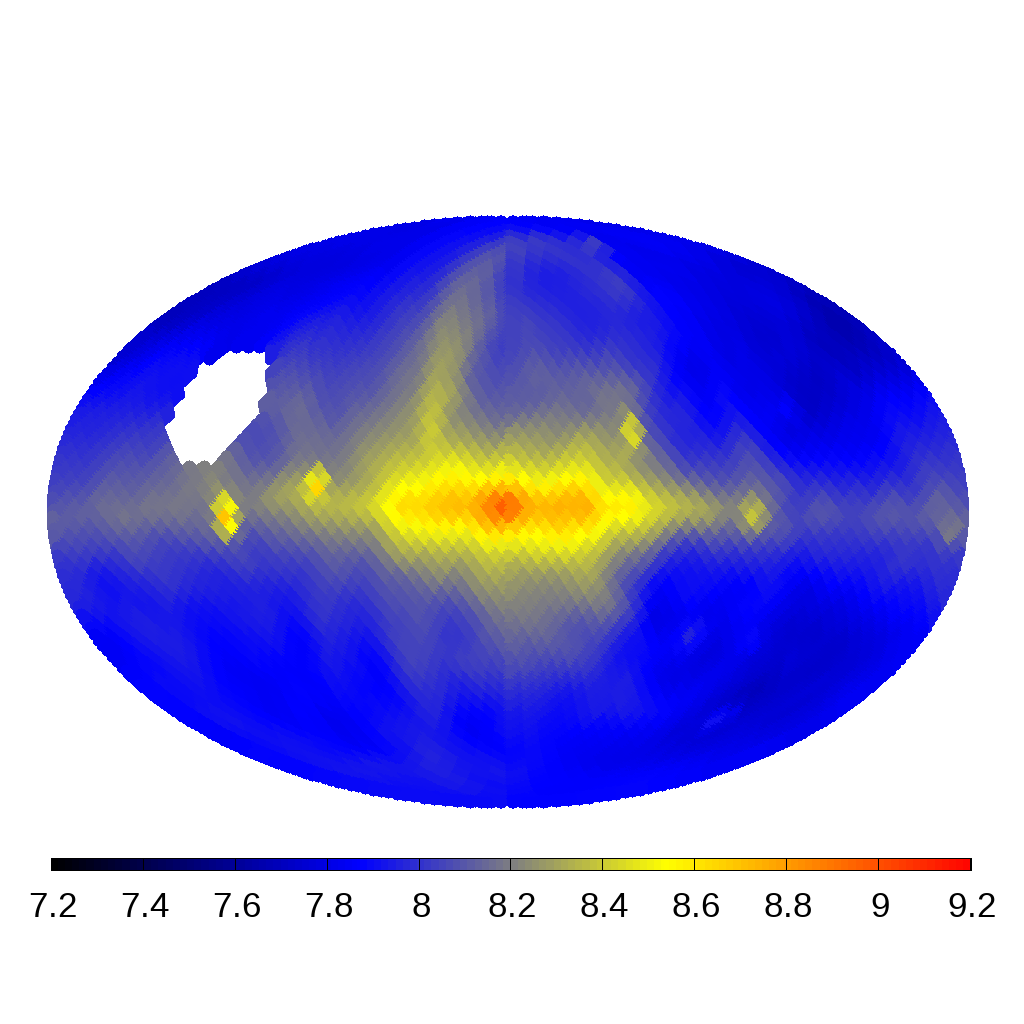}
 \includegraphics[width=0.32\textwidth]{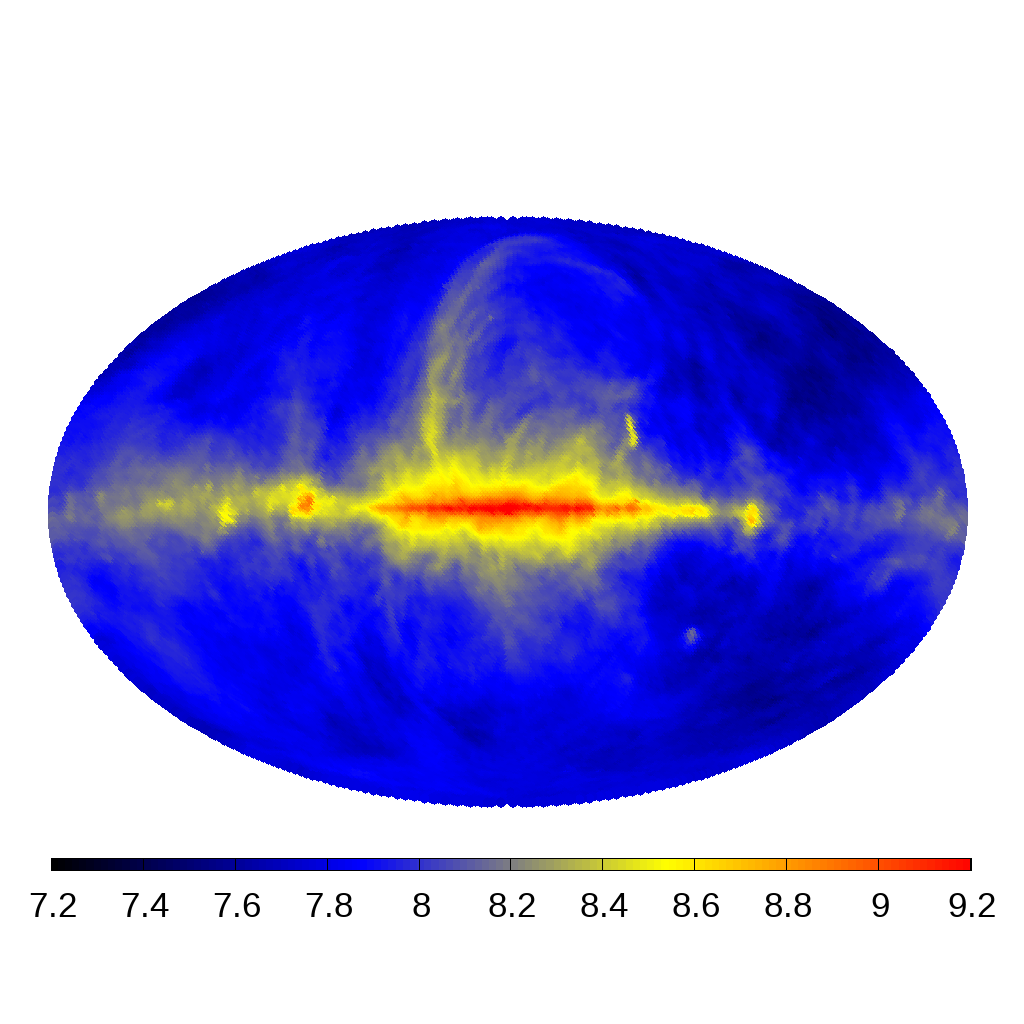}\\
 \includegraphics[width=0.32\textwidth]{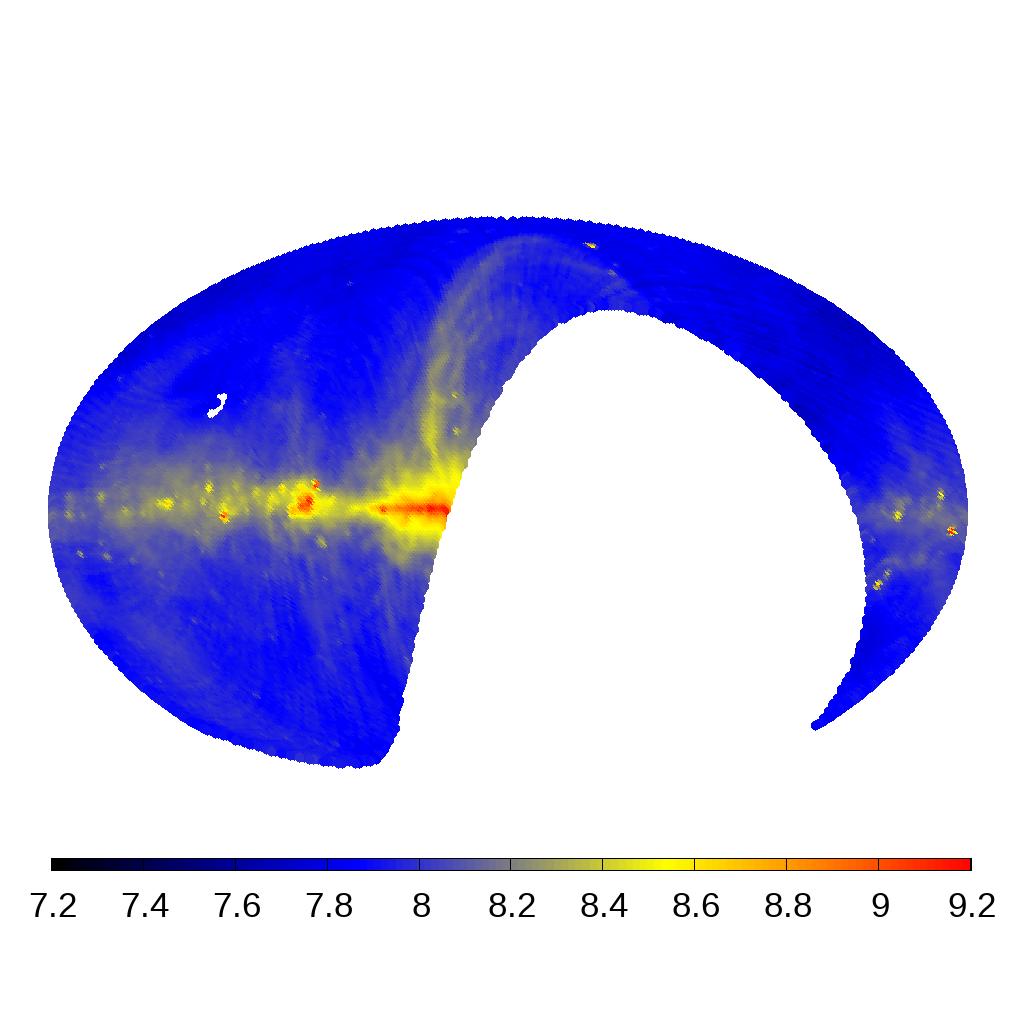}
 \includegraphics[width=0.32\textwidth]{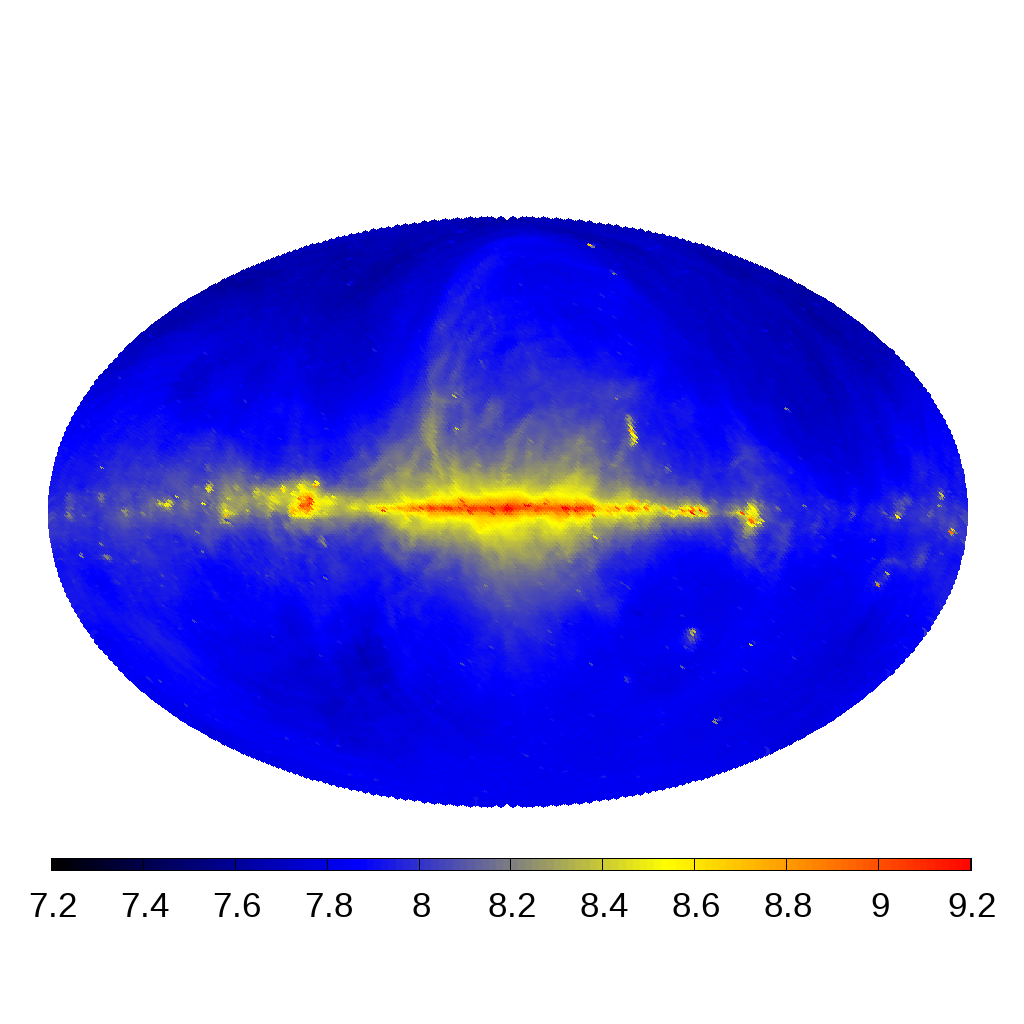}
 \includegraphics[width=0.32\textwidth]{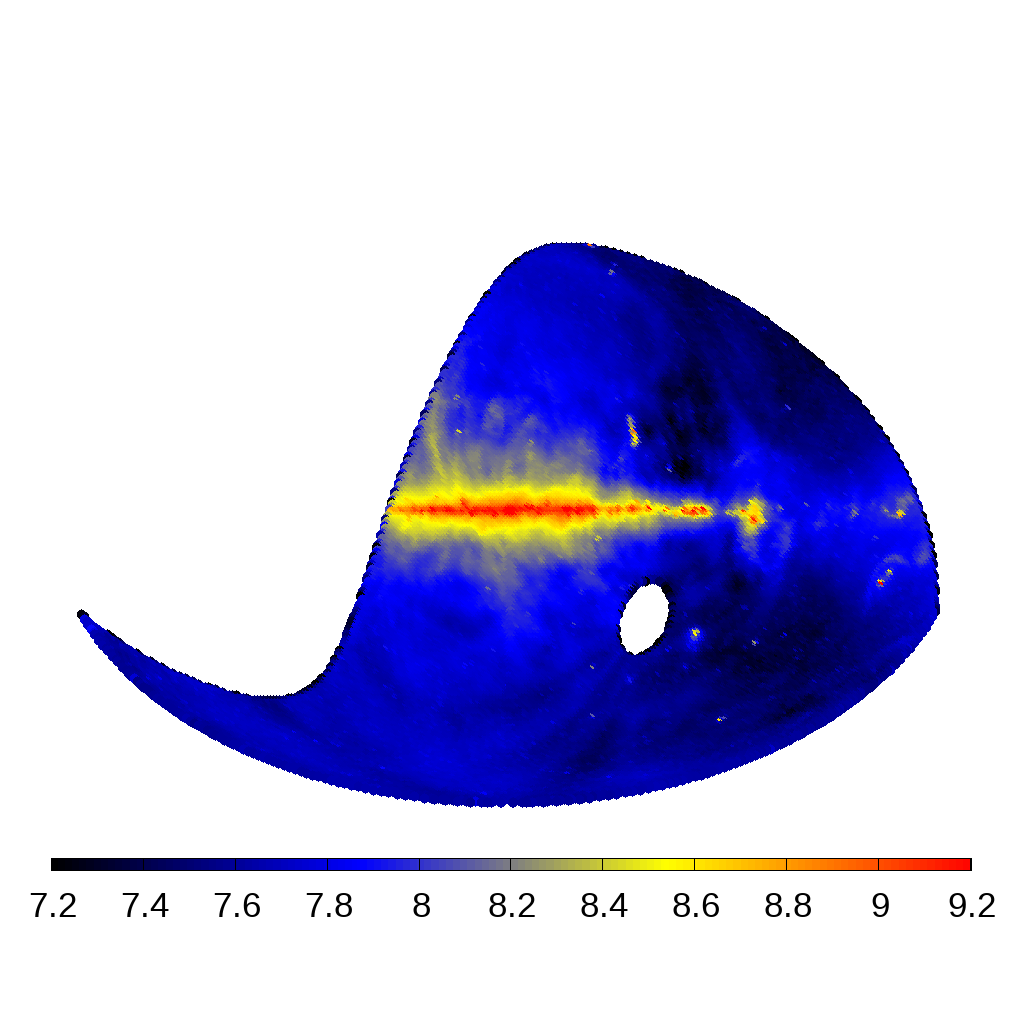}
\caption{ Maps of radio intensity considered in this analysis. Top row, from left to right: 22 MHz, 45 MHz, 408 MHz. Bottom row, from left to right: 820 MHz, 1420 MHz, 2326 MHz. We plot 
$\log(\mbox{T} \nu^{2.5})$ with $T$ measured in Kelvin and $\nu$ in MHz. All the maps are shown at the resolution $N_{\rm side}=64$.
}
\label{surveys}
\end{center}
\end{figure*}

Among the radio datasets available, we then selected the most sensitive ones trying to sample the whole radio frequency band (disregarding very-high radio frequencies which are largely dominated by the CMB).
Since we do not expect a high level of polarization in the extragalactic isotropic component of the sky, we focus on total intensity only.
Properties of the selected maps are summarized in Table~\ref{tab:data}. 
They include surveys of total intensity at 22, 45, 408, 820, 1420, and 2326 MHz. The corresponding
images are shown in Fig.~\ref{surveys} in the HEALPix~\cite{Gorski:2004by} format.
They have been obtained by regridding the original maps into a much finer grid (in order to avoid spurious projection effects) and then filling pixels in the HEALPix tassellation scheme (with a final linear size of pixels close to the original resolution of the survey).

\begin{table}[t]
\begin{center}
\resizebox{\textwidth}{!}{%
\begin{tabular}{|c|c|c|c|c|c|c|}
\hline
Frequency & Angular & rms Noise & Calibration & Zero-level & Fraction & Survey
\tabularnewline
 $[{\rm MHz}]$ & resolution & [K] & error & [K] & of Sky & reference
\tabularnewline
\hline
\hline
22 & $1.1^\circ \times 1.7^\circ/\cos Z$ & 3000 & 5\% & 5000 & 73\% &Roger et al.~\cite{DRAO:22}
\tabularnewline
\hline
\hline
45 & $5^\circ$ & 2300/300 & 10\% & 544 & 96\% & Guzman et al.~\cite{CHILE}
\tabularnewline
\hline
\hline
408 & $0.85^\circ$ & 1.2 & 10\% & 3 & 100\%&Haslam et al.~\cite{Haslam}
\tabularnewline
\hline
\hline
820 & $1.2^\circ$ & 0.5 & 6\% & 0.6 & 51\% & Berkhuijsen~\cite{DWING}
\tabularnewline
\hline
\hline
1420  & $0.6^\circ$ & 0.017 & 5\% & 0.2 (0.5) & 100\%& Reich et al.~\cite{Stockert1,Stockert2,VILLA}
\tabularnewline
\hline
\hline
2326 & $0.33^\circ$ & 0.03 & 5\% & 0.08 & 67\%& Jonas et al.~\cite{Jonas}
\tabularnewline
\hline
\end{tabular} }
\end{center}
\caption{Main parameters of surveys analysed in this work.}
\label{tab:data}
\end{table}

With a single ground-based telescope, it is not possible to survey more than about 70\% of the sky.
All the full-sky radio maps are thus obtained combining observations from different instruments:
this introduces non-uniform noise and zero-level offsets, which can affect the signal in a non-straightforward way when different observational strategies are combined (i.e., different patches observed by different telescopes located in different places).
We have therefore to be conservative in the error estimates.

The map at 22 MHz has been obtained in Ref.~\cite{DRAO:22} reanalyzing past observations performed with the Dominion Radio Astrophysical Observatory (DRAO) in the period 1965-1969.
It is the most complete map at the lowest radio frequency (below which the line-of-sight absorption becomes relevant).
The zero-level is reported to be 5000 K, while the rms noise is not quoted and we assume it to be 3000 K (however, this assumption does not significantly impact our results).

At 45 MHz, a northern and a southern surveys have been recently combined~\cite{CHILE} to form a nearly full-sky map with an angular resolution of $5^\circ$. 
The southern data were observed between 1982 and 1994 thorugh an array of 528 E-W dipoles with a beam of $4.6^\circ \times 2.4^\circ$ and a system noise of 300 K. 
The northern data were obtained in the periods of 1985-1989 and 1997-1999 by means of the Japanese Middle and Upper Atmosphere radar array with a beam of $3.6^\circ$ and a system noise of 2300 K.
The zero-level correction has bees estimated to be 544 K in Ref.~\cite{CHILE}.

A standard reference of full-sky radio map is the Haslam et al.~\cite{Haslam} map at 408 MHz. It is the composition of four different experiments, with data taking extending in the 60's and 70's. The northern celestial polar region and the Galactic anticenter were observed with two telescopes at the Jodrell Bank observatory. The remaining northern part was observed with the Effelsberg 100-m telescope, while the entire southern sky map was carried out with the Parkes 64-m telescope. 
Those observations are combined in a map of resolution of $0.85^\circ$, with an average zero-level estimated to $\pm 3$ K and a (conservative) noise of 1.2 K
\footnote{The map has been downloaded from~\cite{LAMBDA} where the data were processed to mitigate baseline striping and strong point sources.}.

A radio continuum survey of declinations between $-7^\circ$ and $+85^\circ$ at 820 MHz was conducted with the Dwingeloo telescope in the period 1965-1967~\cite{DWING}.
The offset level of the map has been derived to be $\pm 0.6$ K.
Adding up random (0.2 K) and systematic (0.3 K) errors, we considered an overall ``noise'' of 0.5 K.

The most recent map we will be considering is at 1.4 GHz, and is a combination of a survey of the south celestial hemisphere carried out with the Villa Elisa 30-m telescope~\cite{VILLA} and a northern
sky survey ~\cite{Stockert1,Stockert2} made with the 25-m Stockert telescope (the region of overlap is between declinations of $-10^\circ$ and $-19^\circ$).
The two surveys have a similar rms noise $\sim17$ mK. From absolute horn measurements, the zero-level accuracy is derived to be $\pm0.5$ mK. By comparing the map with the 408 MHz survey~\cite{Haslam} it can be however reduced to $\lesssim 0.2$ mK~\cite{Reich:1988}.

At frequencies much above the GHz, the extragalactic component becomes largely dominated by the CMB. Although we know the CMB temperature with very high precision, this adds further complications.
The last map we consider is at 2.3 GHz and consists of scanning observations with the HartRAO 26-m radio telescope~\cite{Jonas}
\footnote{We retrieved the map from Ref.~\cite{Platania:2003ta} where some boundaries between observed and non-observed regions have been smoothed with respect to the original map.}.
The 67\% of the sky was observed in ten periods of observations from 1980 to 1992, with a rms noise of 30 mK.
The observing strategy did not allow the measurement of absolute temperatures and the original map had an arbitrary zero-point. Independent observations at 2 GHz (then rescaled using a spectral index of $-2.75$) were exploited to obtain an absolute temperature scale for the survey. Then the zero level was estimated to be below 80 mK.
We caution however that such rescaling could add a systematic error in our estimate of the extragalactic background at this frequency.
Thus, although we include the map, it has to be considered with a special care.
 
In past works, a full-sky map at 150 MHz~\cite{Landecker:1970} has been used. However, this map was obtained in the 70's combining three surveys, one at indeed 150 MHz covering declination between $-25$ and $+25$, while the rest of the sky was filled by rescaling a 178 MHz (for the northern part) and a 85 MHz (for the southern part) survey by means of an average spectral index.
Since the rescaling is somewhat arbitrary (namely, derived in the overlapping regions and then extrapolated to the rest of the map), it might significantly bias the estimate of the extragalactic term, in particular concerning the 85 MHz map which is not adjacent in frequency.
We therefore decided not to include the 150 MHz map in our sample.

There are further possible artifacts arising in radio maps and associated to the scanning strategy (leading, e.g., to stripes in the map), which are not included in the error estimates mentioned above. In Ref.~\cite{Platania:2003ta} (which focuses on 408 MHz, 1.4 GHz, and 2.3 GHz maps), it was found that, for the majority of the pixels, the temperature variation due to such effects is $\delta T\lesssim 5\%$. In order to account for it, we thus add a systematic 5\% error to all maps.

\section{Synchrotron models}
\label{sec:Models}
As described in the Introduction, we model the total emission measured in the radio maps by means of three contributions:
\begin{equation}
T(l,b)=T_E+T_S(l,b)+T_G(l,b)
\end{equation}
where $l$ and $b$ denote the galactic longitude and latitude.
The isotropic temperature $T_E$ is a constant, and will be the results of our analysis, obtained through the fitting procedures described in Sect. \ref{sec:Fit}.
The technique adopted to estimate the contribution of discrete sources $T_S(l,b)$ will be described in the next Section.
In this Section, instead, we depict the model for the Galactic diffuse emission $T_G(l,b)$.

This is provided by free-free and synchrotron radiations. 
The first can be traced with spatial templates, for which we use the map of the $\mbox{H}\alpha$ recombination line of Ref. \cite{Finkbeiner:2003yt} with a free normalization coefficient. 
We find that the best-fit value of the coefficient is always compatible with zero:
this occurs because the thermal bremsstrahlung emission is mostly located in the region of the Galactic plane, that we actually mask, and peaks at higher frequencies than those adopted in our analysis.
Therefore, the free-free emission is not a critical component in our analysis, and we can include it by fixing a model (i.e., the template of Ref. \cite{Finkbeiner:2003yt}).

The main mechanism we need to analyze is therefore the synchrotron emission, generated by relativistic cosmic-rays electrons and positrons interacting with the Galactic magnetic field.
CR can be described as particles propagating in a fixed Galactic interstellar medium, i.e. by disregarding, in first approximation, the back-reaction of CRs themselves. 
The transport equation for a particle species $i$ given by~\cite{Berezinskii:1990}:
\begin{eqnarray}
{\partial n_i (\vec r,p,t) \over \partial t} = 
   && \vec\nabla \cdot ( \Dxx\vec\nabla n_i - \vec v_c \,n_i )
   + \ddp\, p^2 \Dpp \ddp\, {1\over p^2}\, n_i                  
   - {\partial\over\partial p} \left[\dot{p}\,n_i
   - {p\over 3} \, (\vec\nabla \cdot \vec v_c )\,n_i\right] + \nonumber \\
   && + q(\vec r, p, t)+\frac{n_i}{\tau_f}+\frac{n_i}{\tau_r}
\label{Eq:transport}
\end{eqnarray}
where $n_i(\vec r,p,t)$ is the number density per particle momentum, $q(\vec r, p, t)$ is the source term
(both in general function of position $\vec r$, momentum $p$ and time $t$), $\Dxx$ is the spatial diffusion coefficient along the regular magnetic field lines, $\vec v_c$ is the velocity of the Galactic wind, $\Dpp$ is the coefficient of the diffusion in momentum space, $\dot{p}$ is the momentum loss rate, and $\tau_f$ and $\tau_r$ are the time scales for fragmentation loss and radioactive decay, respectively. In the following we will actually disregard convection and reacceleration (i.e., $\Dpp$ and $\vec v_c$ will be set to zero) and we will assume steady-state.

The transport equation is solved numerically in 4D (3 spatial dimensions plus momentum) by means of the GALPROP code~\cite{Strong:1998pw}, in a cylindric box with boundaries along the Galactic plane at $x_h=20$ kpc and $y_h=20$ kpc and a vertical half-thickness $L$ (for which we will consider few different cases as described below).
Our main goal is to compute the equilibrium distribution of electrons and positrons below few tens of GeV (at larger energies they don't significantly contribute to the GHz emission). This
is given by the sum of primary electrons (we assume primary positrons to be negligible at these energies) and secondary electrons and positrons produced by decays of charged pions produced in the interactions of primary CRs (mainly protons) with the ISM.

The total synchrotron emissivity $j_{\rm syn}(\vec r,\nu)$ at a given frequency $\nu$ and position $\vec r$ is then obtained by folding the $e^+$$e^-$ number density $n_e(\vec r,E)$ with the total radiative emission power $P_{\rm syn}$ \cite{Rybicki}:
\be
j_{\rm syn}(\vec r,\nu)=\int dE\,P_{syn}(\vec r,E,\nu)\, n_e(\vec r,E)
\label{eqjsynch}
\ee
with:
\be
P_{\rm syn} (\vec r,E,\nu)= \frac{\sqrt{3}\,e^3}{m_e c^2} \,B(\vec r) F(\nu/\nu_c)\;,
\label{eqjsynch2}
\ee
and where $m_e$ is the electron mass, the critical synchrotron frequency in a magnetic field $B(\vec r)$ is defined as $\nu_c \equiv  3/(4\,\pi) \cdot {c\,e}/{(m_e c^2)^3} B(\vec r) E^2$,  and $F(t) \equiv t \int_t^\infty dz K_{5/3}(z)$ is the function setting the spectral behavior of synchrotron radiation.
Then the flux density can be simply estimated integrating the emissivity along the line of sight (since, in the range of frequencies and for the diffuse emission considered here, absorption is negligible) \footnote{The computation of the synchrotron emission is performed by means of the GALPROP routines.}.

The three key ingredients in the computation of the radio flux are therefore: the CR injection distribution, the propagation setup, and the total magnetic field.

\subsection{Source term}
\label{sourceterm}

Primary Galactic CRs are thought to be accelerated to GeV energies mainly by their scattering with the strong shock wave fronts produced by supernova remnants (SNRs) in the circumstellar medium. Neglecting discreteness and time variation effects (which are unlikely to be relevant at the low energies of interest here), the spatial part of the source function can be thus described following the mean 
SNR distribution in the Galaxy (which is also roughly traced by the gas distribution).
The latter can be in turn derived from pulsar surveys and we will consider the models of Ref.~\cite{Lorimer:2006qs,Strong:2010pr}.

Sources are mostly confined to the Galactic plane, and the distribution along the vertical direction $z$ is described to scale proportionally to $\exp(-|z|/z_s)$ with $z_s=0.2$~kpc.
We verified that taking larger $z_s$ (but still satisfying the condition $z_s< L$) does not significantly affect our results, see Sect.~\ref{comparison}.
In other words, for a reasonable range of $z_s$, the vertical scale of the equilibrium distribution is mostly dictated by the size of the diffusion box rather than the initial profile.

For what concerns the source function normalizations, we match the CR distributions after propagation to the local measurements at a reference energy.
In details, we set the local electron+positron flux $\phi_e=4\cdot 10^{-10}\,{\rm cm^{-2} sr^{-1} s^{-1} MeV^{-1}}$ at 34.5 GeV and local proton flux $\phi_p=5\cdot 10^{-9}\,{\rm cm^{-2} sr^{-1} s^{-1} MeV^{-1}}$ at 100 GeV.

For relativistic CR particles, the theory of first-order Fermi acceleration at astrophysical shocks predicts a power-law spectrum with spectral index at injection $\beta_{\rm inj} \sim 2$ in the limit of strong shocks~\cite{Blandford:1987pw}.  
Since the low-frequency radio emission approximately scales as $\nu^{-\alpha}$ with $\alpha\simeq2.5$, the spectral index of the equilibrium distribution of $e^+$$e^-$, which is roughly given by $\beta_e\sim 2\alpha-3$, needs to be around $\beta_e\simeq2$ in order to fit the data, as it has been noticed in Ref. \cite{Strong:2011wd}.
This is a rather hard spectrum if compared to the local measured one at higher-energy, $\beta_e\gtrsim3$, which is instead in fair agreement with higher frequency radio data pointing towards $\alpha\simeq3$ (see, e.g., Ref. \cite{Strong:2010pr}). On the other hand, the inference of the local GeV-sub GeV 
$e^+e^-$ spectrum is limited by the presence of poorly known solar modulation effects: at such low-energies, a value of $\beta_e\simeq2$ is therefore in principle fully viable.

Assuming the secondary production to be subdominant (we will comment about that in the next Section), the spectral index after propagation is approximately linked to the spectral index of injection by $\beta_e\simeq\beta_{\rm inj,e}+1+(\delta-1)/2$, where the $+1$ comes from energy losses and $\delta$ being the spectral index of the diffusion term. 
The primary spectral index of injection which is required to fit the data is thus rather hard: $\beta_{\rm inj,e}\lesssim1.5$.
This is an extreme value if thought in the context of shock acceleration, but a discussion on this subject is beyond the goal of this work, and in the following we will assume for definiteness $\beta_{\rm inj,e}=1.2$.

The CR spectrum at energies above 10 GeV is less affected by solar modulation. However, the two most recent experiments, Fermi-LAT and AMS-02, show some level of disagreement.
Since the AMS-02 data are still preliminary \cite{AMS}, we choose to tune our models to fit the Fermi-LAT spectrum~\cite{Ackermann:2010ij}. This translates, for the propagation setup chosen, into $\beta_{\rm inj,e}\simeq 2.3$ above the break.
This range of energy is not crucial for the low-frequency synchrotron emission considered in this work, but we verified with an explicit example that taking a model that fits the  AMS-02 data (up to 100 GeV, while at higher energy, the contribution of local sources can be significant) our results in the derived isotropic radio emission are unchanged. In the example, the spectral index above 7 GeV is taken to be $\beta_{\rm inj,e}=2.6$ (with the solar modulation potential being $\phi=900$ MV).

\begin{table}[t]
\begin{center}
\begin{tabular}{|c|c|c|c|c|c|c|}
\hline
code name & $L$   & $D_0$                         & $\beta_{\rm inj,nuc}$&$\beta_{{\rm inj},e}$& $B_0$ &color coding\\
 & [kpc] & [$10^{28}\,{\rm cm^2s^{-1}}$] & &  & [$\mu$G] &~
\tabularnewline
\hline
\hline
L1&1&0.75 &1.80/2.3&1.20/2.3& 12&\color{red} red
\tabularnewline
\hline
\hline
L2&2&1.7&1.80/2.3&1.20/2.35&8.0& \color{blue} blue
\tabularnewline
\hline
\hline
L4&4&3.4&1.80/2.3&1.20/2.35&6.0/7.0 & \color{green}green
\tabularnewline
\hline
\hline
L8&8&5.8&1.80/2.3&1.20/2.35&4.6/4.7& \color{orange}orange
\tabularnewline
\hline
\hline
L16&16&8.0 &1.80/2.3&0.5/2.35&4.0/4.7& \color{cyan}cyan
\tabularnewline
\hline
\hline
L25&25&8.1&1.80/2.3&0.5/2.35&3.9&\color{maroon}maroon
\tabularnewline
\hline
\hline
L40&40&8.3 &1.80/2.3&0.5/2.35&3.8& \color{brown}brown
\tabularnewline
\hline
\end{tabular}
\end{center}
\caption{Benchmark models of CR propagation. The diffusion coefficient is described by \mbox{$D_{xx}=D_0\,(\rho/\rho_0)^{0.5}$}, where $\rho$ is the rigidity and $\rho_0=4$~GV.
The spectral index $\beta_{\rm inj,nuc}$ for nuclei has a break at 9 GeV (and the two
values below/above the break are reported), while the spectral index $\beta_{\rm inj,e}$ of electrons has a break at 4 GeV. The adopted models assume no reacceleration or convection.
The normalization $B_0$ of the random magnetic field-strength is reported for 
models $a$/models $b$, i.e. models with $z_B=L$/$z_B=2$~kpc. When only one value is reported, it refers to model $a$.}
\label{tab:BM}
\end{table}

\subsection{Propagation setup}
\label{sec:propagation}

We are not interested in performing a full scan to estimate confidence intervals for the propagation parameters, but rather, we want to investigate how they can impact the high-latitude radio emission.
To this aim we consider the simplest model that can accommodate CR data, i.e., plain diffusion (see, e.g., Ref. \cite{Strong:2007nh} for a review).
This is also motivated by the fact that reacceleration, which is sometimes included to improve the fit to CR data, seems to be in tension with radio maps at low frequencies.
Indeed models with reacceleration tend to increase the flux of secondary $e^+e^-$ at low energy. 
Once the proton spectrum is fixed to fit local proton data, and even taking an extremely hard spectrum for the primary electrons (as mentioned in the previous Section), the contribution of secondaries makes the final spectrum of $e^+e^-$ too soft, as already noticed in Ref. \cite{Strong:2011wd}.
This is not the case for pure diffusion, where the secondary contribution is less important.
The vertical scale of the diffusion box is the propagation ingredient with the largest impact on the high-latitude behaviour of the CR emission.
We will consider a conservative broad range of $L$,  from 1 to 40 kpc. The extreme values might be strongly constrained by CR data like unstable secondaries.
In particular, the cases with $L\gg10$ kpc are probably not realistic, but we nevertheless include them to confidently assess the maximum high-latitude contribution that can come from Galactic emissions.

\begin{figure*}[t]
\begin{center}
 \includegraphics[width=0.32\textwidth]{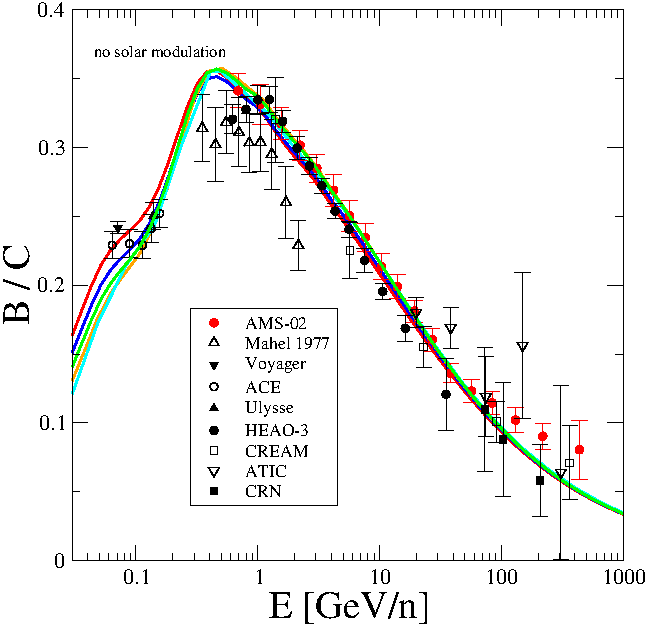}
\includegraphics[width=0.32\textwidth]{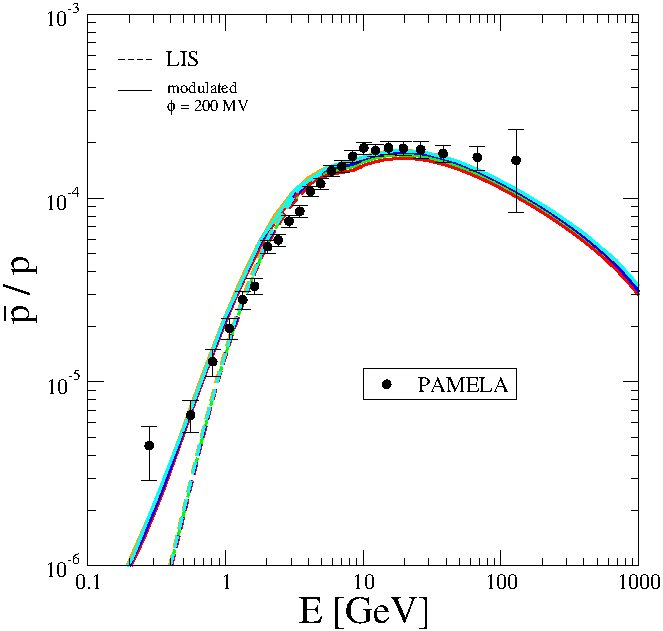}
 \includegraphics[width=0.32\textwidth]{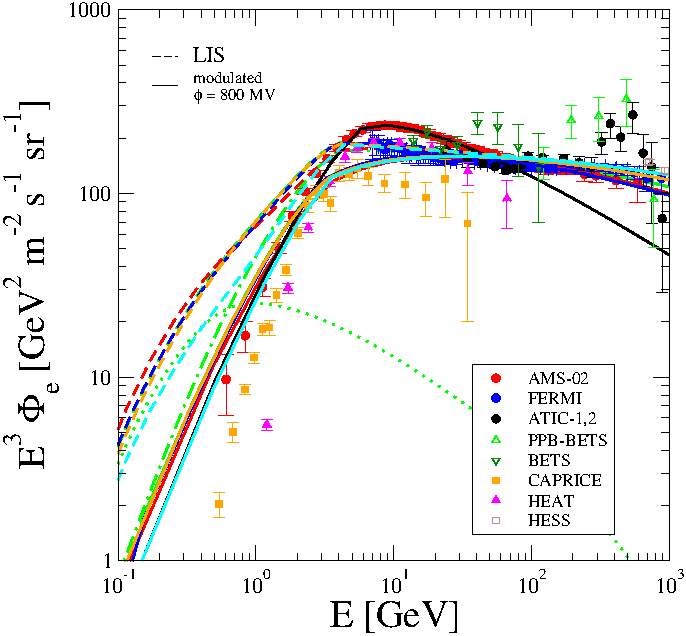}
\caption{Comparison between the predictions of the CR models adopted in this work and the measured local spectra of B/C (left), $\bar p/p$ (central), and $e^++e^-$ (right). The curves adopt the color coding of Table \ref{tab:BM} to differentiate among the models. The model L25 and L40 basically overlap with L16 and are not shown.
A model of $e^++e^-$ with $L=4$ kpc, which fits the AMS-02 results~\cite{AMS} (see text), is shown in black.
For illustrative purposes, in the right panel, we show primary and secondary components with dashed-dotted and dotted lines, respectively, in the L4 model.} 
\label{fig:CR1}
\end{center}
\end{figure*}

We follow the propagation setups described in Ref. \cite{Strong:2010pr} (see their Table 1 for plain diffusion), which have been built to be in (approximate) agreement with CR and gamma-ray data, and extend them to the broader range of cases we consider here. The parameters adopted in the benchmarks models used in our analysis are quoted in Table~\ref{tab:BM}. 
Fig.~\ref{fig:CR1} shows the various models predictions together with available CR data, to demonstrate that they are able to properly explain CR observations\footnote{Some of the data have been downloaded from \cite{Maurin:2013lwa}.}.

\subsection{Magnetic fields}
\label{sec:magnetic}

Nowadays, thanks to great efforts in measuring its various components, and despite many issues are still open, we can draw a reasonable description of the Galactic magnetic field. 
In this work, the recent estimate presented in Refs.~\cite{Jansson:2012rt,Jansson:2012pc} will be our reference model (see Refs. \cite{Pshirkov:2011um,Sun:2010sm,Sun:2007mx} for different models).
It includes both large-scale coherent and small-scale random fields. The first is composed by a disk, a toroidal-halo and an ``X-field'' components, while a disk and a halo components make up the random part.
A striated component (with orientation aligned with the regular field, but with strength and sign varying on small scales) is also included.

The model has been constrained by a simultaneous fit of extragalactic Faraday rotation measures and of the 22-GHz WMAP7 polarized and total intensity synchrotron emission maps~\cite{Gold:2010fm}.
We can safely take the results and independently model the CR sources and propagation since we analyze maps at much lower frequency, involving $e^+e^-$ at much lower energy: this implies that there is no interference between the assumptions in Ref. \cite{Jansson:2012rt,Jansson:2012pc} and the CR models we are going to consider.
The magnetic field model described above has then been implemented in the GALPROP code.

Although the model of Ref. \cite{Jansson:2012rt,Jansson:2012pc} is very accurate, we have to take into account that there are still possible systematics (namely, one can consider a different viable model) which may slightly modify some of the components. 
In particular, since we are only interested in mid-high latitudes emissions, the halo component of the magnetic field (which is mainly given by the random term) is the most relevant.
In order to have a robust analysis we will include some flexibility for this term.
We will keep the coherent component of the model from Refs.~\cite{Jansson:2012rt,Jansson:2012pc} as fixed, while the random component will be described by a double-exponential law: $B(R,z)=B_0\,\exp[-(R-R_T)/R_B]\,\exp(-|z|/z_B)$, where we set $R=\sqrt{x^2+y^2}$ ($x$ and $y$ being orthogonal coordinates in the Galactic plane), $R_T=8.5$ kpc, and $R_B=30$ kpc, $B_0$ is determined through our fit, and we will considered few different benchmark cases for $z_B$. 

A first set of models is such that, for any fixed propagation setup, the scale height of the random magnetic field is taken to be equal to the vertical scale of the diffusion box: i.e. $z_B=L$. These models are labeled as models $a$. This assumption comes from the fact that diffusion is indeed due to scattering of CR particles with hydro-magnetic turbulences. However, a mismatch between these two scales is possible, and in particular in connection with the spectrum of turbulences: e.g., for isotropic turbulences in the quasi-linear approximation, one typically has $z_B=\delta\,L$ with $\delta$ being the spectral index of the diffusion coefficient. We therefore investigate also models with $z_B<L$.
Specifically, we fix the scale $z_B$ to 2 kpc for the cases with $L=4,8,16$ kpc, and label them as models $b$.

It is clear that, in the estimate of the extragalactic emission, the $z$-scaling represents the main source of uncertainty related to the magnetic field modeling.
As we will see, it actually has only a very minor impact making some appreciable but nevertheless mild differences only in the $L=16$ kpc case, i.e., when $L\gg2$ kpc. 
Therefore we don't expect other (less crucial) modifications in the model of $B$ to significantly affect our conclusions.

\section{Fitting procedure}
\label{sec:Fit}

\begin{figure*}[t]
\begin{center}
 \includegraphics[width=0.32\textwidth]{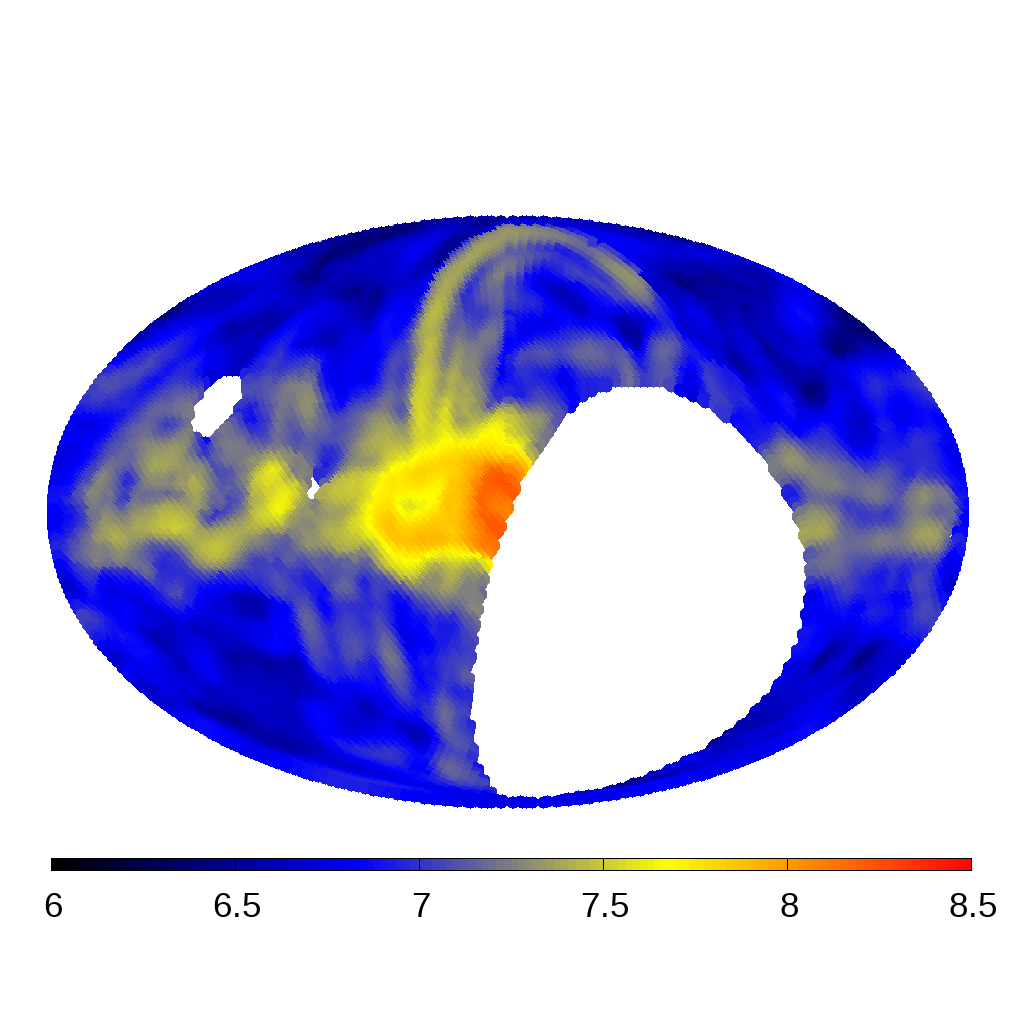}
 \includegraphics[width=0.32\textwidth]{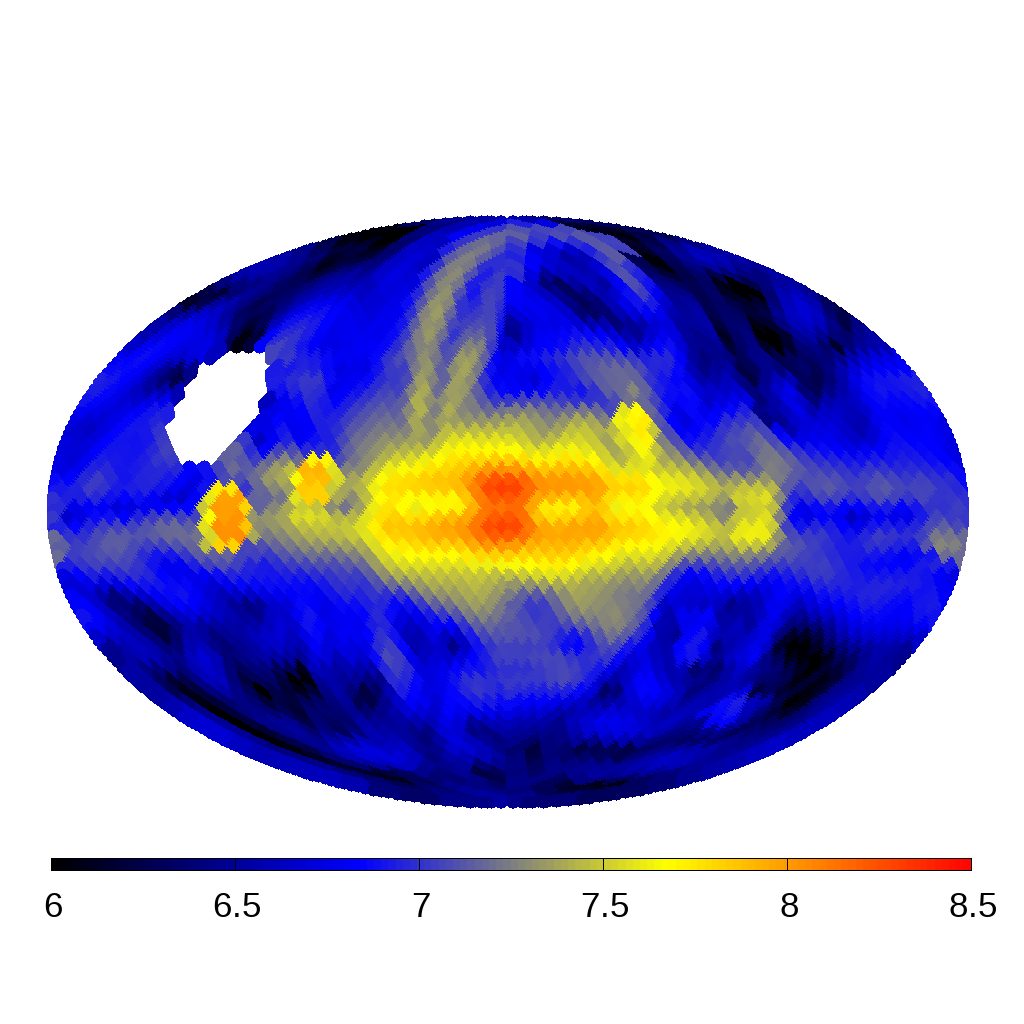}
 \includegraphics[width=0.32\textwidth]{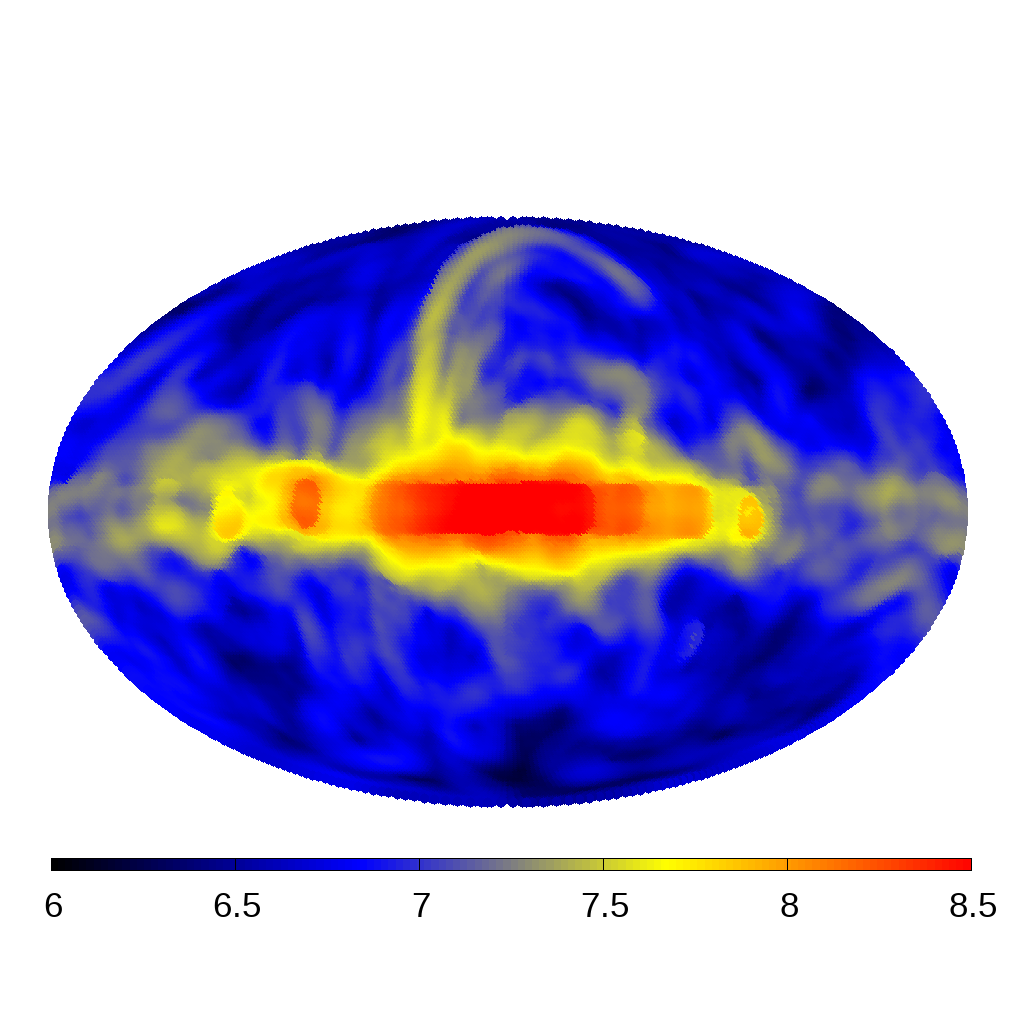}\\
 \includegraphics[width=0.32\textwidth]{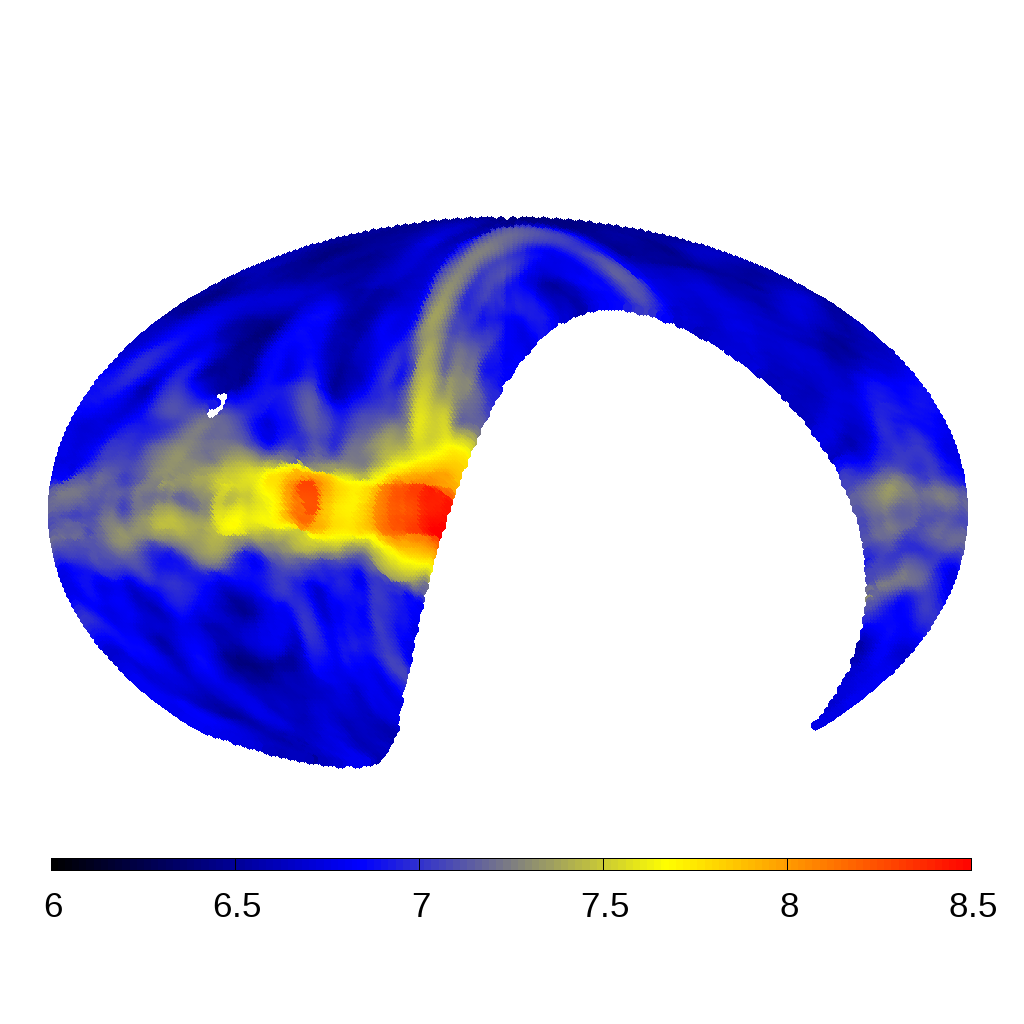}
 \includegraphics[width=0.32\textwidth]{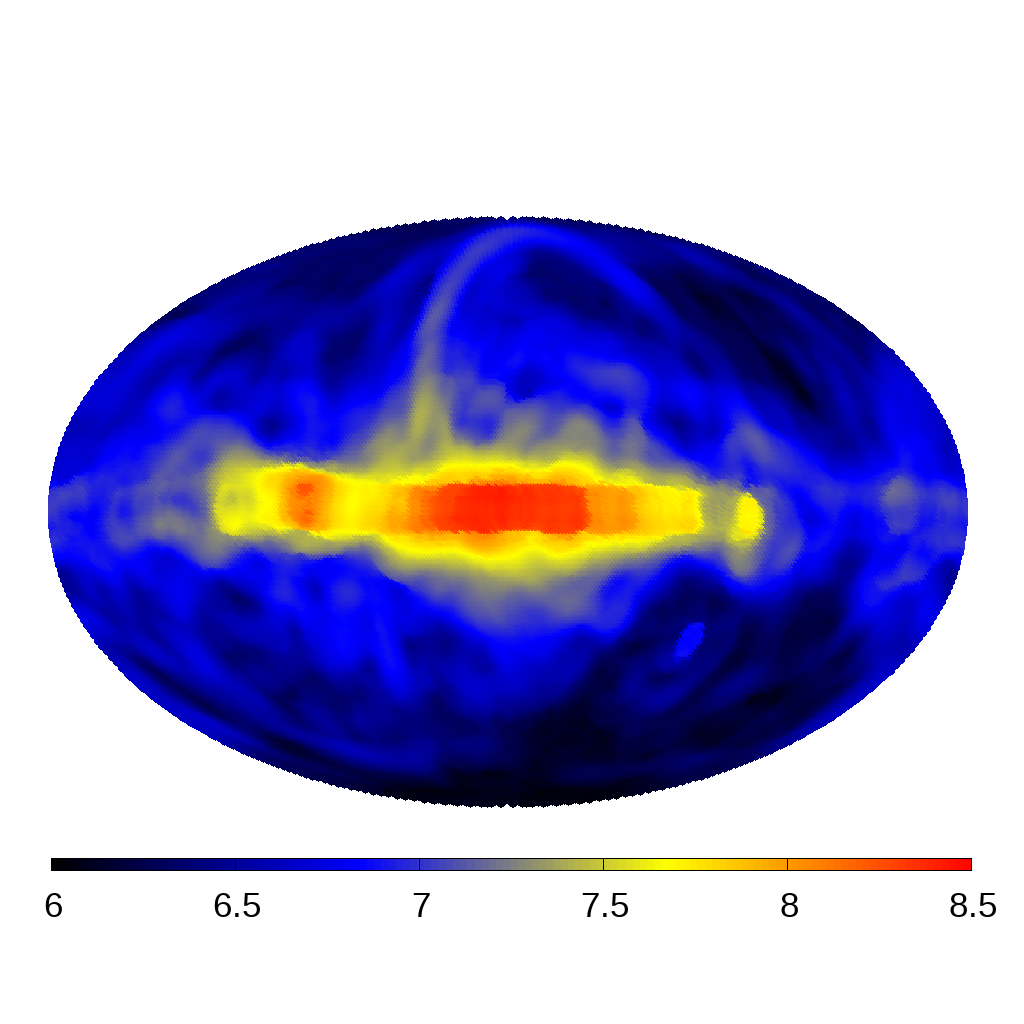}
 \includegraphics[width=0.32\textwidth]{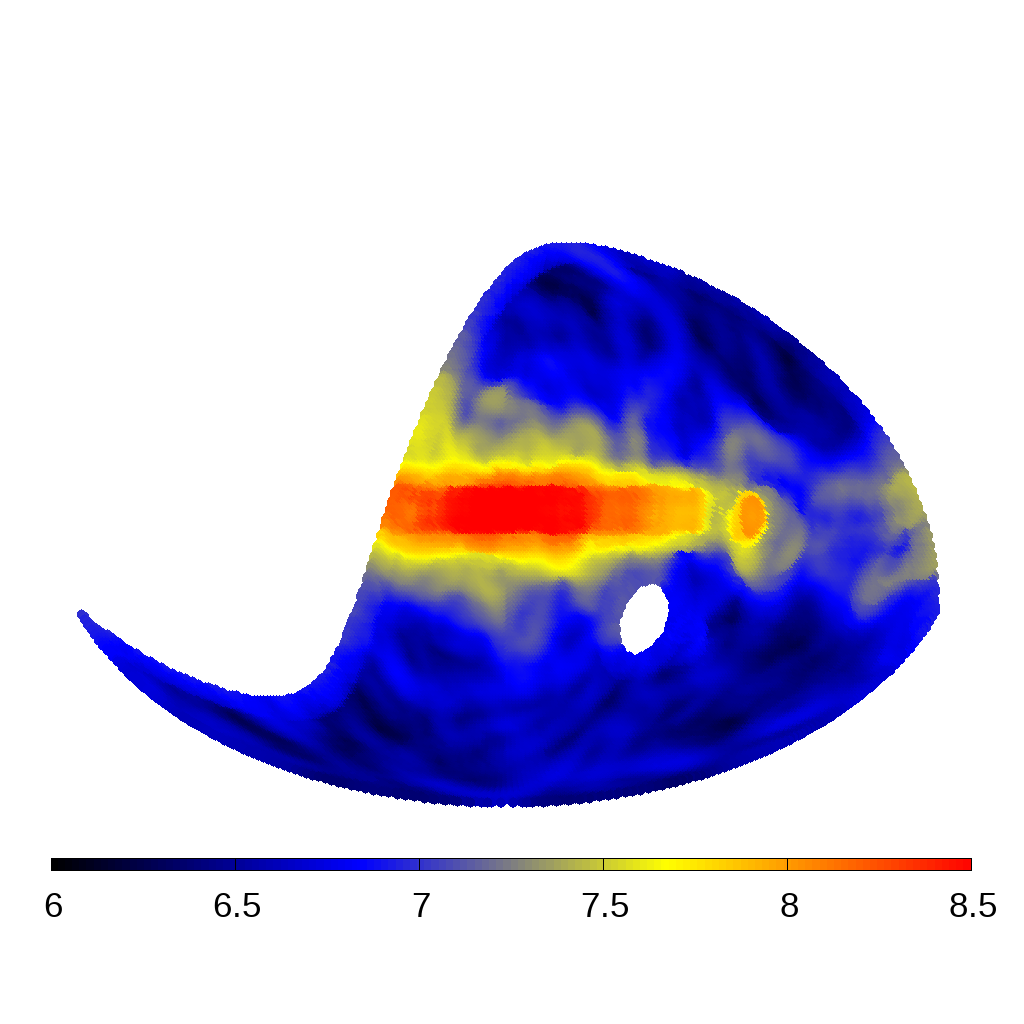}
\caption{Maps of fluctuations $\sigma_i$, as defined in Sect.\ref{sec:Fit}. The top row, from left to right refers to: 22MHz, 45 MHz,408 MHz. The bottom row reports, from left to right: 820 MHz, 1420 MHz, 2326 MHz. We plot $\log(\mbox{T } \nu^{2.5})$ with T in [K] and $\nu$ in MHz. All the maps are shown at the resolution $N_{\rm side}=64$.}
\label{sigma}
\end{center}
\end{figure*}

The observed radio sky at frequencies below a few GHz is the sum of the isotropic extragalactic background
and the Galactic emission. The former contains the CMB monopole. In the rest of the analysis we subtract the CMB brightness temperature from the radio maps that we analyze (taking $T_0=2.72548 \pm 0.00057$ K~\cite{Fixsen:2009cmb} for the CMB thermodynamic temperature).
The Galactic emission includes the diffuse synchrotron radiation produced by cosmic-rays electrons spiraling in the Galactic magnetic field (GMF), as mentioned in Sect.~\ref{sec:Models}. Therefore radio maps carry information both on the propagation of the interstellar CR electrons and the structure of the GMF. As discussed in Sect. \ref{sec:magnetic}, the GMF is composed by a large scale regular field, with a coherence length of order of $\mathcal{O}(\mbox{kpc}),$ and a random component, which varies its direction and intensity on scales of $\mathcal{O}(100\mbox{ pc})$ or smaller.
The strength of the regular GMF inferred by Faraday Rotation measurements of extragalactic sources and radio polarization data, is not enough to explain the observed intensity of radio maps, at least at intermediate and high galactic latitudes. This suggests that the bulk of the diffuse Galactic synchrotron emission is induced by the random component of the GMF.
The stochastic nature of the random magnetic field introduces a variance in the intensity of the radio maps at scales larger than its coherence length. Obviously, this is not taken into account by simply modeling the random field with its RMS value.
For this reason is appropriate to compare the data with a model of the Galactic radio emission on scales where the fluctuations due to the turbulence of the random GMF have been averaged.
Moreover, the size of these fluctuations can be directly inferred from the variance of the intensity of the radio map in a given angular region. This allows to estimate the level of accuracy at which a model of Galactic synchrotron emission is expected to reproduce the observations.

It is not obvious which is the most correct angular scale at which this course-graining of the radio maps should be performed. Indeed, the observed radio flux arises from the contribution of all the electrons along the line of sight emitting synchrotron radiation. Therefore the coherence length of the random magnetic field is mapped into different angular scales for emitting volumes located at different distances (for the angular power spectrum of Galactic magnetic turbulences in the radio frequency range considered here, see \cite{Regis:2011}).
Typically one expect that the turbulence of the magnetic field impacts the radio maps an scales of tens of degrees or below.

In this analysis we have averaged the radio maps on an angular scale of about 15 degrees. As we will see later, this corresponds to a conservative assumption for what concerns our results. We also repeat the analysis for a smaller angular scale in Sect.~\ref{comparison}. 

More specifically, in the following we downgrade the radio maps described in Sect.~\ref{sec:Surveys} and the models of Galactic synchrotron emission presented in Sect.~\ref{sec:Models} to a HEALPix resolution $N_{\rm side}=4,$ which corresponds to a total number of pixel in the map $N_{\rm pix}=192$, with a size of $14.7$ degrees each.
The observational data and the models are then compared, as we discuss in more detail in the following Sections, computing the $\chi^2:$
\begin{equation}
\chi^2=\sum_i \frac{ (T_i^{\rm data} -T_i^{\rm model})^2}{\sigma_i^2},
\label{eq:chi2}
\end{equation} 
where the index $i$ runs over the pixels.
The maps of fluctuations $\sigma_i$ are determined by combining experimental uncertainties and fluctuations due to magnetic turbulences.
For each survey, considered at its original angular resolution reported in Table~\ref{tab:data}, we estimate the variance $\sigma_{i}^B$ induced by the turbulence of the magnetic field in each pixel by taking the temperature variance in an angular region of diameter $14.7$ degrees centered around the pixel.
Then, for each pixel in the map, we combine (summing in quadrature) the experimental errors $\sigma_{i}^{\rm exp}$ summarized in Table~\ref{tab:data}, including the calibration error and the rms noise, but not the zero level uncertainty, since this corresponds to an overall rescaling of the survey. 
The maps of fluctuations $\sigma_{i}=\sqrt{(\sigma_{i}^B)^2+(\sigma_{i}^{exp})^2}$, downgraded to $N_{\rm side}=4$, are shown in Fig.~\ref{sigma}.

As mentioned in Sect.~\ref{sec:Models}, the temperature of the model $T(l,b)$, is a linear combination of the isotropic background, Galactic synchrotron radiation, free-free emission and contributions from sources (all modeled with the same resolution $N_{\rm side}=4$). We introduce coefficients weighting the different components: these coefficients are then fixed by minimizing the $\chi^2$ of Eq. (\ref{eq:chi2}).
The synchrotron radiation and thermal bremsstrahlung have been described in Sect.~\ref{sec:Models}.
In addition to their contributions, also unsubtracted point-like and extended sources make up the total radio emission.

Galactic point sources are mostly located in the Galactic disk and could account for a large fraction of the emission at low galactic latitudes. 
For this reason in our analysis we mask the region $|b|<10$ degrees.

Several ring-like extended features, called radio loops, have been detected in radio maps (see for instance Ref. \cite{Hasmal-Kahn, Berkhuijsen, Berkhuijsen-Haslam}).
The most prominent one, Loop I (also known as North Polar Spur), is located in the northern hemisphere, extends over 100 degrees and can be clearly recognized in the maps of Fig. \ref{surveys}. These emissions are also observed at other frequencies, including microwave, X-rays and gamma-rays.
Radio loops are commonly associated with shells of old SNRs. In particular, the four major radio loops directly visible with radio observations (Loop I-IV) are believed to be originated from nearby sources, $r\lesssim 0.5-1$ kpc. More distant old SNRs also contribute to the Galactic radio emission. An attempt to model the emission from this population of sources has been done in Ref. \cite{Mertsch:2013pua}. As explained above, distant Galactic radio sources impact the radio sky at small latitudes, therefore we expect that masking the region $|b|<10$ degrees should remove most of the emission from non-local shells of SNRs.

Still, extended local sources and other high-latitude sources should be taken into account in the modeling of the radio sky. We adopt two different methods to this purpose, namely we either mask or model such sources. The first method is discussed in Sect. \ref{sect:mask}, where we describe how the masked region is found by means of iterative procedures. In this case, in the non-masked part of the map, we assume no contribution from sources. 
In Sect. \ref{sect:template}, instead, we model the emission from sources with a spatial template, which is obtained from a polarization map.

Before explaining in more details these two options, we shall comment about how we have chosen the models of Galactic synchrotron emission.
As mentioned in Sect. \ref{sec:Models}, although we set the CR and magnetic field parameters according to data, there are still two parameters, which are crucial for our purposes, that are poorly constrained: the electrons spectral index at energies below few GeVs and the normalization of the random magnetic field $B_0$.
The former basically determines the spectral index of the diffuse synchrotron radiation at radio frequencies, but the presence of solar modulation limits its knowledge at the local position. 
The normalization of the emission, on the other hand, strongly depends on the intensity of the magnetic field.

In principle, one can include those two parameters in the fit, but this would involve a large number of GALPROP runs to find the best-fit values.
We instead fix the two parameters to benchmark values which, approximately, already provide a good fit. Then, at all frequencies, we introduce a coefficient in the fitting procedure, which normalizes the Galactic template.
Ratios of coefficients at different frequencies different from one would account for a mismatch in the electron spectral index, while their overall absolute normalization allows to adjust for a mismatch in the magnetic field intensity. We will see that the best-fit values for such coefficients come out to be very close to one, meaning that we are indeed using synchrotron templates which agree with data.
We also notice that allowing the normalization to vary for different maps, we are also effectively taking into account possible offsets due to different calibrations or different sky coverages among the various experiments.

\subsection{Masks}
\label{sect:mask}

\begin{figure*}[t]
\begin{center}
 \includegraphics[width=0.32\textwidth]{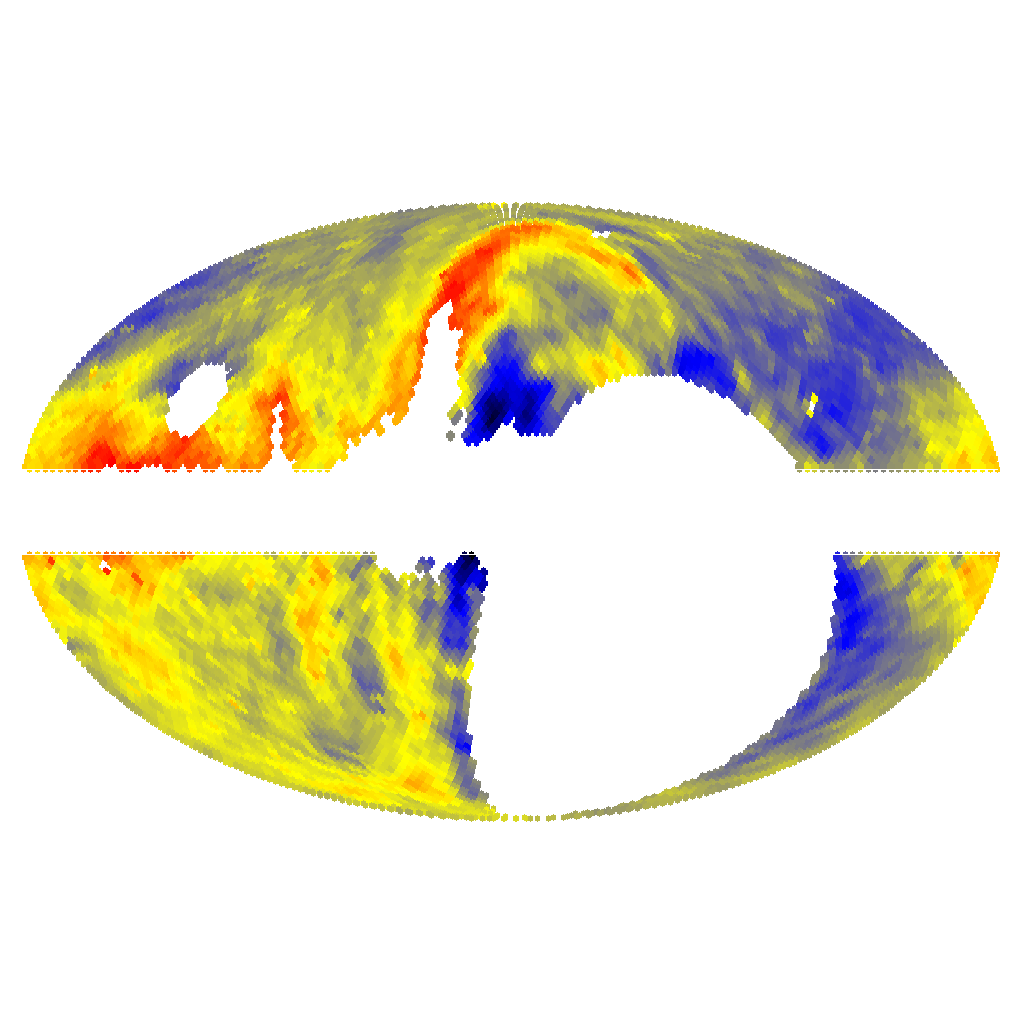}
 \includegraphics[width=0.32\textwidth]{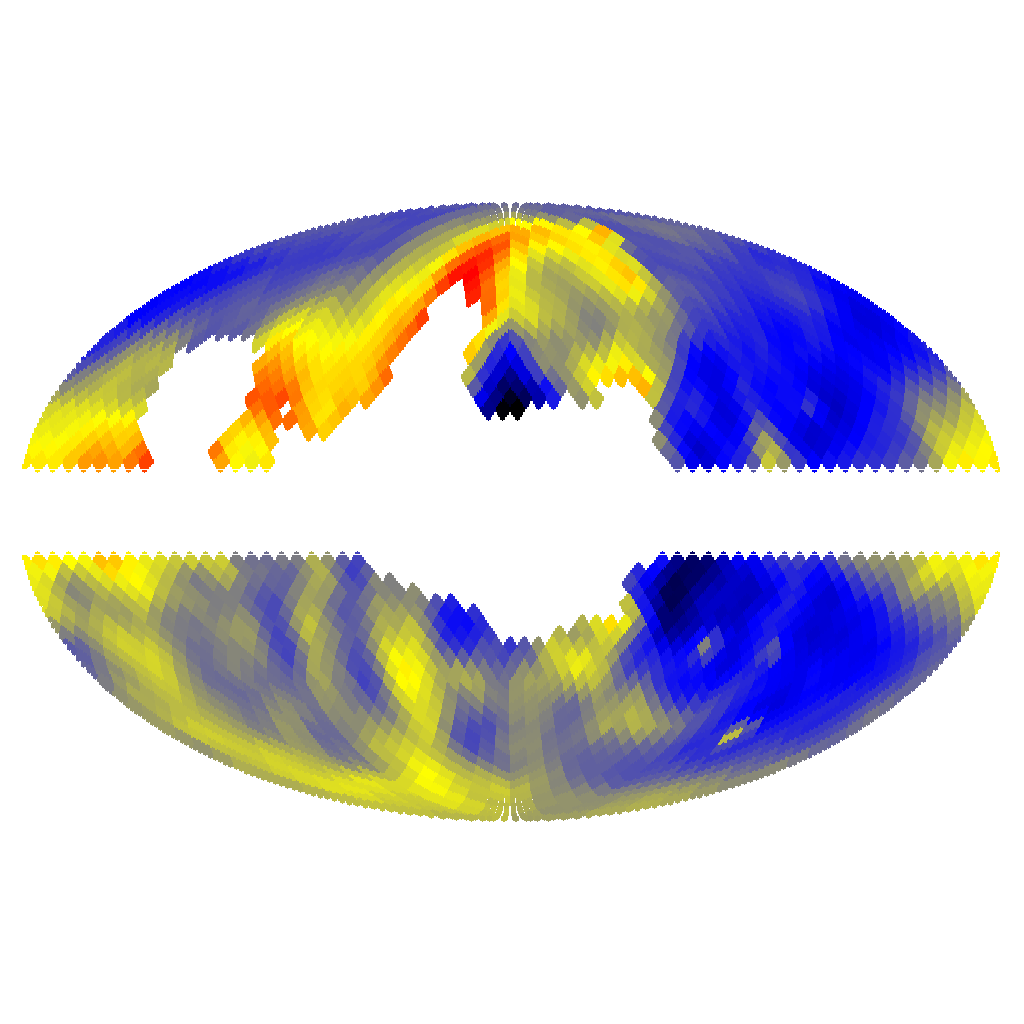}
 \includegraphics[width=0.32\textwidth]{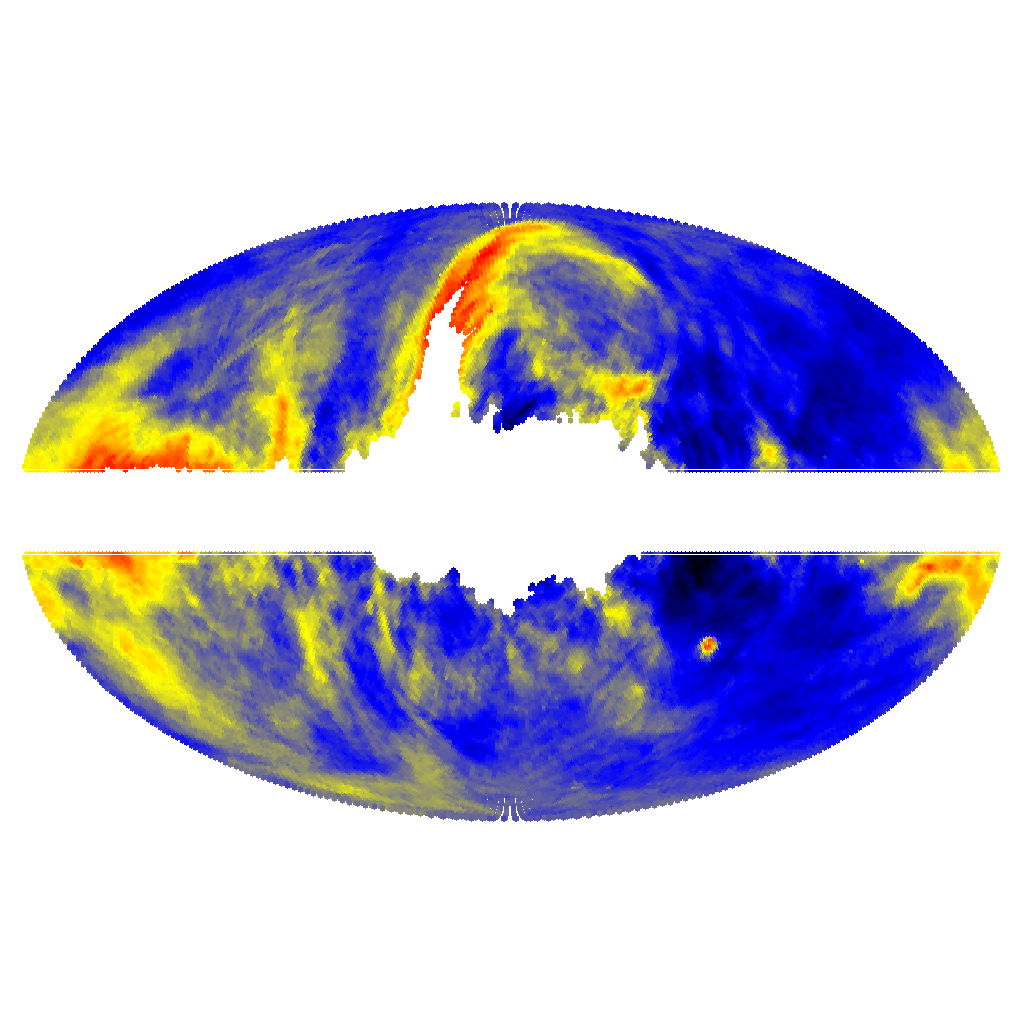}\\
 \includegraphics[width=0.32\textwidth]{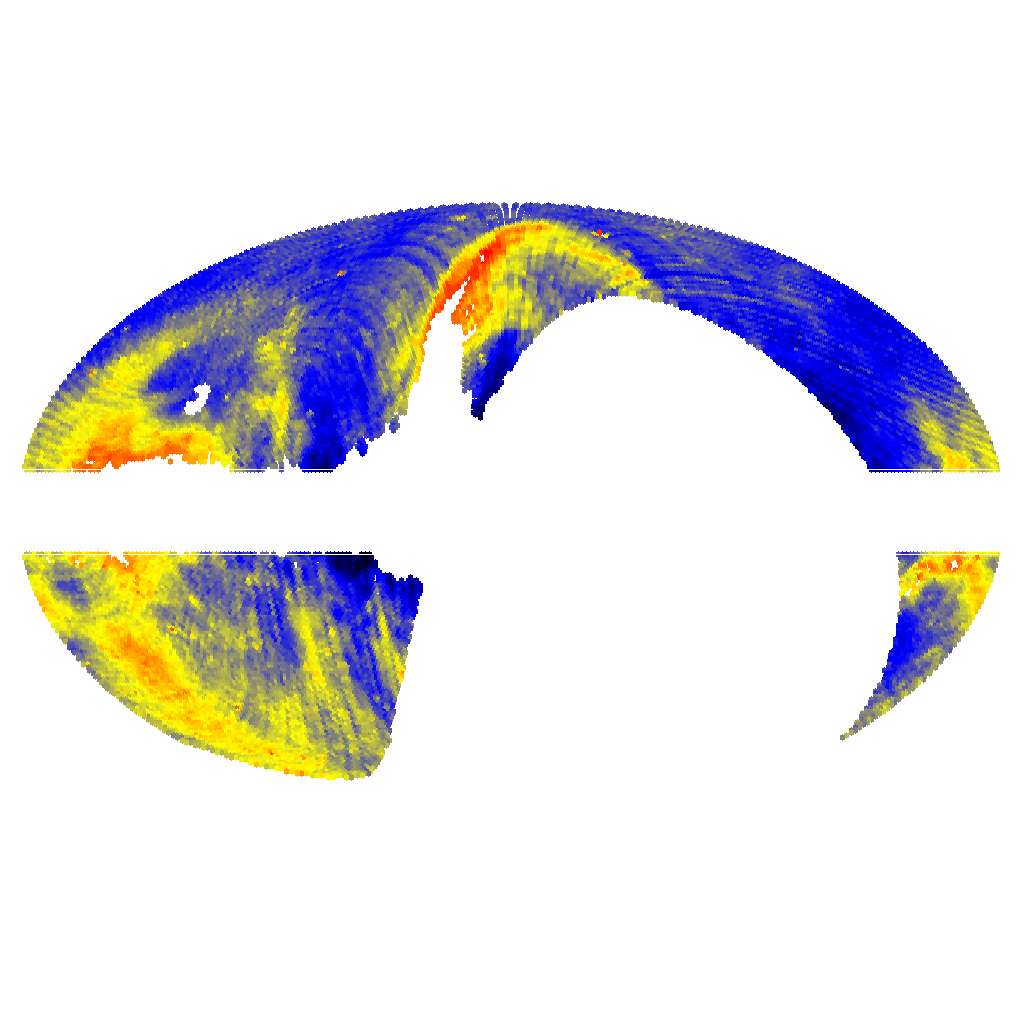}
 \includegraphics[width=0.32\textwidth]{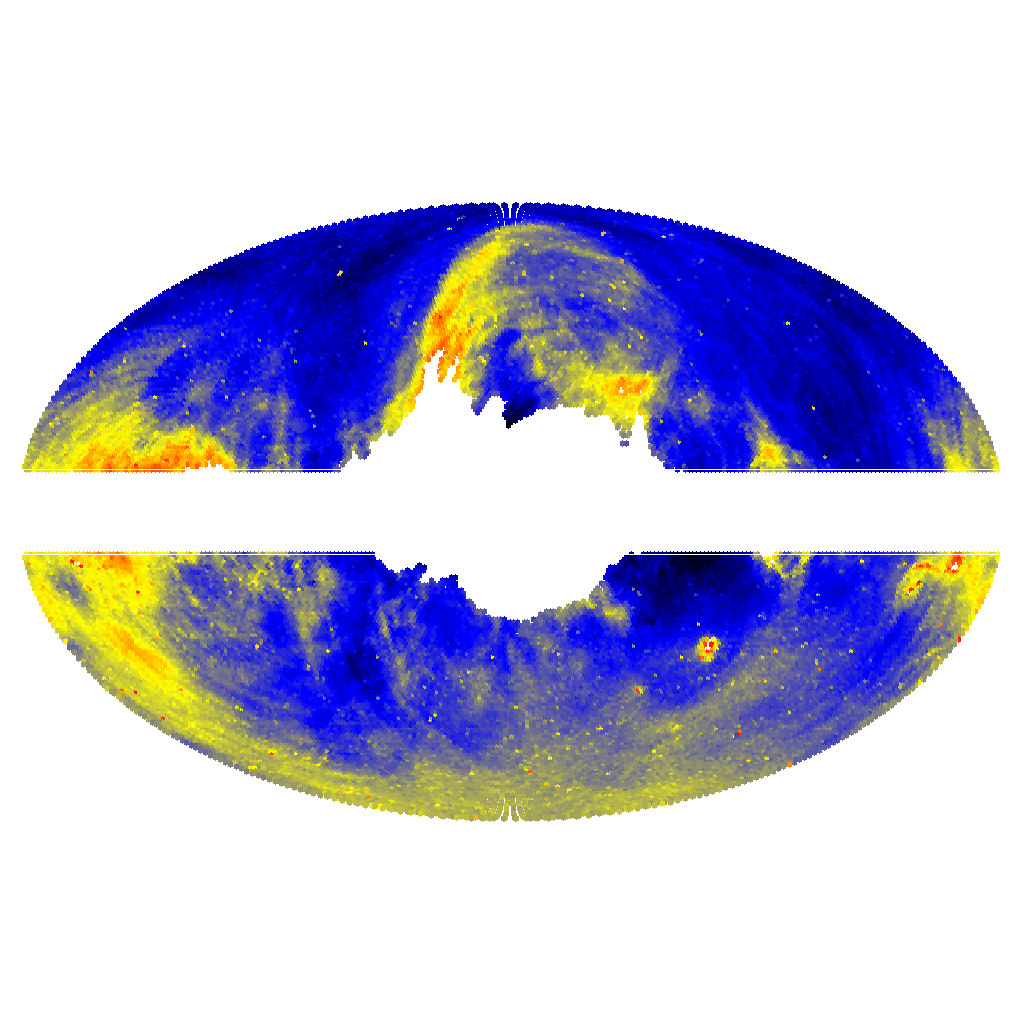}
 \includegraphics[width=0.32\textwidth]{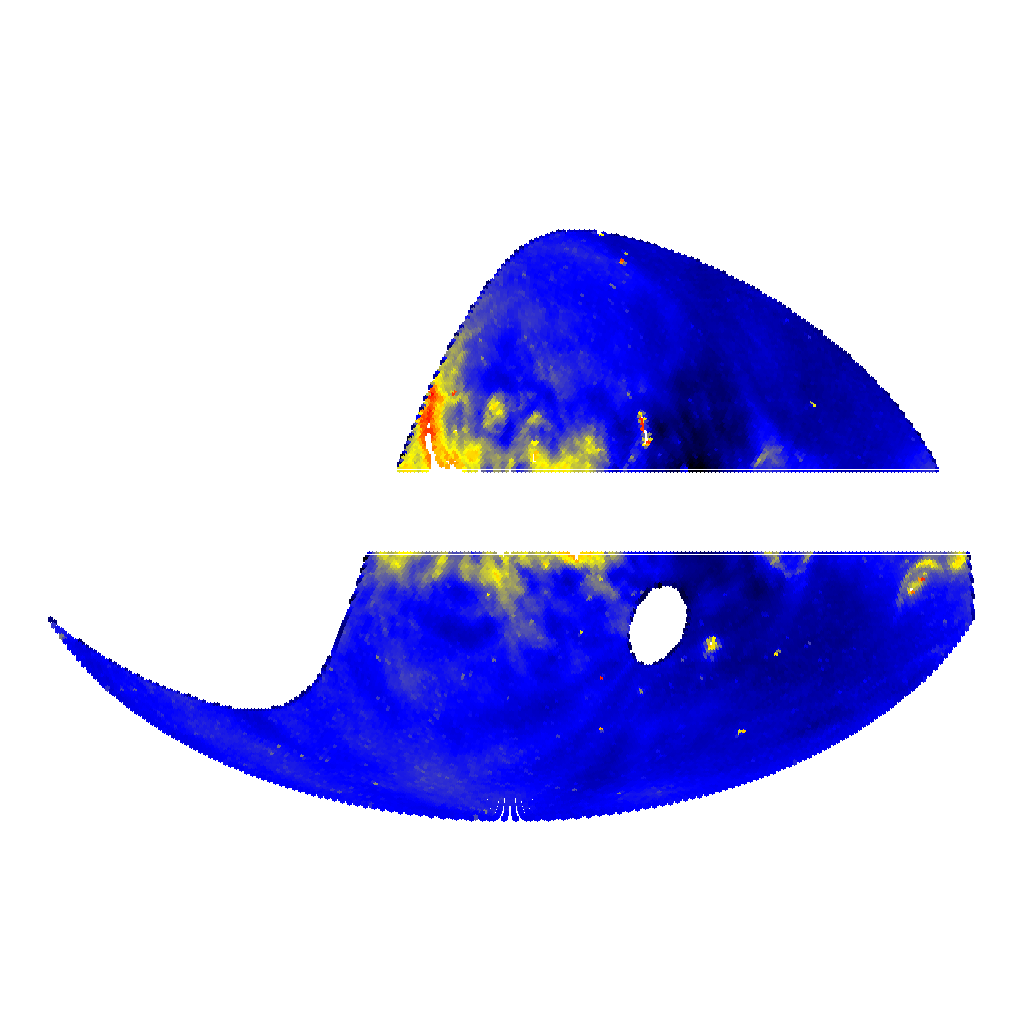}
\caption{Masks (white areas) in the model L8a obtained through the iterative method described in Sect. \ref{sect:mask}. The maps of Fig.~\ref{surveys} are shown in the background. 
Top row and from left to right: 22MHz, 45 MHz,408 MHz. Bottom row and from left to right: 820 MHz, 1420 MHz, 2326 MHz.}
\label{Fit3mask}
\end{center}
\end{figure*}

In this Section, we attempt to identify and mask the regions of the sky which contain bright radio sources. 
To this aim, we implement an iterative method. The idea is that outside the source regions, the emission is well described by the diffuse emission of the model depicted above. 
As a first trial, we fit the coefficients of the diffuse emission without any mask.
Comparing this model to the observational map, we identify as sources those regions where the model largely underestimates the data.
We mask these regions and repeat the fit of the coefficients in the remaining part of the sky, in order to better tune the models against the data. 
Then, we define sources as discussed above (i.e. considering the full map) but with the new model. We repeat the previous steps, and 
after a certain number of iterations the method converges: when this occurs, then the mask extracted from the map remains constant.
In more details, the scheme of the method is the following:

\begin{enumerate}

\item We take the radio map at its original resolution and we minimize the $\chi^2$ in 
Eq. (\ref{eq:chi2}) with:
\be
T_i^{\rm model}=T_E +c_{\rm gal} T_i^{\rm gal,synch} + c_{\rm brem} 
T_i^{\rm gal,brem}.
\ee
The quantities to be fitted are $T_E$ (the isotropic radio background) and the coefficients $c_{\rm gal}$ (one for each frequency) of the Galactic synchrotron model $T^{\rm gal,synch}$ (we include also the free-free contribution $T_i^{\rm gal,brem}$ but we find it to be negligible, as already mentioned). The index $i$ runs over the pixels, excluding the region $|b|<10$ degrees.

\item We compute the residuals for the unmasked pixels $R_i:$
\be
R_i=T_i^{\rm data} -T_i^{\rm model}
\ee
with $T^{\rm model}$ being the best-fit model {\rm obtained in Step 1.}
From the residuals, we compute the mean temperatures $T_{R,i}$ and the standard deviations 
$\sigma_{R,i}$ of the residuals in a region of 50 degrees around the pixel $i$.
\footnote{We choose a region of 50 degrees such that it encompass large structures, like Loop I, but still with a limited spatial variation of the diffuse emission in the region (to avoid a possible identification of brightest diffuse regions as sources).}
The mask is defined as those pixels where:
\be
R_i>T_{R,i}+5 \sigma_i	
\ee
where $\sigma_i$ is obtained adding in quadrature the experimental uncertainties (rms noise and calibration error) to $\sigma_{R,i}$.

\item We perform again the fit, now excluding the pixels in the mask defined at Step 2 (as well
as the region $|b|<10$ degrees).

\item We go back to Step 2.
\end{enumerate}

The iterations stop when the mask stabilizes, remaining the same as in the previous iteration. This method should allow the model to adjust to the data in the regions of the sky where the presence of unaccounted sources is minimal.
The mask depends on the survey under investigation and on the Galactic model employed.
One example of the results, which refers to the model L8$a$ (described in Sect.~\ref{sec:propagation} and Table \ref{tab:BM}), is shown in Fig. \ref{Fit3mask}. 
White areas denote the derived masks.
The masks that we obtain follow some of the features of the radio sky, in particular Loop I, as expected. 

Once the masks are defined, we use them to perform the final fit as explained
in Sect. \ref {sec:Fit} with Eq. (\ref{eq:chi2}). For the fit, we downgrade the surveys, the templates and the maps of $\sigma_i$ to the resolution $N_{\rm side}=4$, as discussed above. 
We remark that there is some uncertainty in the definition of masks. For instance, we could obtain larger or smaller masks considering a different thresholds in Step 2, like e.g. by adopting $3 \sigma_i$ or $7\sigma_i$.
Our choice is a compromise between two requirements: the need to fully mask the regions of the sky contaminated by sources and, at the same time, to still keep a large fraction of the sky in the analysis, in order to properly tune the Galactic model.
Note also that downgrading the maps to a smaller resolution actually enlarges the mask used in the analysis (see Fig.~\ref{Fit3res}). The one depicted here is thus a quite conservative procedure to exclude regions around bright sources.

\begin{figure*}[t]
\begin{center}
 \includegraphics[width=0.32\textwidth]{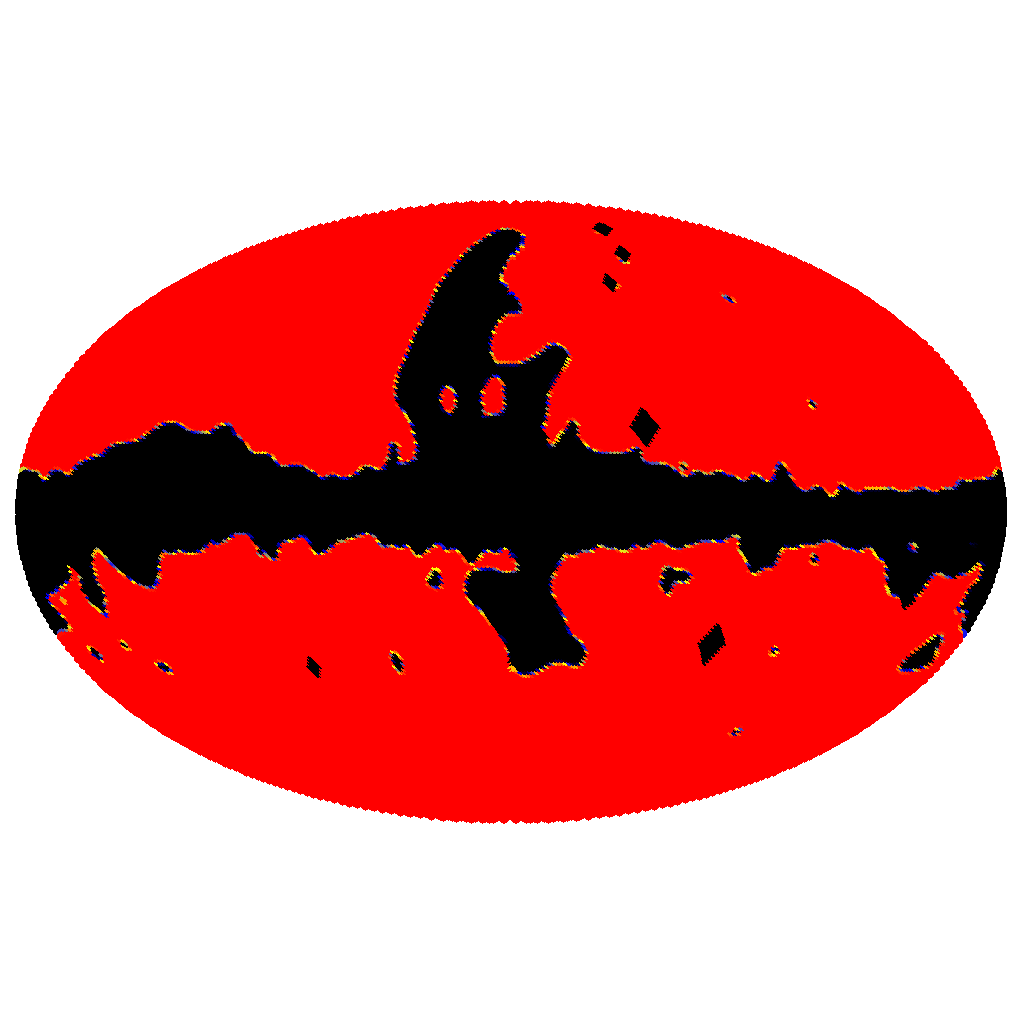}\\
 \includegraphics[width=0.32\textwidth]{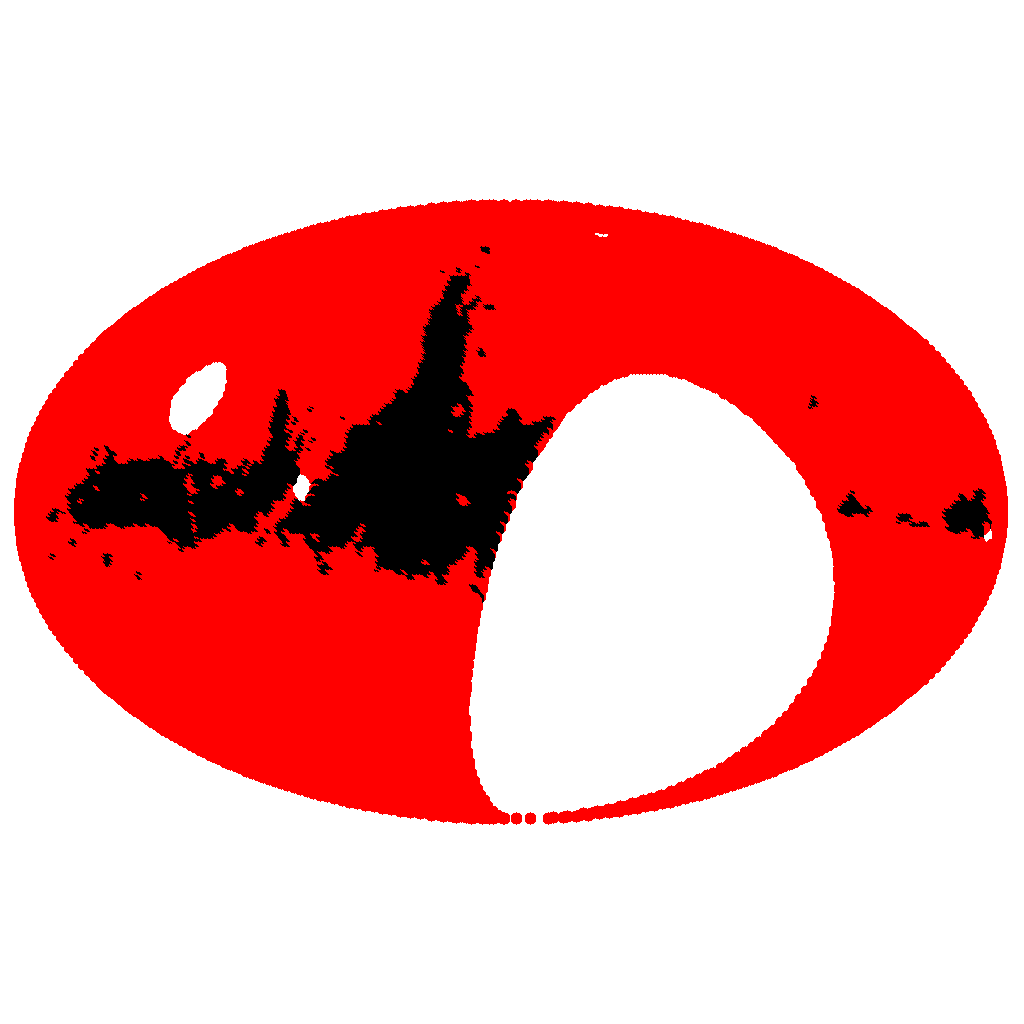}
 \includegraphics[width=0.32\textwidth]{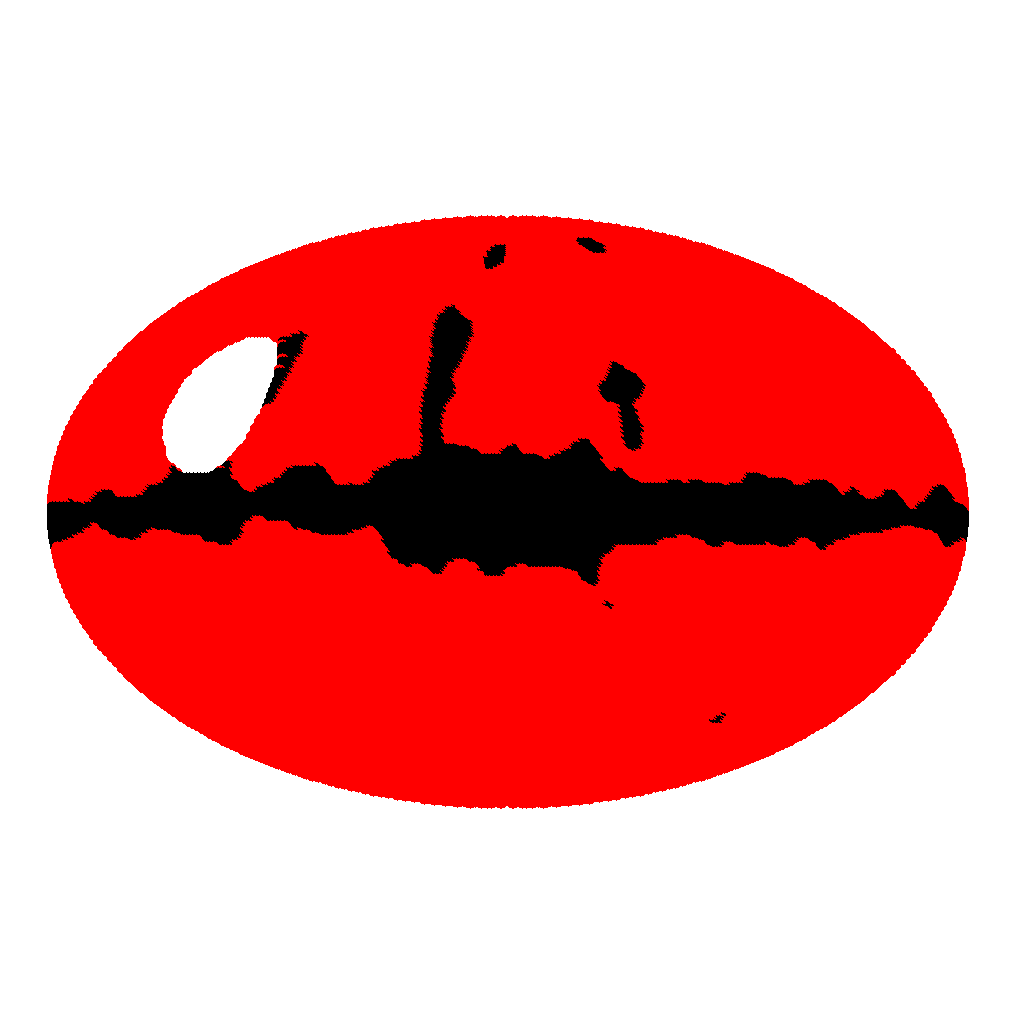}
 \includegraphics[width=0.32\textwidth]{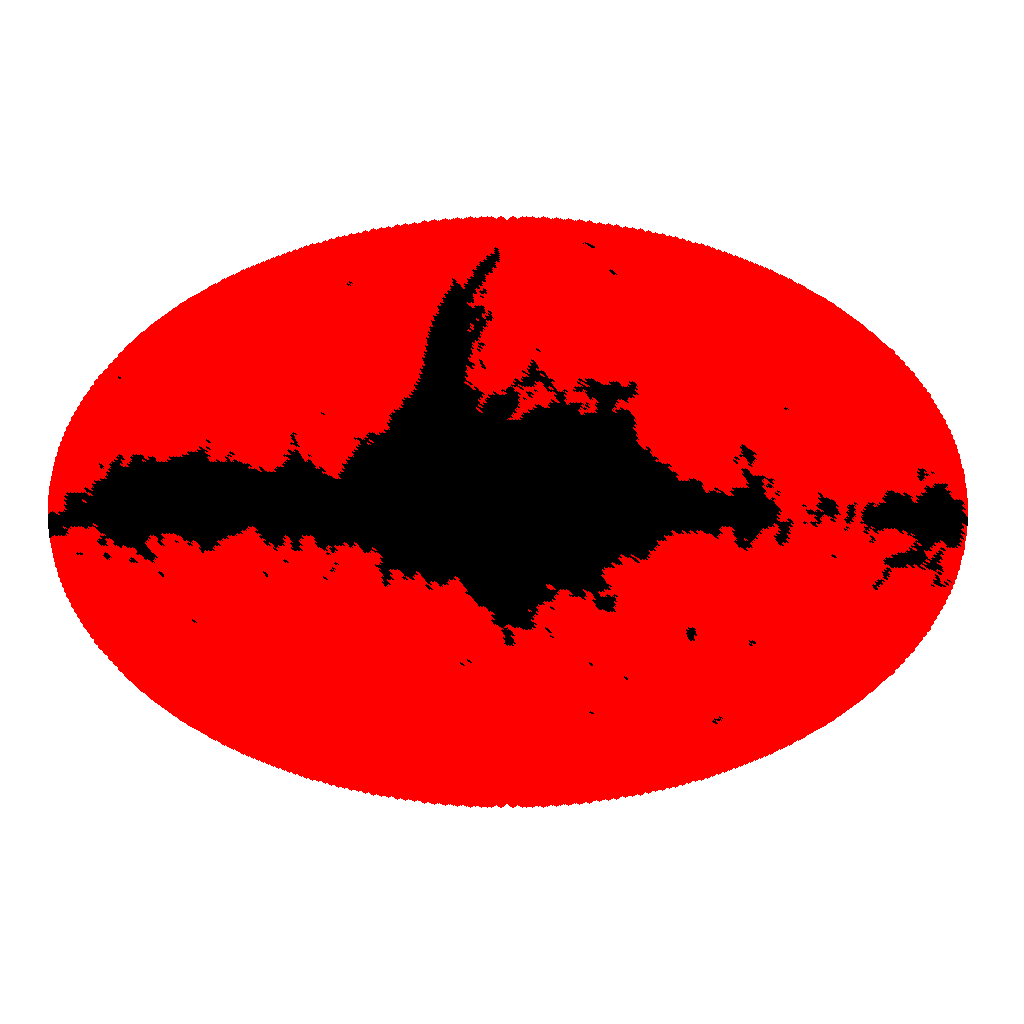}\\
 \includegraphics[width=0.32\textwidth]{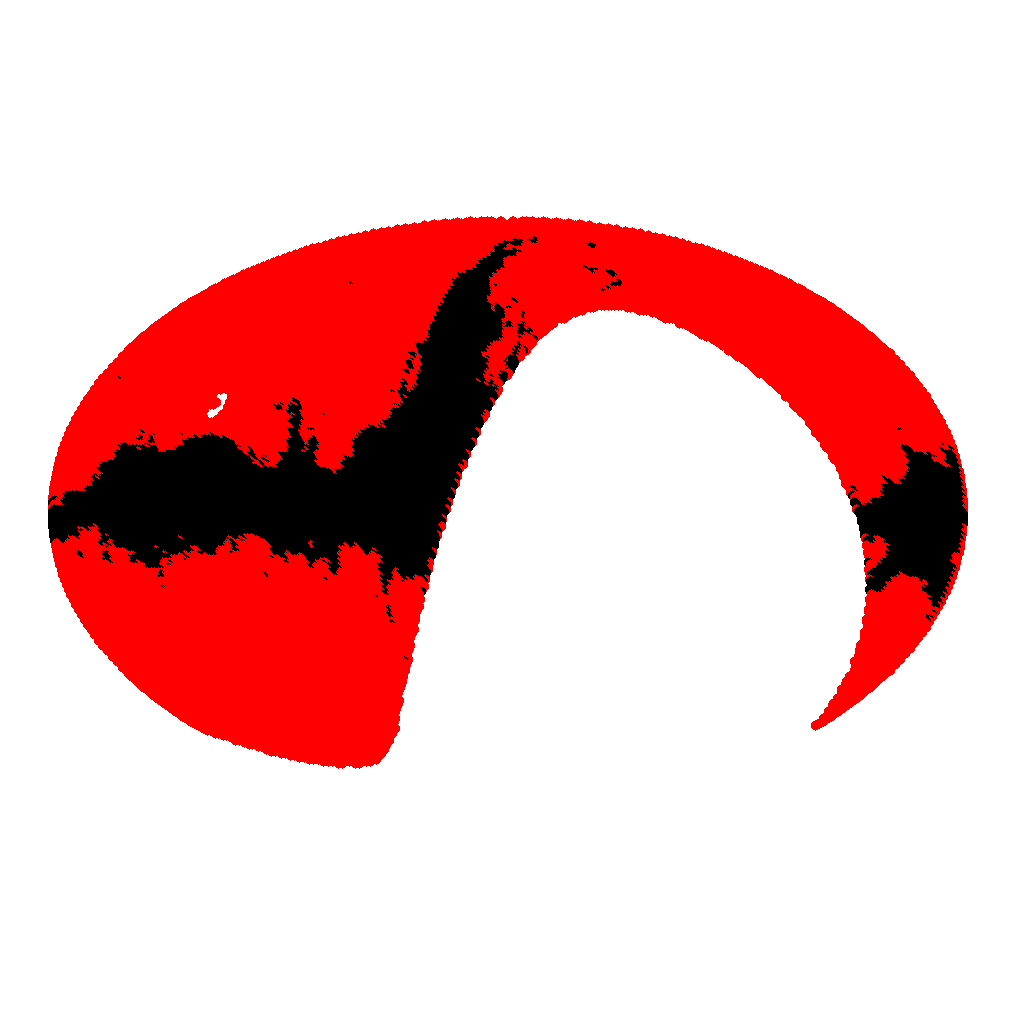}
 \includegraphics[width=0.32\textwidth]{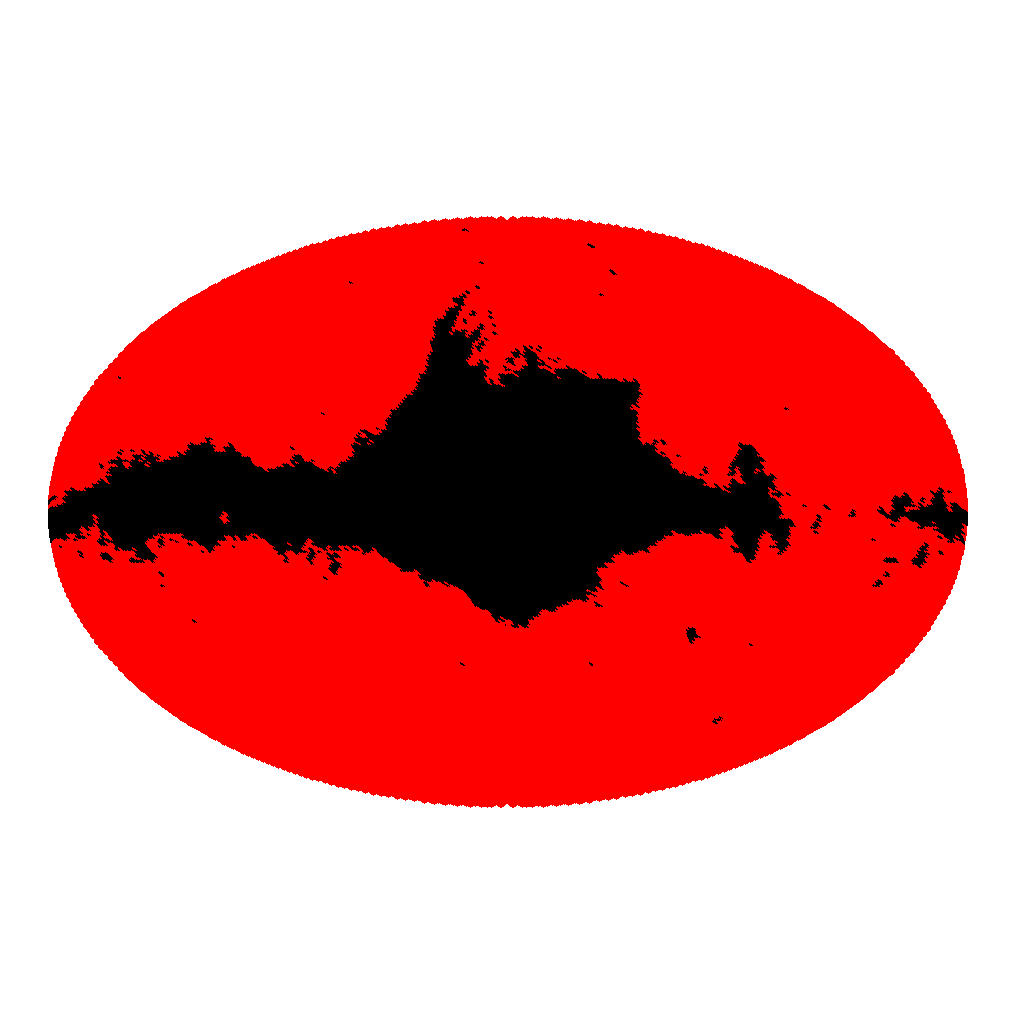}
 \includegraphics[width=0.32\textwidth]{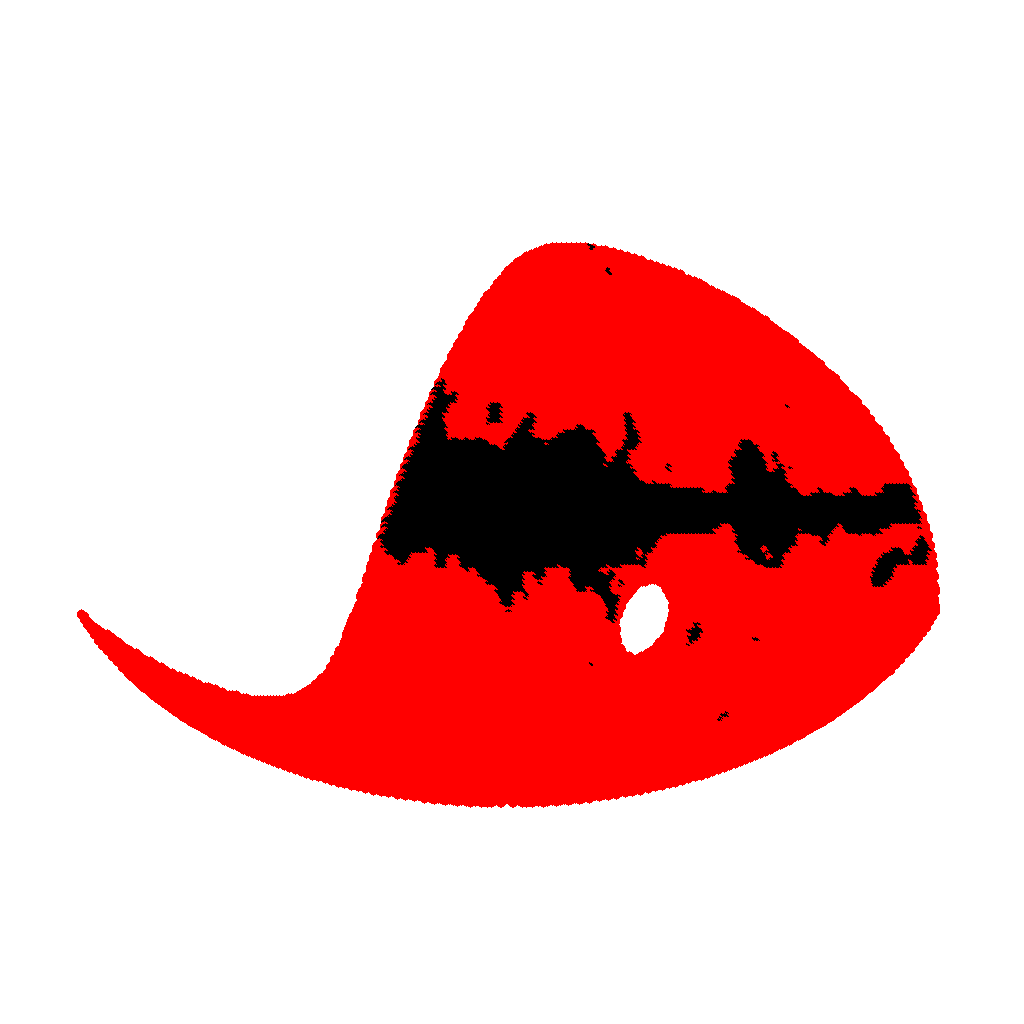}

\caption{{\sl Top row}: Mask (in black) from WMAP \cite{wmapmask}. {\sl Middle and bottom rows}: Masks (in black) from Sextractor from 22 MHz to 2326 MHz. All the maps are at the resolution $N_{\rm side}=64$.}
\label{mask}
\end{center}
\end{figure*}

However, one can wonder how much the result will change if considering different masks, in particular covering larger portions of the sky.
To this aim, we perform two additional analysis by adopting two different sets of masks.
In one case, we use the mask employed by the WMAP7 collaboration based on data at 22 GHz\footnote{Specifically, we use the mask wmap$\_$polarization$\_$analysis$\_$mask$\_$r9$\_$9yr$\_$v5$.$fits of Ref. \cite{wmapmask} and described in Ref. \cite{Bennett:2012zja}}, shown in 
Fig. \ref{mask}, which covers about 27\% of the sky. 
In the second case, we generate, for all the surveys analyzed, a mask using of the publicly available software SExtractor~\cite{Bertin:1996fj}. 
It is a package for source detection, which proceeds through segmentation by identifying groups of connected pixels that exceed some threshold (which is set in term of the rms) above the background. 
The background and rms noise maps are obtained iteratively, splitting the original map in regions of a certain size (we choose 50 degrees) and computing the mean and the standard deviation of the distribution of pixel values in such regions. This computations is repeated many times, each time discarding the most deviant values, until all the remaining pixel values are within 3-$\sigma$ from the mean.
Thresholding is then applied to the background-subtracted map to isolate connected groups of pixels, and we set the detection threshold at 5-$\sigma$ above the local background.
We take the map of sources constructed in such a way as our mask.

It is clear that the SExtractor algorithm is quite similar to the one we discussed above. The main difference is that in SExtractor the background is taken to be constant over a certain region and computed as the mean in that region (discarding the brightest pixels), while in our algorithm it can vary and comes from a physical model, i.e. the Galactic synchrotron emission.

The WMAP and SExtractor masks are similar among themselves, although larger than those computed with out iterative method.
In all cases, we will obtain quite similar results for the determined extragalactic isotropic emission, as we will discuss in Sect.~\ref{sec:Results}.

\subsection{Template}
\label{sect:template}

\begin{figure*}[!ht]
\begin{center}
 \includegraphics[width=0.48\textwidth]{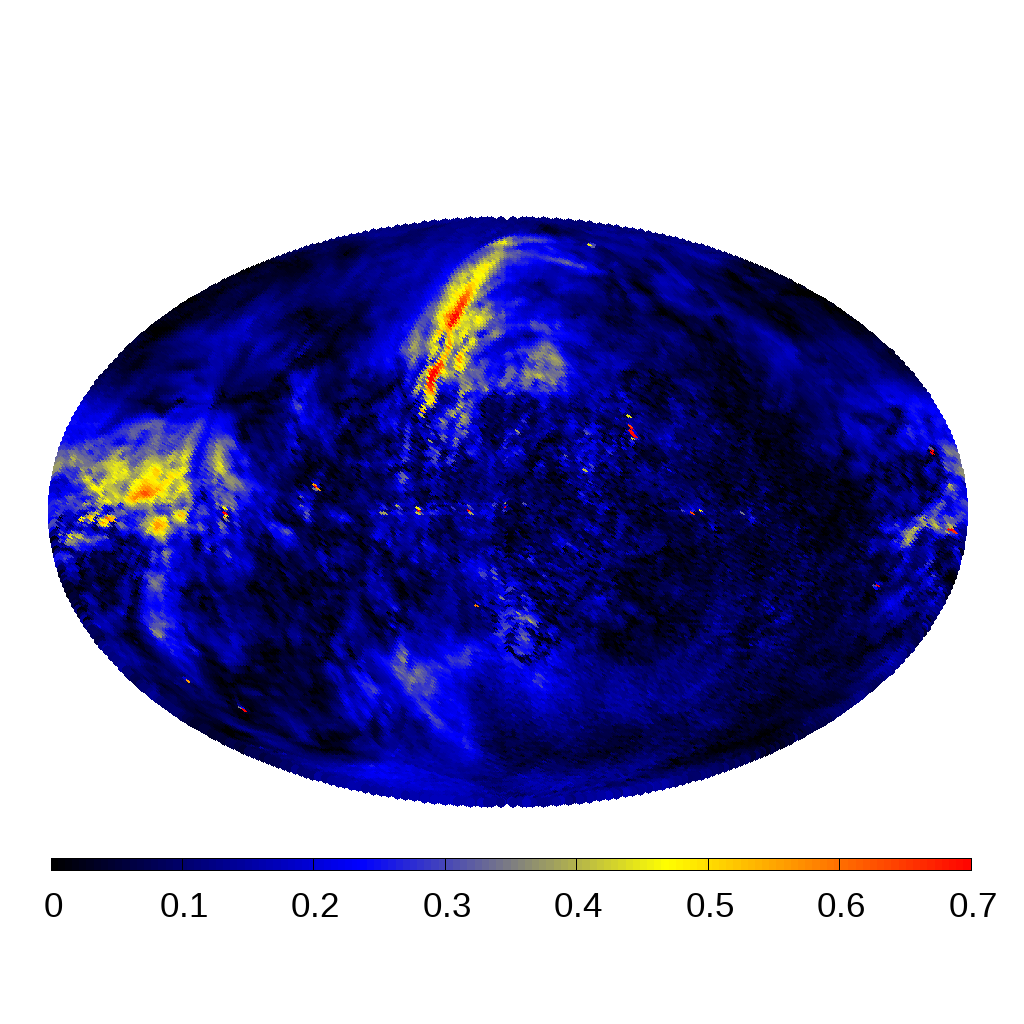}
 \includegraphics[width=0.48\textwidth]{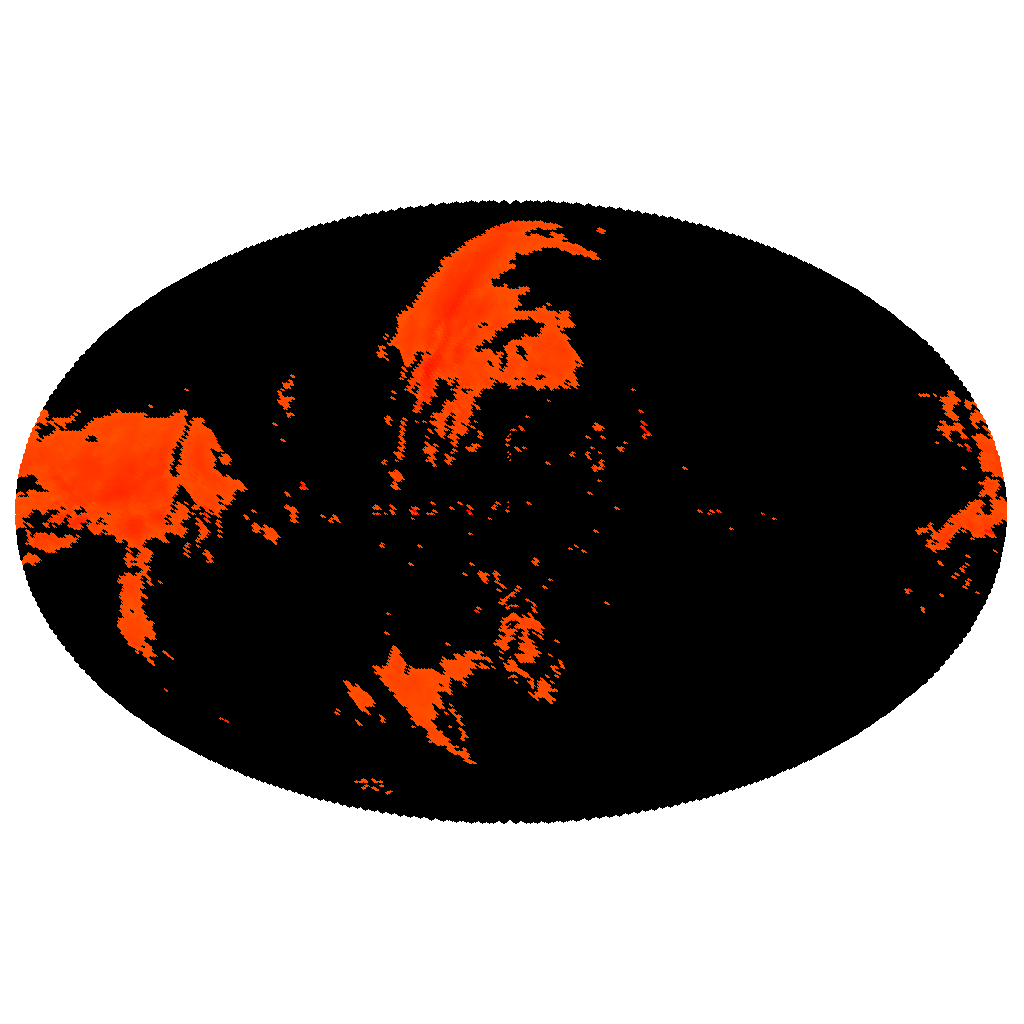}
\caption{{\sl Left}: Map of polarization at 1420 MHz in K. {\sl Right}: For the same map reported in the left panel, we show in red the pixels with $T>0.225$ K: this gives the region of our source template. 
See Sect. \ref{sect:template} for details. All the maps are at the resolution $N_{\rm side}=64$.}
\label{PI}
\end{center}
\end{figure*}

At the frequency we investigate, the majority of sources in the sky are synchrotron sources.
They typically show a high degree of polarization.
The extended Galactic radio loops go in this category as well.
Indeed, supernova explosions accelerate particles (including electrons) and compress the surrounding medium, amplifying the magnetic field, with a resulting polarized synchrotron radiation.
Polarization surveys have been extensively used to reveal and study SNR radio loops (see, e.g.,
Ref. \cite{Wolleben:2007pq} and references therein).

We attempt to trace the most intense synchrotron sources by means of a template based on the full sky polarization map at 1420 MHz obtained combining very recent observations at DRAO~\cite{DRAO:1420} and Villa Elisa~\cite{VILLA:pol} telescopes. The two maps have similar rms noise ($\sim15$ mK) and angular resolution ($\sim36'$), and agree well in the region where thry overlap, once the zero level is adjusted. The zero-level accuracy is estimated to be 30 mK~\cite{VILLA:pol}, so we take a total error of 45 mK.
The map is shown in Fig.~\ref{PI}a.
Note, for instance, that the prominent Loop I is clearly visible.

We build the template by taking only the brightest sources in the polarization map.
Indeed, the synchrotron Galactic diffuse emission can be polarized as well, but it is already included as a separate component of our model, 
so including all the pixels we would end up in, e.g., double-counting the Galactic monopole.
We select the pixels with polarized intensity $T_p> 5\sigma$ with $\sigma=45$ mK being the experimental error of the map, with the resulting template shown in Fig.~\ref{PI}b.
It is clear that, qualitatively, the regions which are mostly contaminated by sources (as the positions of the major radio loops) are matched, and the structure of the right panel is similar to the masks of Figs.~\ref{Fit3mask} and \ref{mask}.
 
On the other hand, Faraday depolarization is particularly effective in the Galactic disk, and low latitudes show a low level of polarization.
Therefore sources at low latitudes are hardly captured by this template. However, remind that we will be masking the $|b|<10$ degrees region in our analysis.

We will fit the radio maps at a resolution $N_{\rm side}=4$ including this template. Thus, we will consider:
\be
T_i^{\rm model}=T_E +c_{\rm gal} T_i^{\rm gal,synch} + c_{\rm brem} T_i^{\rm gal,brem} + c_{\rm PI} T_i^{\rm PI},
\ee
with $c_{\rm PI}$ are the coefficient that normalize the polarization template described above (one for each frequency).

Although the source template seems to be able to capture the salient features of the brightest extended sources present in the maps, we caution that both the spatial shape of the template and the flux ratios among different source regions are fixed (we only allow for a different overall normalization at the various frequencies). 
These two properties are instead likely to be frequency dependent.
On the other hand, such possible mismatches are softened by the fact that in our analysis we downgrade the maps to a resolution of about 15 degree, so a very fine description is in fact not essential.

\section{Results}
\label{sec:Results}

\begin{figure*}[t]
\begin{center}
 \includegraphics[width=0.34\textwidth]{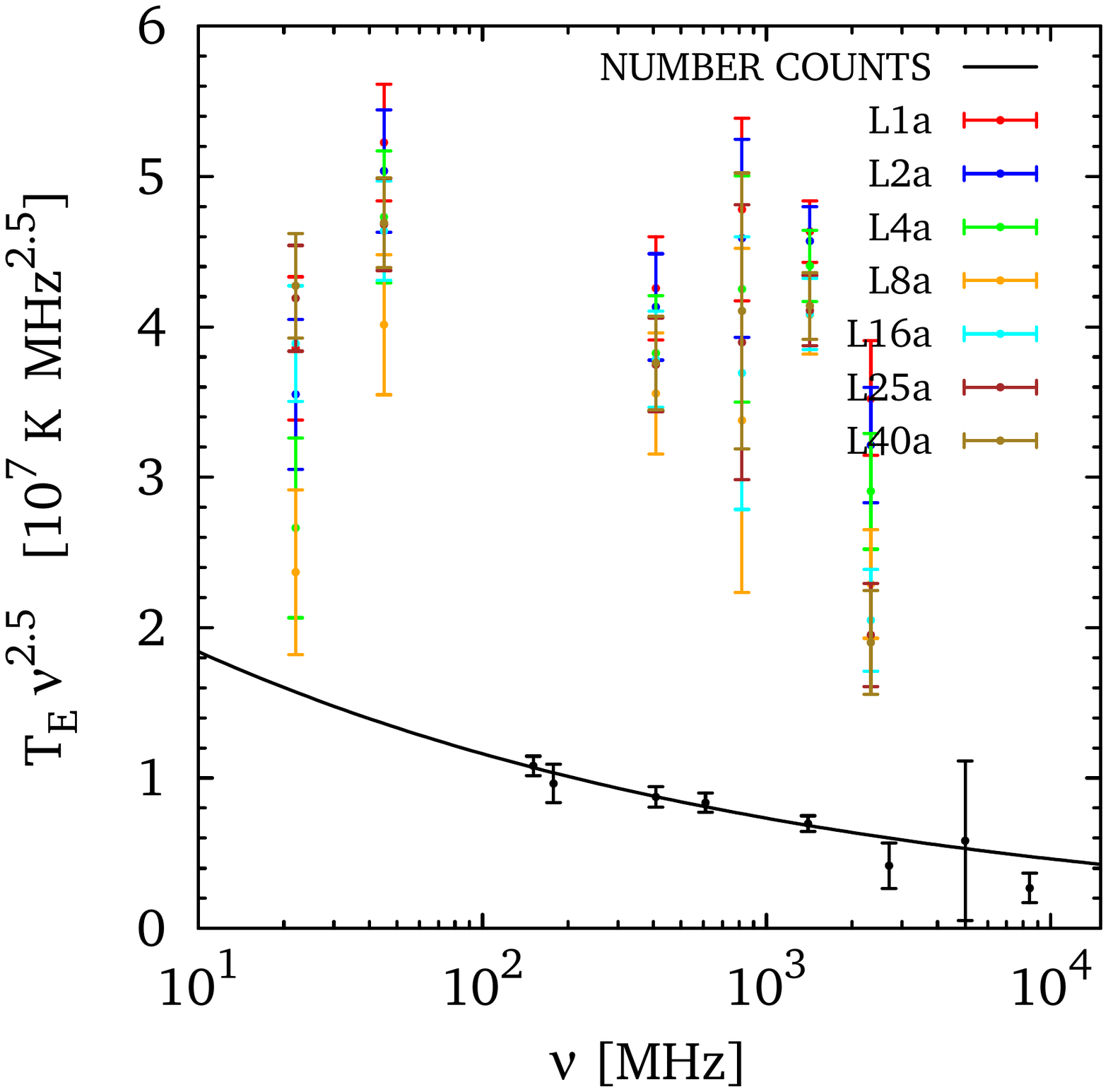}\hspace{-0.4cm}
 \includegraphics[width=0.34\textwidth]{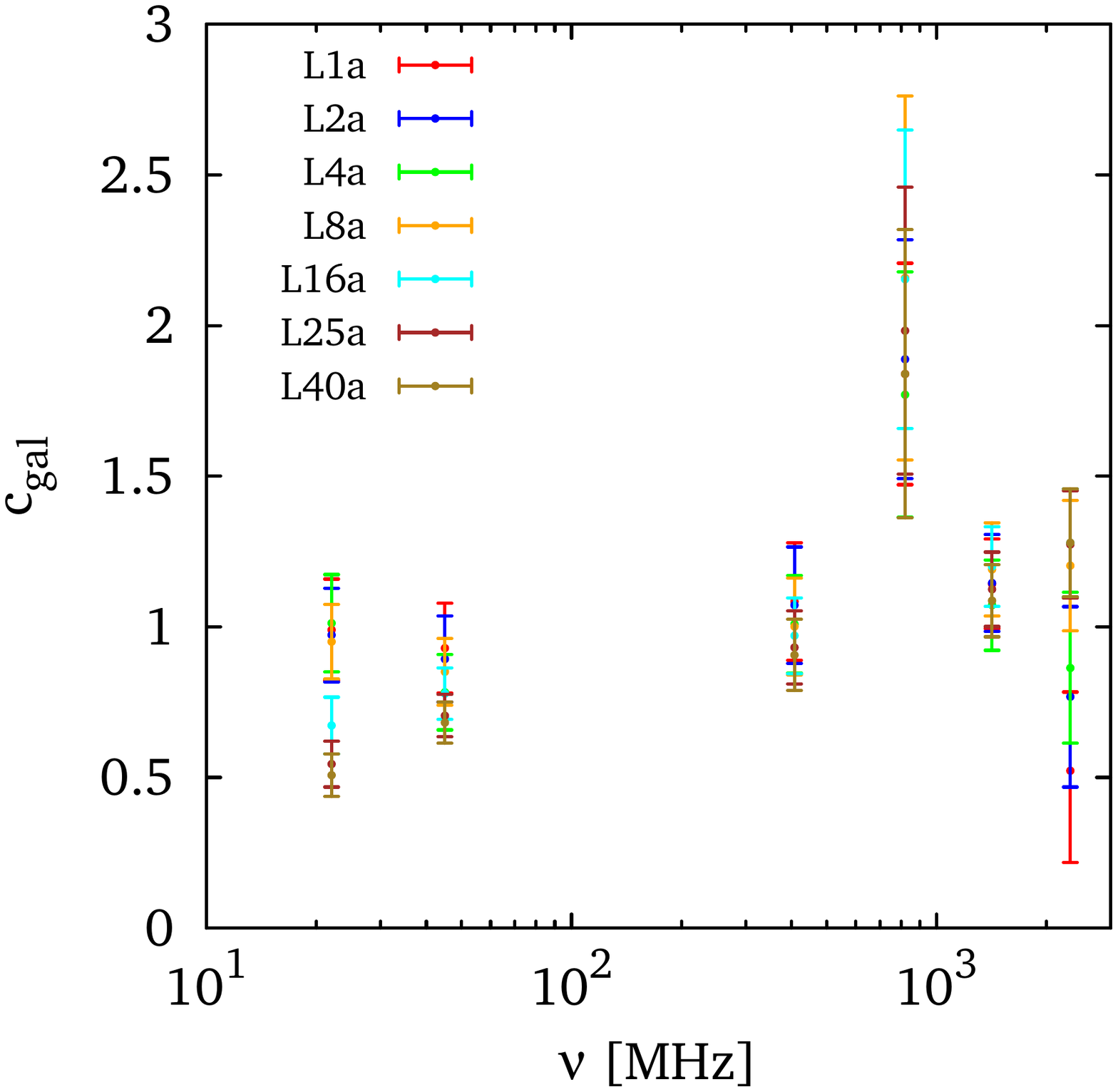}\hspace{-0.4cm}
 \includegraphics[width=0.34\textwidth]{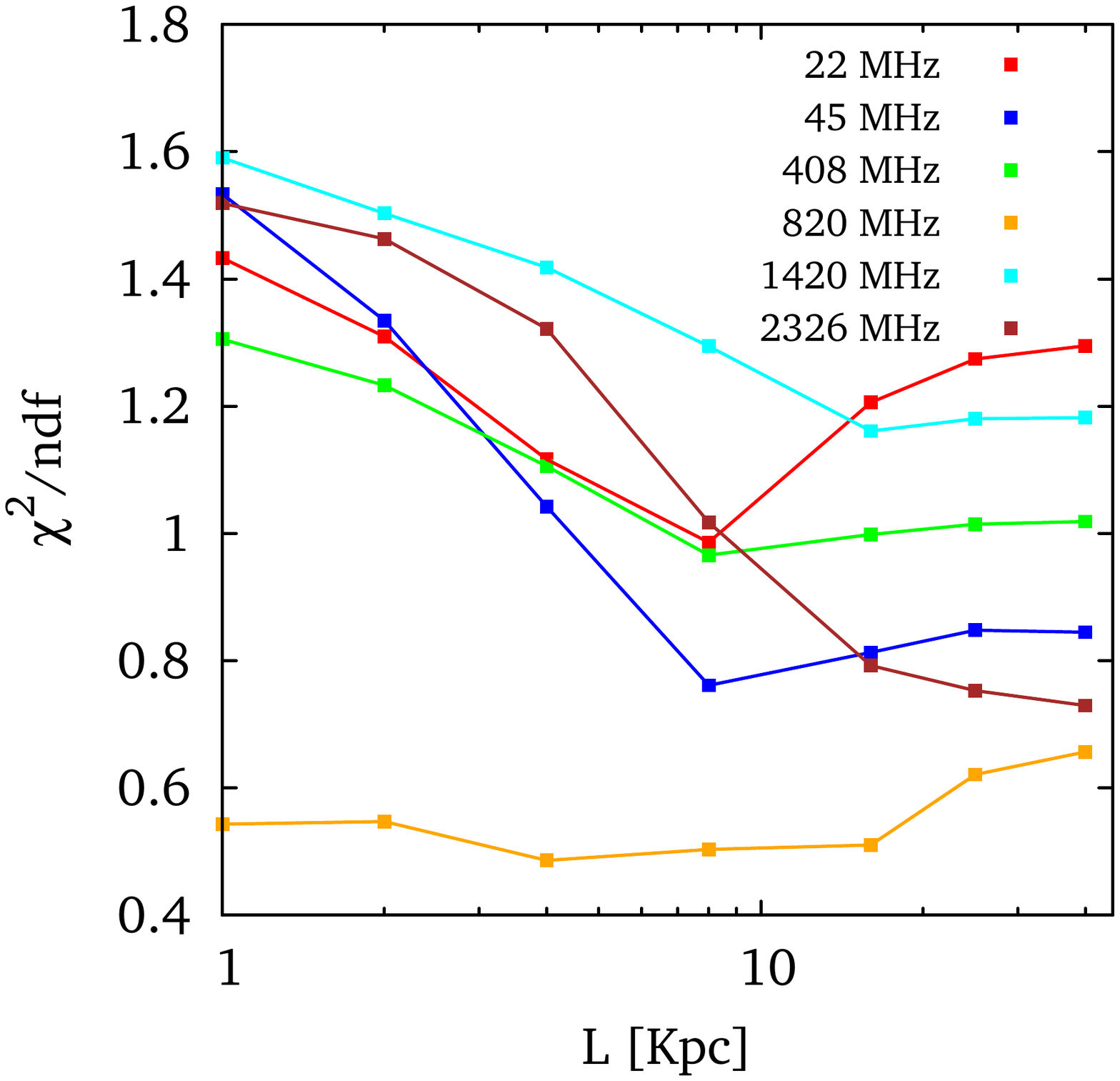}\\ \vspace{-2cm}
 \includegraphics[width=0.34\textwidth]{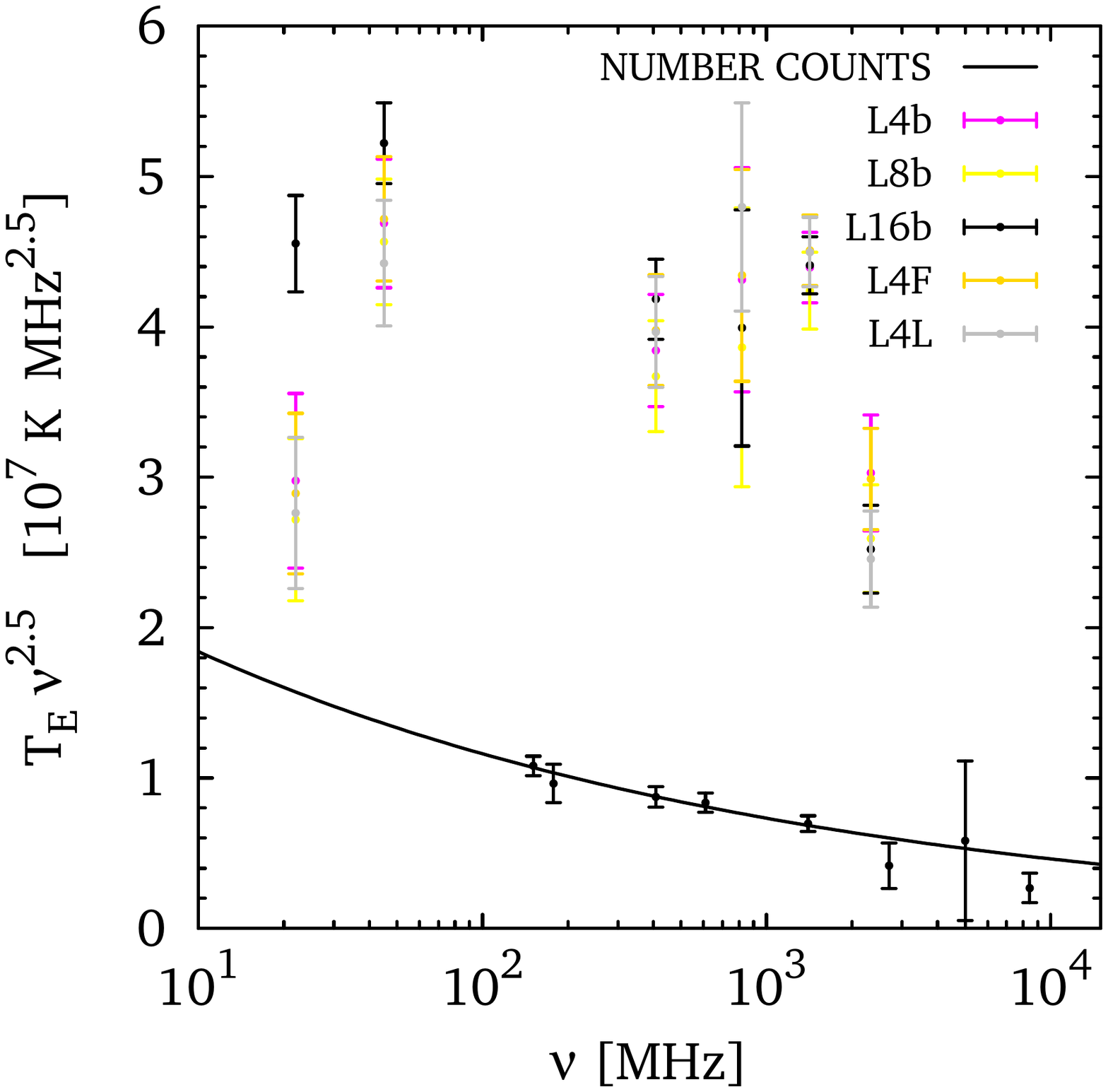}\hspace{-0.4cm}
 \includegraphics[width=0.34\textwidth]{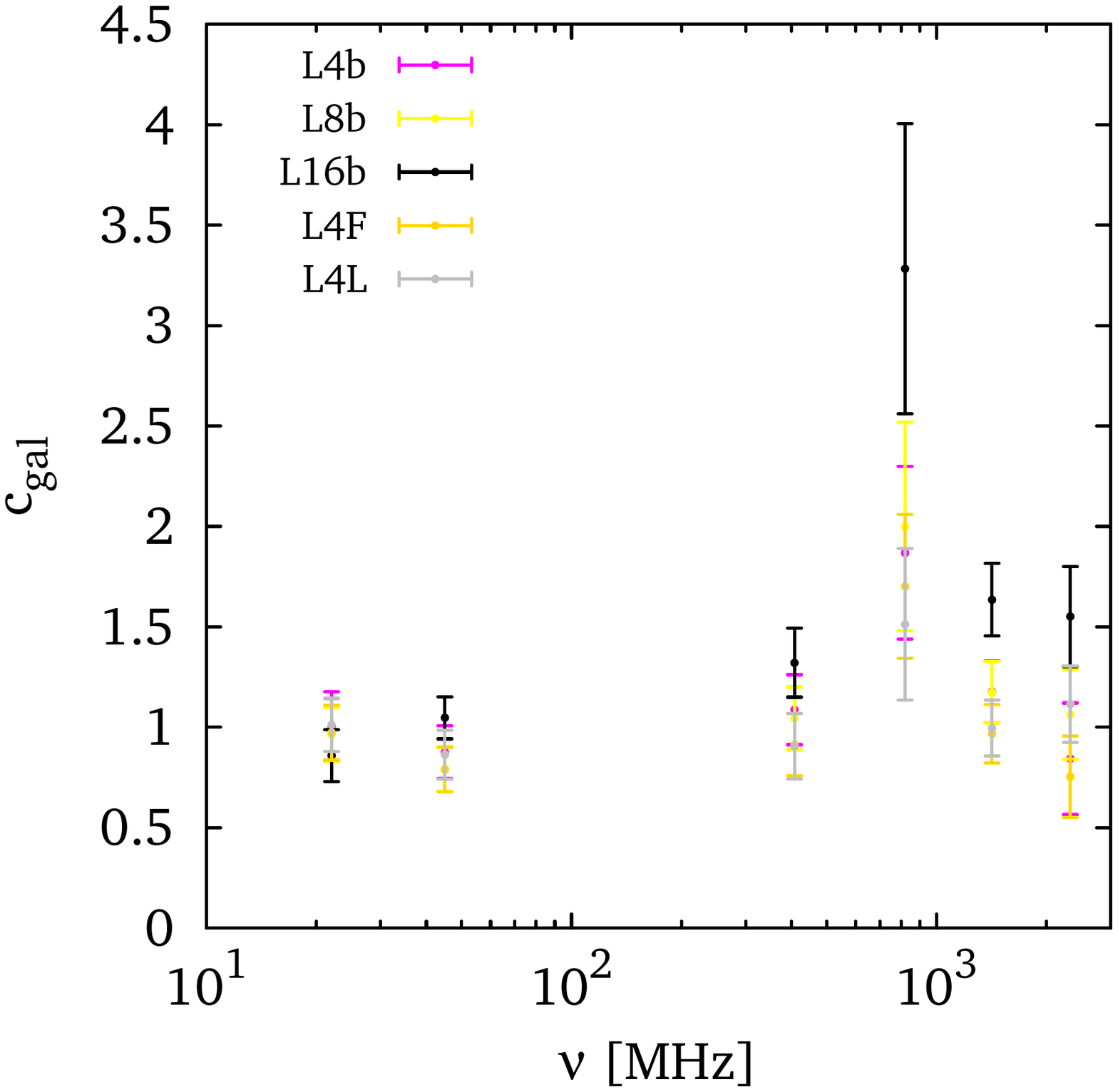}\hspace{-0.4cm}
 \includegraphics[width=0.34\textwidth]{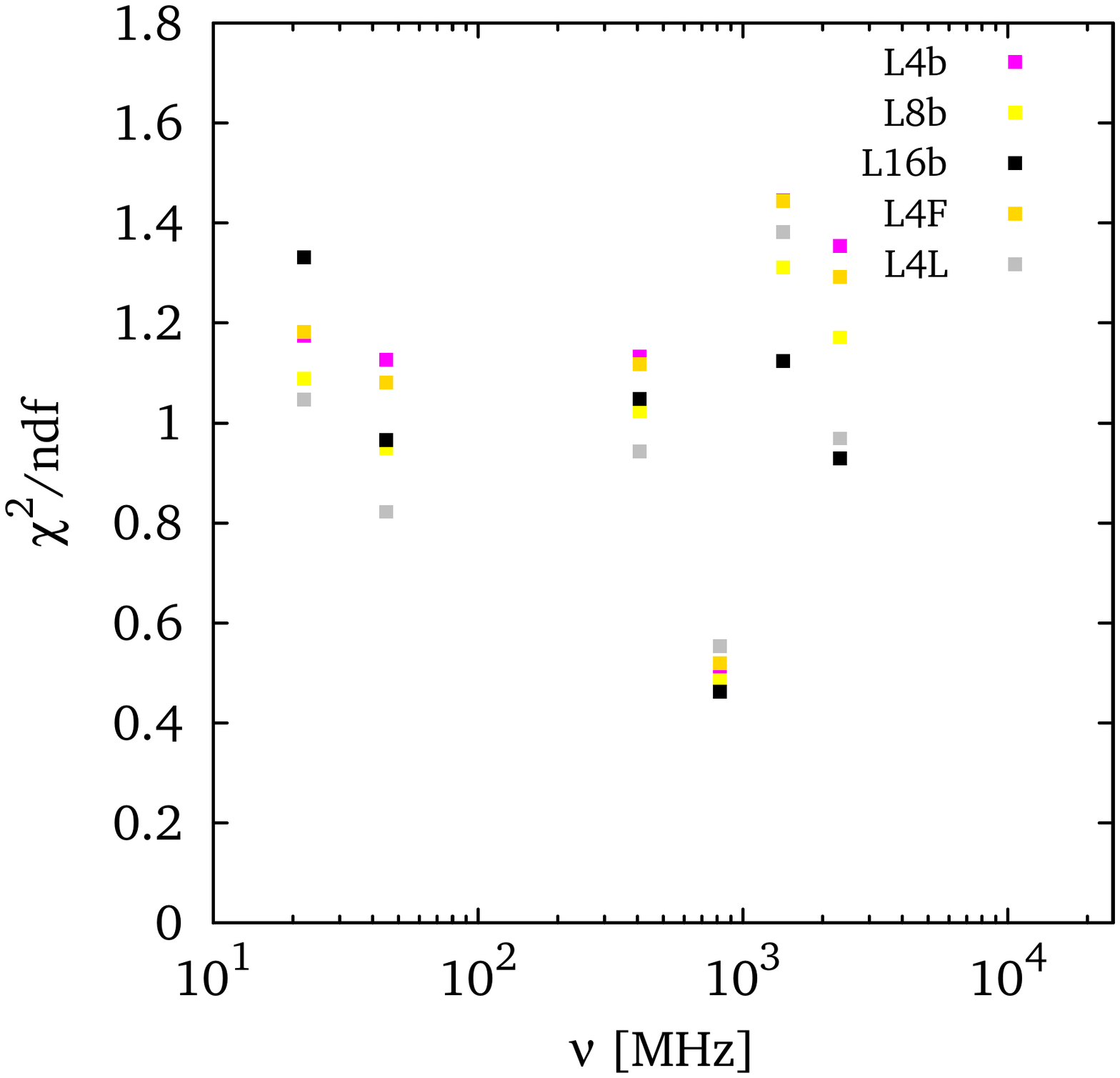}\\ 
\caption{Results of the fits obtained following the method of Sect. \ref{sect:mask}, i.e. when iteratively-derived masks are adopted.
The left column shows the best-fit values for the isotropic temperature $T_E$ (multiplied by $\nu^{2.5}$) vs. the map frequency $\nu$ for the various Galactic models adopted
in the analysis; the lower black points \cite{Gervasi:2008rr} and the solid line which fits them show the extragalactic temperature expected from number counts.
The central column reports the values of the normalization coefficients $c_{\rm gal}$ for the
Galactic contributions. The right column shows the $\chi^2/{\rm ndf}$ for our best-fits: in the upper panel, each line refers to a different frequency, and shows the values of the 
$\chi^2/{\rm ndf}$ as a function of the extension $L$ of the cosmic rays confinement volume; 
the lower panel shows $\chi^2/{\rm ndf}$ vs. the frequency $\nu$. 
Models labelled with $a$ and $b$ have been defined in 
Sect. \ref{sec:magnetic} and Table \ref{tab:BM}. Models L4F and L4L are discussed in Sect. \ref{comparison}.
}
\label{Fit3a}
\end{center}
\end{figure*}

\begin{figure*}[t]
\begin{center}
 \includegraphics[width=0.34\textwidth]{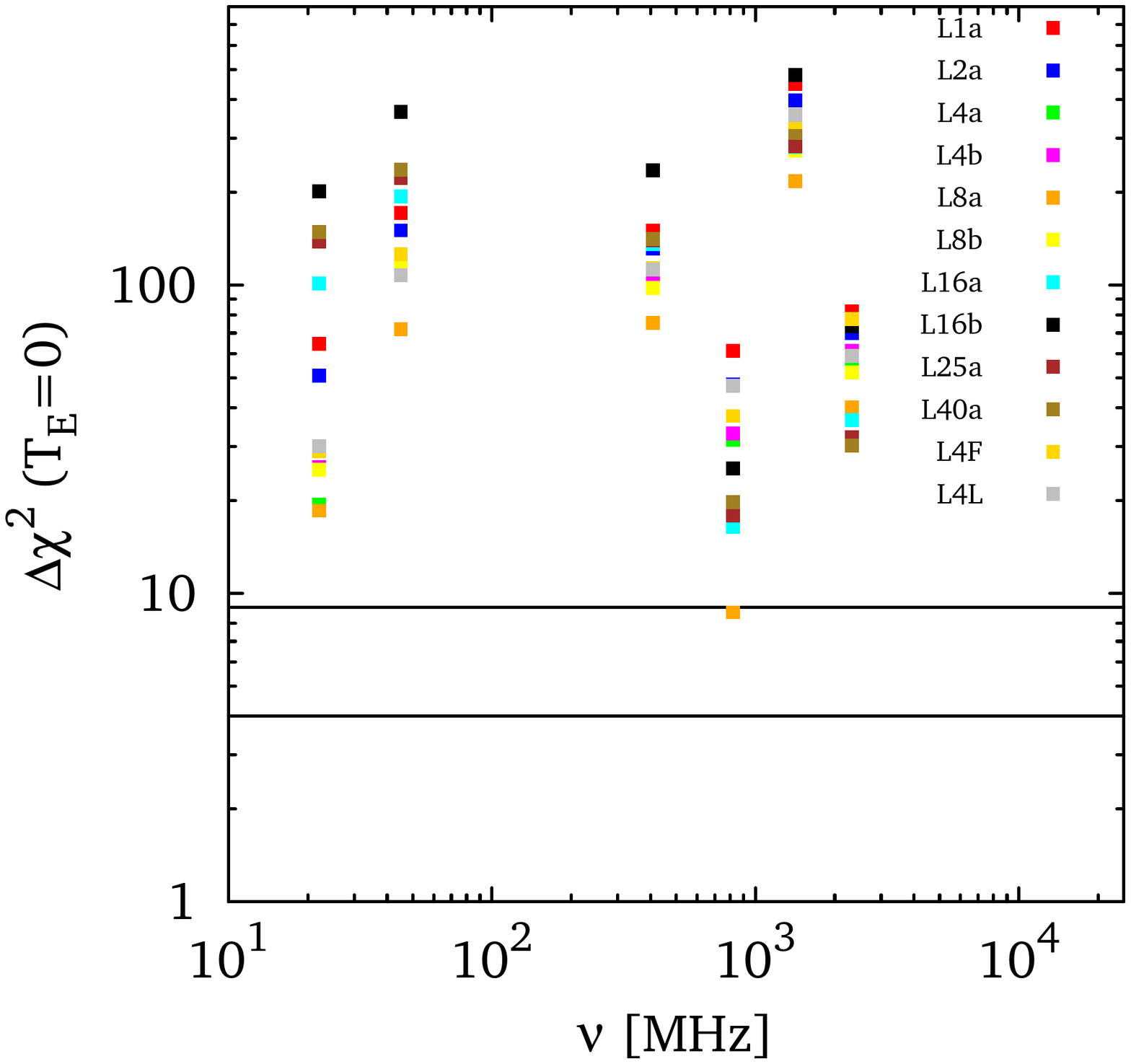}\hspace{-0.3cm}
 \includegraphics[width=0.34\textwidth]{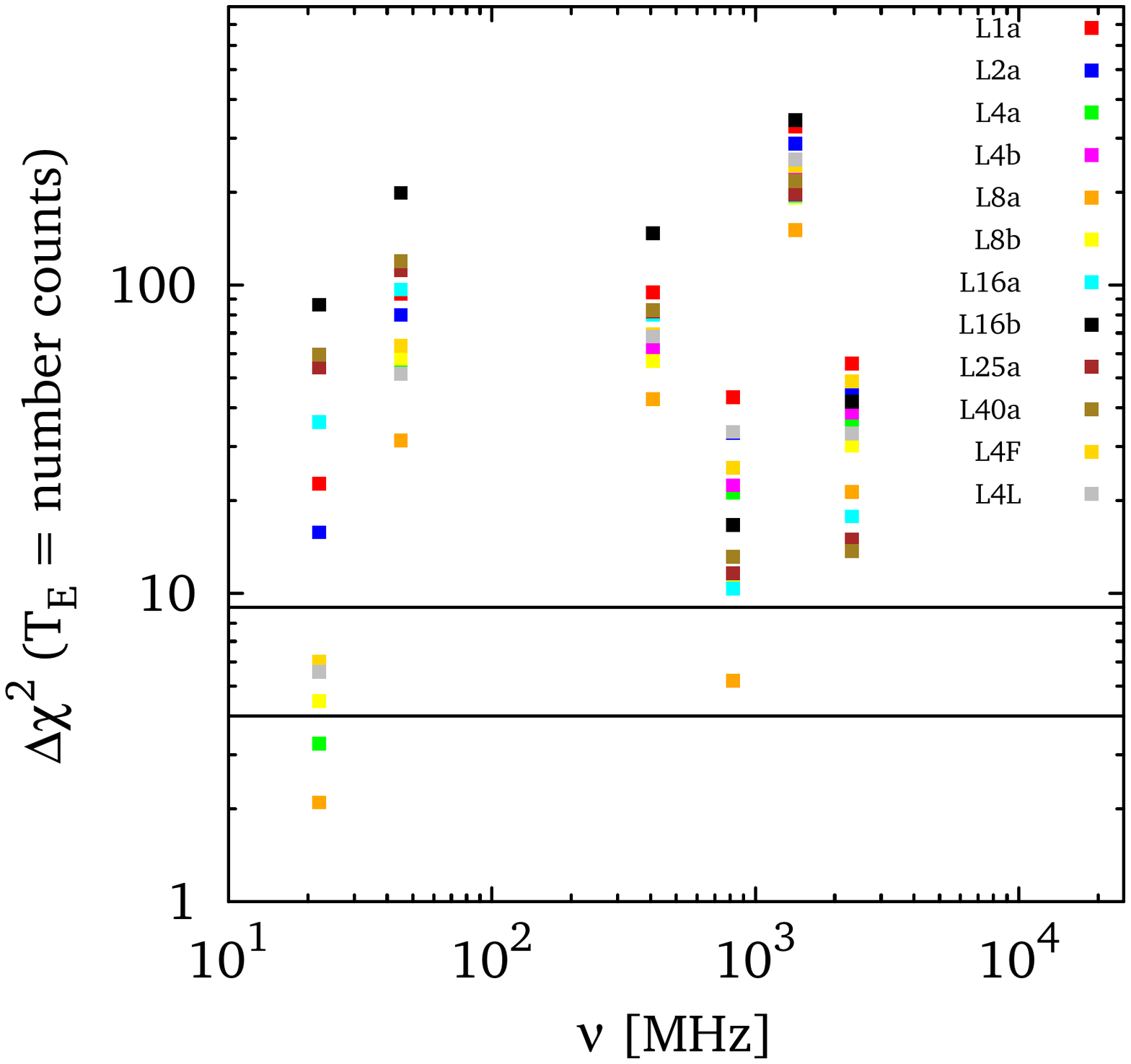}
\caption{Results of the fits obtained following the method of Sect. \ref{sect:mask}, i.e. when iteratively-derived masks are adopted. The left panel shows the increase of $\chi^2$ when one assumes that the extragalactic temperature $T_E$ is zero;
the right panel refers to the case when one assumes the $T_E$ value deduced from radio number counts. The two horizontal lines set the $2\sigma$ and $3\sigma$ C.L. 
Models labelled with $a$ and $b$ have been presented in 
Sect. \ref{sec:magnetic} and Table \ref{tab:BM}. Models L4F and L4L are discussed in Sect. \ref{comparison}.
}
\label{Fit3b}
\end{center}
\end{figure*}

Now that we have defined the modeling for the Galactic emission and the various methods
adopted to
deal with sources (either through masks or templates), we can perform the fit
of our models to the radio maps, and derive information on the isotropic
temperature $T_E$. For each of the many different cases under study, we perform 
fits according to the procedure outlined in Section \ref{sec:Fit}.

In Fig.~\ref{Fit3a}, we summarize the results obtained following the method described in Sect.~\ref{sect:mask}, i.e. by adopting masks created with an iterative algorithm. 
We show the results for all the Galactic synchrotron models that we have considered in our analysis. 
The left column shows the best-fit values for the isotropic temperature $T_E$ (multiplied by $\nu^{2.5}$) at the various frequencies.
The central column reports the values of the normalization coefficients 
$c_{\rm gal}$ for the
Galactic contributions. The right column shows the $\chi^2/{\rm ndf}$ for our best-fits: in the upper panel, each line refers to a different frequency, and shows the values of the 
$\chi^2/{\rm ndf}$ as a function of the extension $L$ of the cosmic rays confinement volume; 
the lower panel shows $\chi^2/{\rm ndf}$ at the various frequencies. 
The error bars refer to a $1\sigma$ C.L. for a marginal confidence interval (i.e., obtained by marginalizing over the other parameters).

The coefficients $c_{\rm gal}$ of the Galactic models are all consistent with one, with two exceptions. 
The first is at low frequencies for models with a very large vertical scale $L$ of the diffusion box (models L16a, L25a and L40a).
In this case the secondary electron/positron production becomes relatively more important at low energy with respect to the primary flux, and this makes the $e^++e^-$ spectrum softer: in turn, this induces a larger radio flux at low frequencies, hence a smaller
normalization coefficients $c_{\rm gal}$ is required. Even taking an unrealistically large primary spectral index for the electrons, the problem persists being the spectrum of electron/positron secondaries related to primary CR nuclei, which are in turn fitted to data.  
Still, this does not seem to have any impact on the estimation of the isotropic radio background, which is the main focus of this work. 

\begin{figure*}[t]
\begin{center}
 \includegraphics[width=0.32\textwidth]{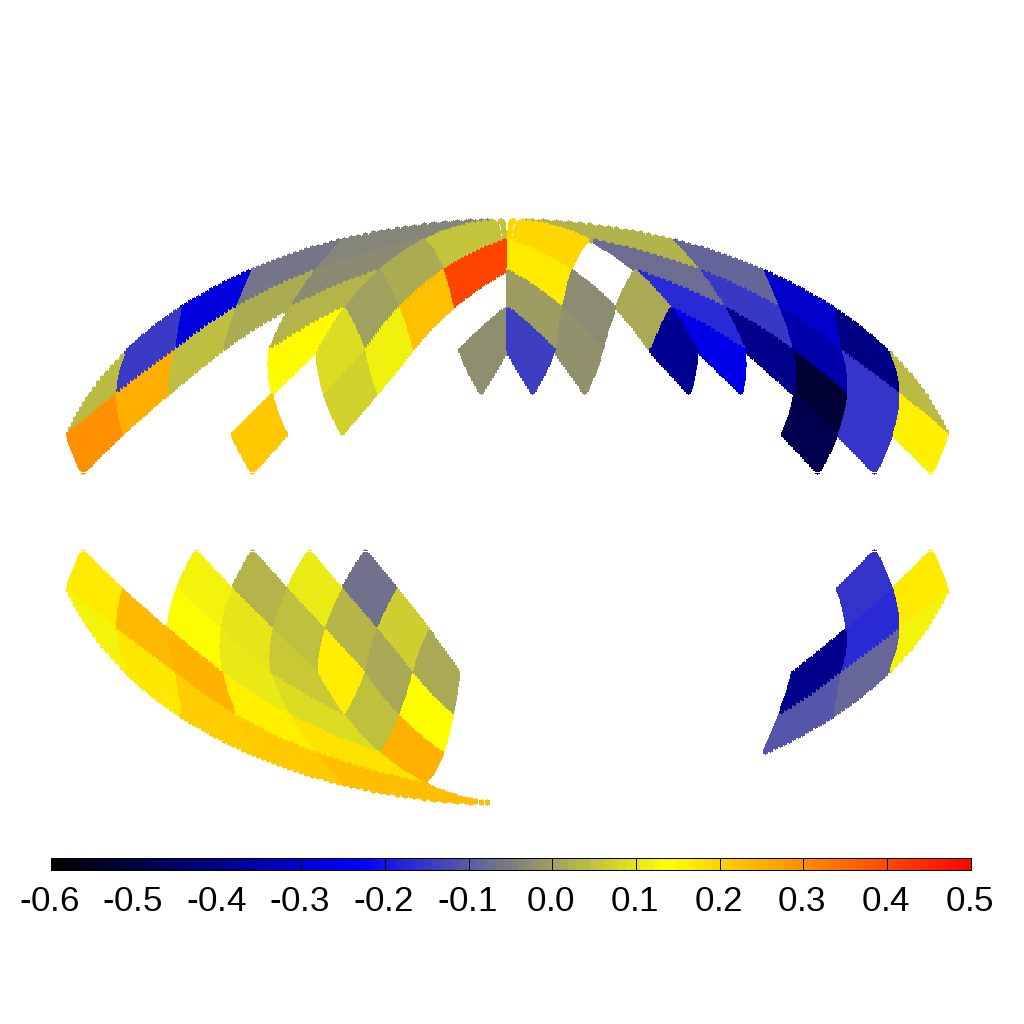}
 \includegraphics[width=0.32\textwidth]{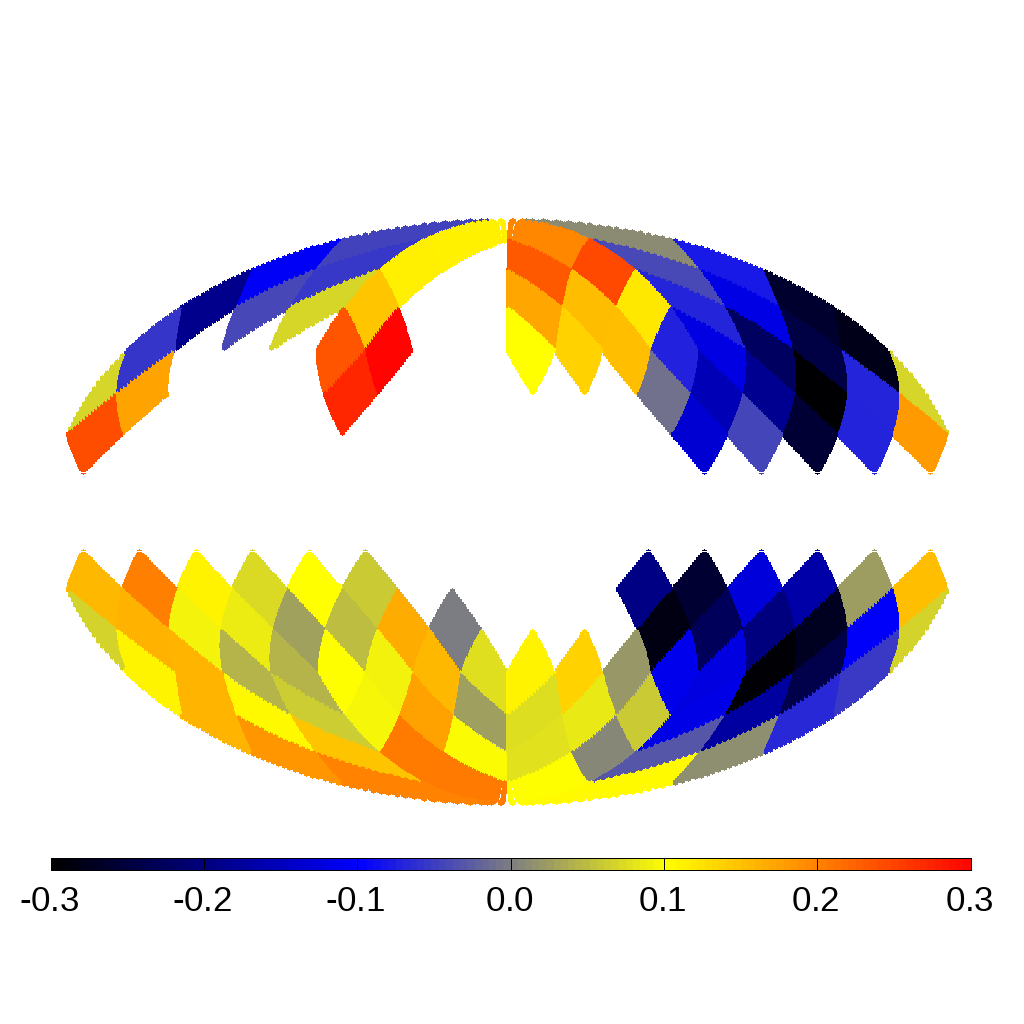}
 \includegraphics[width=0.32\textwidth]{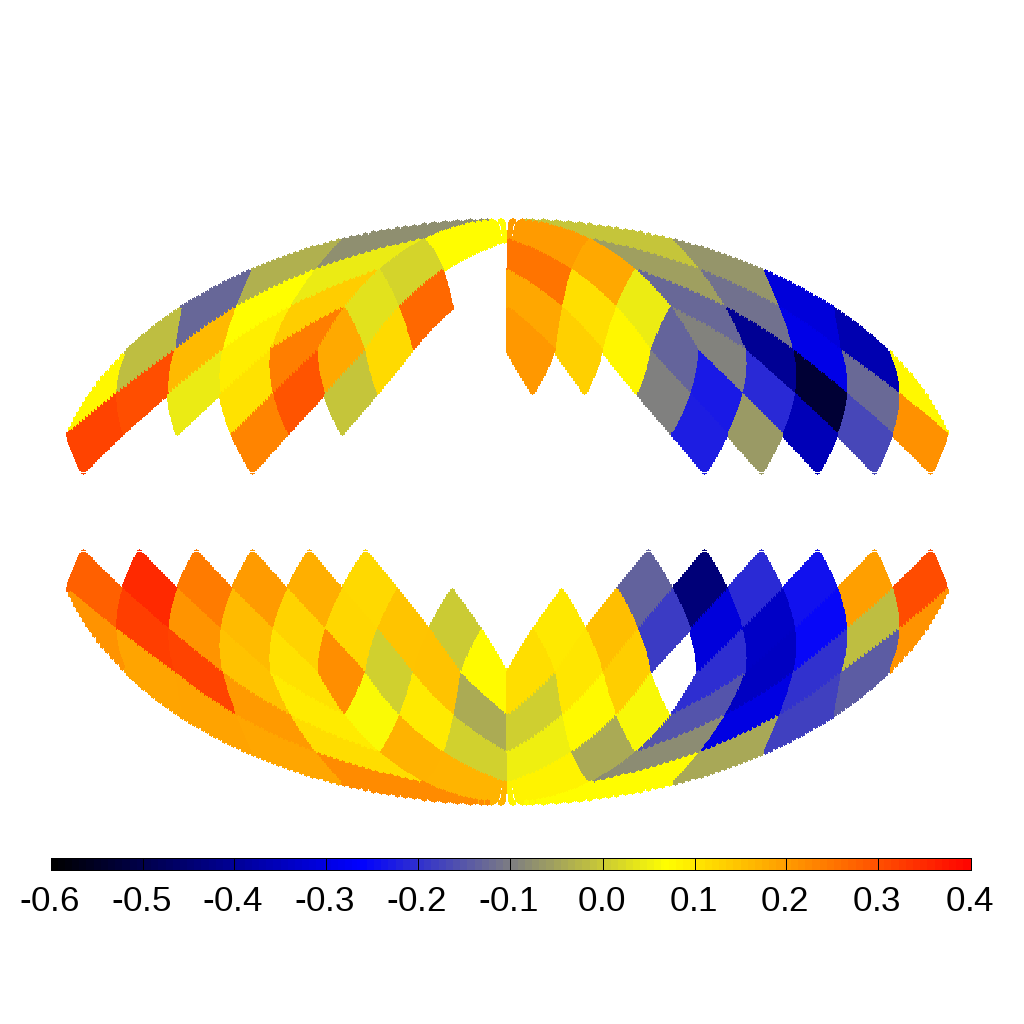}\\
 \includegraphics[width=0.32\textwidth]{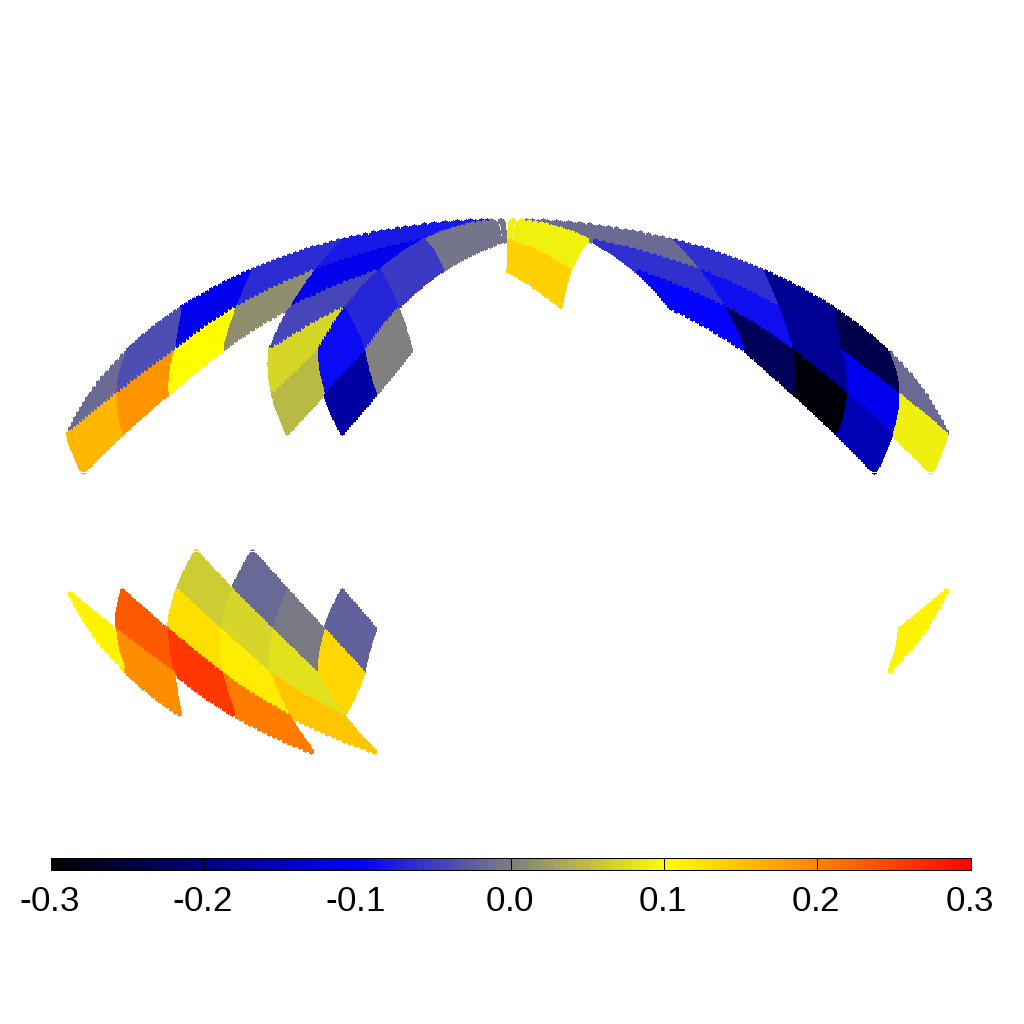}
 \includegraphics[width=0.32\textwidth]{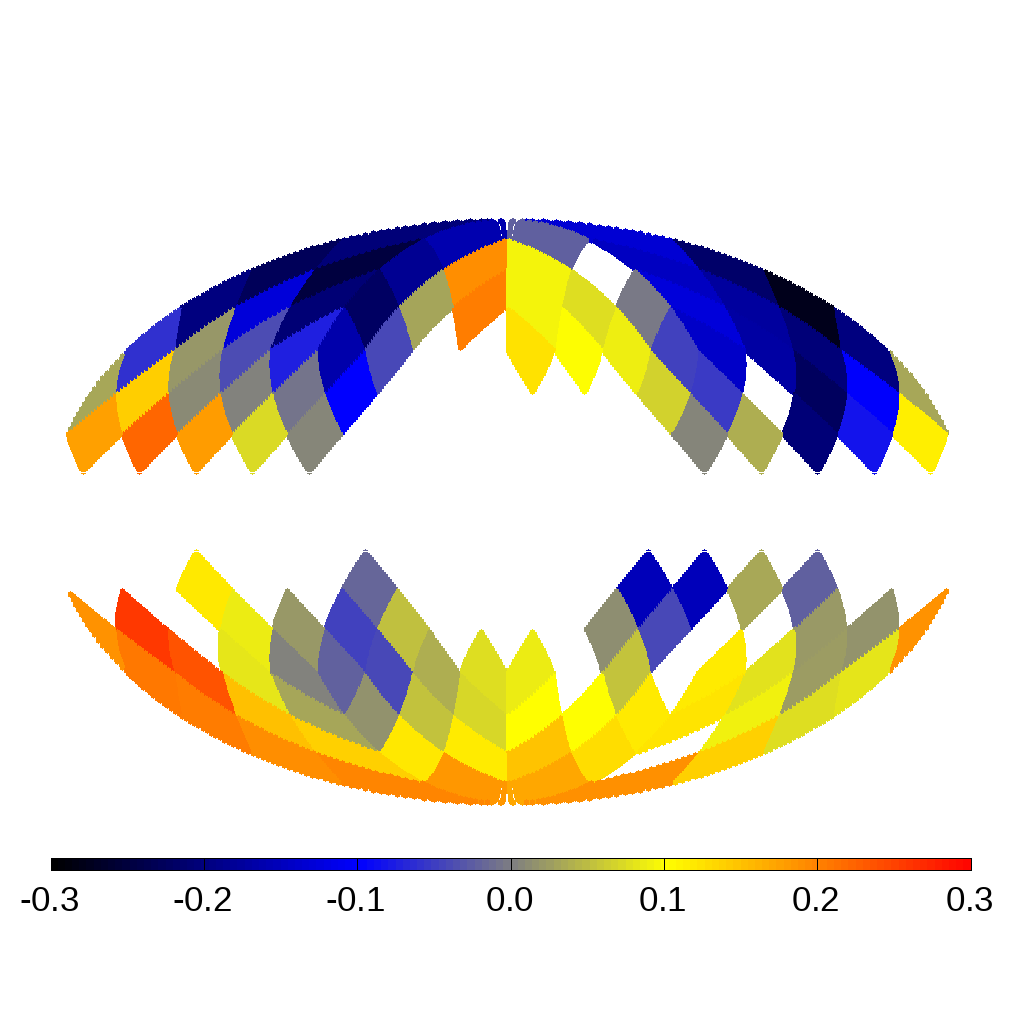}
 \includegraphics[width=0.32\textwidth]{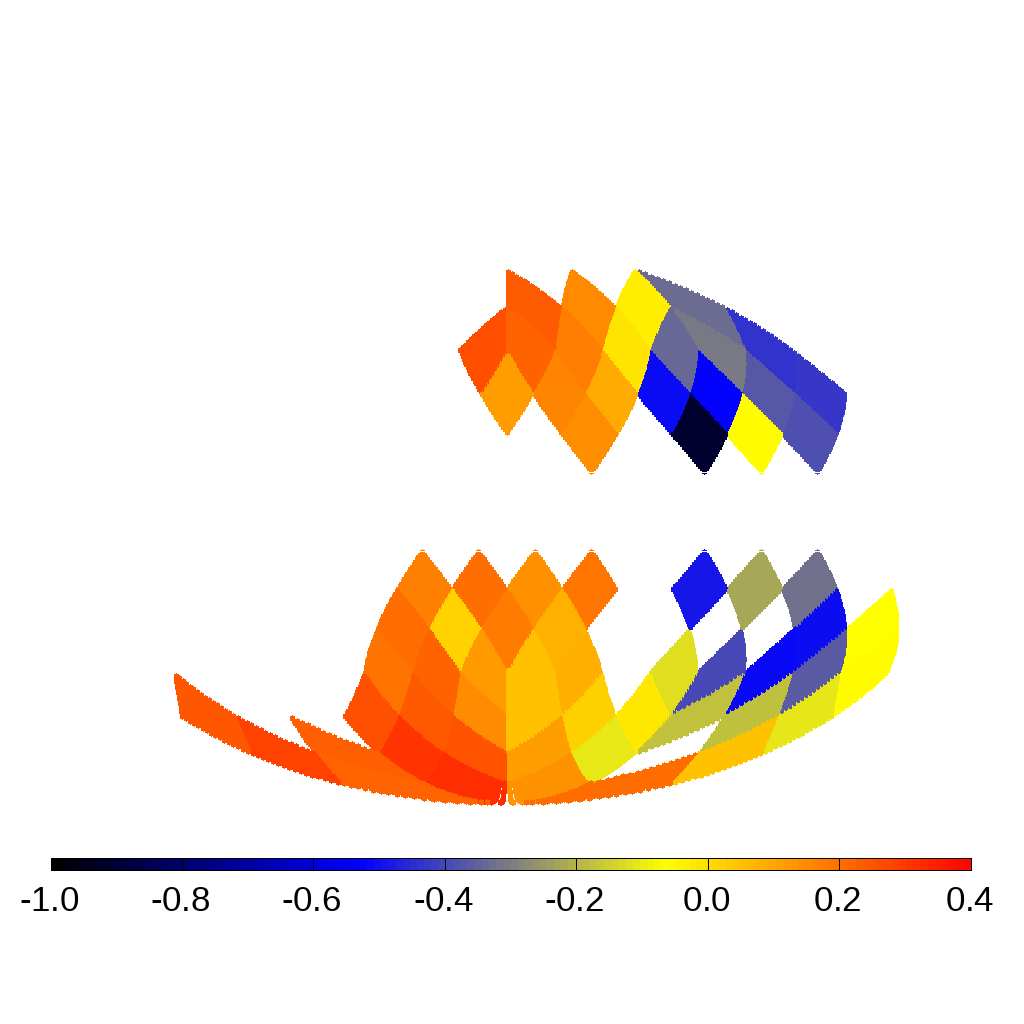}
\caption{Fractional residuals of the model L8a, defined as $(T_i^{\rm data} -T_i^{\rm model})/T_i^{\rm data}$, for the iterative source-masking method of Sect. \ref{sect:mask}.}
\label{Fit3res}
\end{center}
\end{figure*}

The second exception is at 820 MHz. Here the coefficient of the Galactic component is typically larger than one. 
This might be due to calibration issues of the survey or to the limited fraction of the sky available at that frequency (see Fig.~\ref{Fit3res}). 
Indeed, the smaller is the available map the more the Galactic and extragalactic components become degenerate and the less information we have (this fact can possibly lead to peculiar fluctuations with large error bars). This is also testified by the anomalously low reduced $\chi^2$ obtained at 820 MHz.
The most solid results are instead obtained at the frequencies where the largest portion of the sky is available, thus in particular at 45 MHz, 408 MHz and 1420 MHz.

We note that the reduced $\chi^2$ are around one in most of the cases. This suggests that the Galactic emission models are detailed enough to describe the observational data. In particular, this can be considered as an a posteriori check that the scale assumed for the averaging of magnetic turbulences is appropriate. We checked that with different scales we would get a reduced $\chi^2$ far away from one.

In agreement with previous analysis \cite{Bringmann:2011py,DiBernardo:2012zu,Orlando:2013ysa}, we find that models with small $L$ are disfavored by radio data, as can be understood by comparing the $\chi^2$ of the various cases. This occurs because the latitude profile of the synchrotron emission is too steep when $L\lesssim$ kpc.
The values of the $\chi^2$ also give an indication of the values of $L$ preferred by radio data, which is around 8 kpc. However, a precise determination of its value would require a more complete scan of models and parameters. This is beyond the scope of this paper, which is instead focused on the extragalactic estimate.

The temperature of the isotropic background, that we have obtained from the fits, has some dependence on the model of Galactic synchrotron emission considered: the left panels in Fig. \ref{Fit3a} show some scatter
among the derived values of $T_E$ at each frequency, for the different Galactic models. However, the variation of the results due to Galactic modeling is quite limited at all frequencies, typically within a factor of 2, and it is especially small for those frequencies where the radio maps have a large fraction of the sky available. This is the main result of the paper, telling us the the estimates of the isotropic radio background $T_E$ is robust.

To get more insight on this result, we computed the mean values of $T_E$ and the corresponding standard deviations $\sigma_{T_E}$, among all the models and methods we employed.
Although this has not a specific statistical relevance, it nevertheless provides a qualitative understanding. We found $\bar T_E=(1.55\cdot10^4,\,3.30\cdot10^3,\,11.1,\,2.19,\,0.551,\,0.098)$ K and $\sigma_{T_E}=(3.3\cdot10^3,\,310,\,0.8,\,0.19,\,0.030,\,0.016)$ K, at the frequencies 22 MHz, 45 MHz, 408 MHz, 820 MHz, 1420 MHz, and 2326 MHz.
Notice that the standard deviations are a factor of few smaller than the total uncertainty band one could derive by folding in all the errors of all the models (this can be understood from Fig. \ref{Fit3a}, but will be shown and discussed later on, in the left panel of Fig. \ref{TEB}).
This means that the uncertainty is predominantly due to observational limitations rather than from a scatter of the central values obtained with the different models: the latter is actually moderate.
This is another reason to believe that the isotropic estimate is robust against the variation of the Galactic models. 

Another crucial results is that the values of the isotropic emission we obtain is 
significantly and systematically larger than what is inferred from the number counts of extragalactic sources. 
The latter has been computed by several groups, with a good agreement between different estimates (among the recent ones, see Ref. \cite{Vernstrom:2011xt,Gervasi:2008rr}). In the left panels of Fig.~\ref{Fit3a}, we show with black points the results of Ref. \cite{Gervasi:2008rr}, as long as an analytic fit to these data (more precisely: FIT1 in Table 5 of Ref. \cite{Gervasi:2008rr}). These points lie significantly below our estimates of the isotropic radio background.

In order to better quantify this excess, we perform two additional fits to the radio maps: in one case we fix the temperature of the isotropic background to the value suggested by the number counts, taken from Ref. \cite{Gervasi:2008rr}; in the second case, we
have assumed a vanishing isotropic emission temperature. The results are shown in Fig.\ref{Fit3b}. In the former case, the increase of the $\chi^2$ with respect to the previous best-fits, quantifies the statistical significance of the excess: namely, it tells us the confidence at which we can reject the null hypothesis, i.e. that isotropic background is equal to the number counts estimate. In the latter case, we can establish at which confidence level we can affirm that the radio extragalactic background is non-zero. Fig. \ref{Fit3b} shows that the presence of an isotropic background is established with large significance at all frequencies.
An excess with respect to the estimates from number counts is above $5\sigma$ C.L. at 45 MHz, 408 MHz and 1420 MHz; it is less significant but still solid at 820 MHz and 2326 MHz, while the picture is more uncertain at 22 MHz. 
Note that the largest significance is for the maps which cover the largest fraction of the sky,
so for the most robust cases (which also turn out to be the most stable in fitting the Galactic dominated part).
This result tells us again that the uncertainty in the extragalactic estimate is dominated by the scarcity of available data rather than by a scatter induced from considering different Galactic models.

To have an indication of the quality of the fits, we compute the residuals. Fig.~\ref{Fit3res} shows the fractional residuals, i.e. $R_i^{\%}=(T_i^{\rm data} -T_i^{\rm model})/T_i^{\rm data}$, for model L8a, one of the models which better fit the data.
The average of $|R_i^{\%}|$ in the map is 0.15, 0.13, 0.16, 0.11, 0.11, 0.22 at the frequencies 22 MHz, 45 MHz, 408 MHz, 820 MHz, 1420 MHz, and 2326 MHz, respectively.

\begin{figure*}[t]
\begin{center}
 \includegraphics[width=0.34\textwidth]{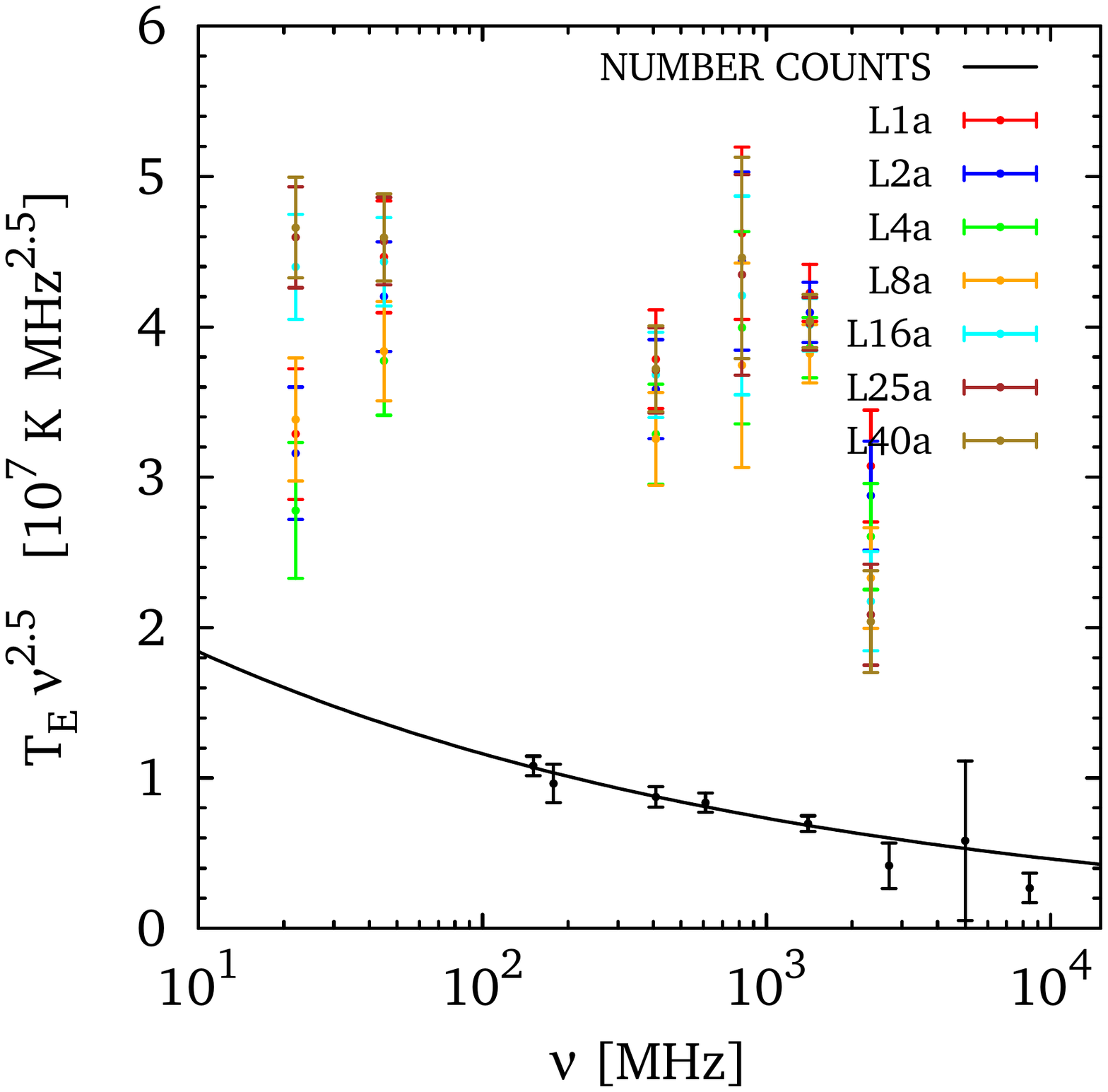}\hspace{-0.4cm}
 \includegraphics[width=0.34\textwidth]{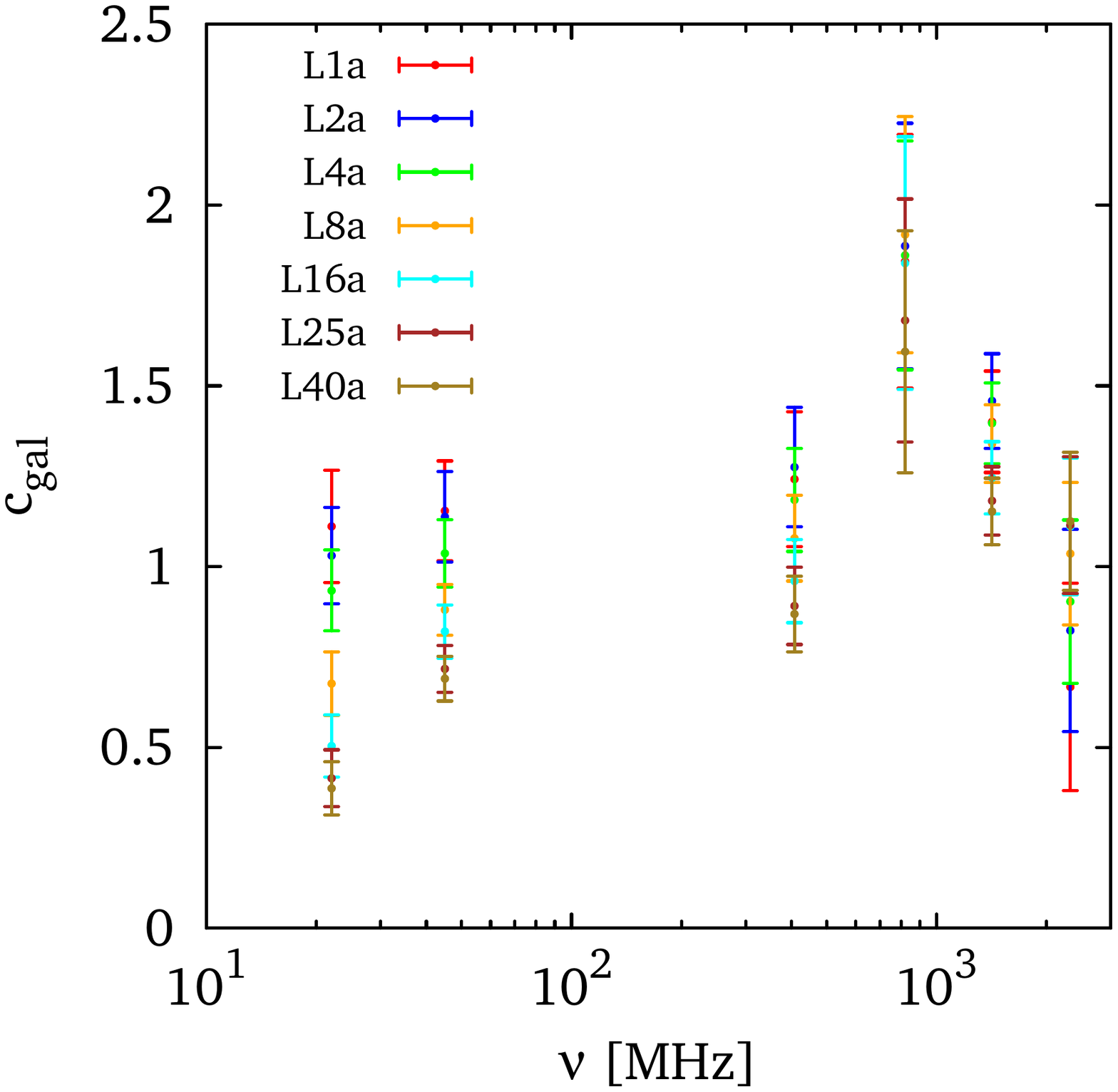}\hspace{-0.4cm}
 \includegraphics[width=0.34\textwidth]{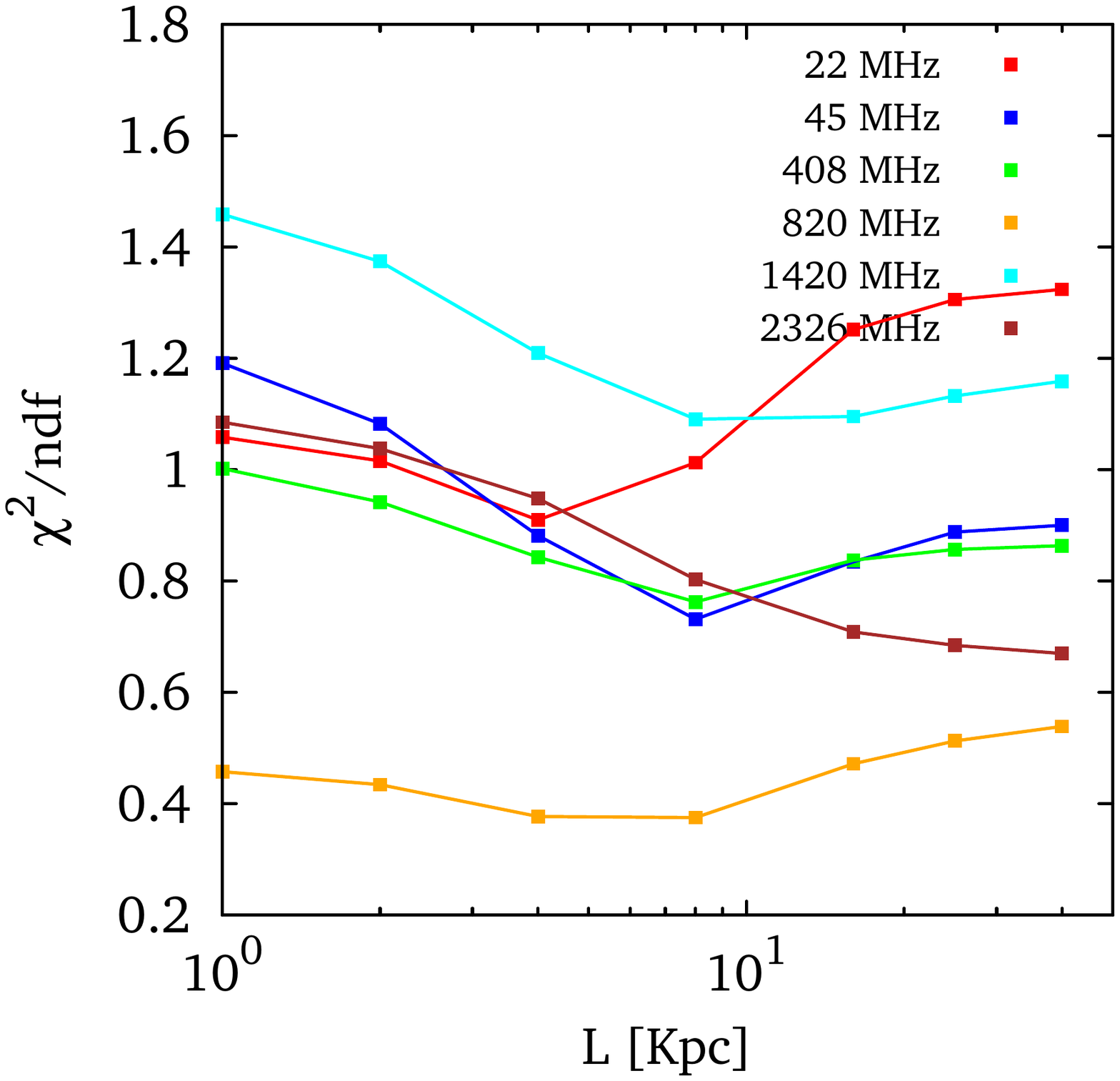}\\ \vspace{-2cm}
 \includegraphics[width=0.34\textwidth]{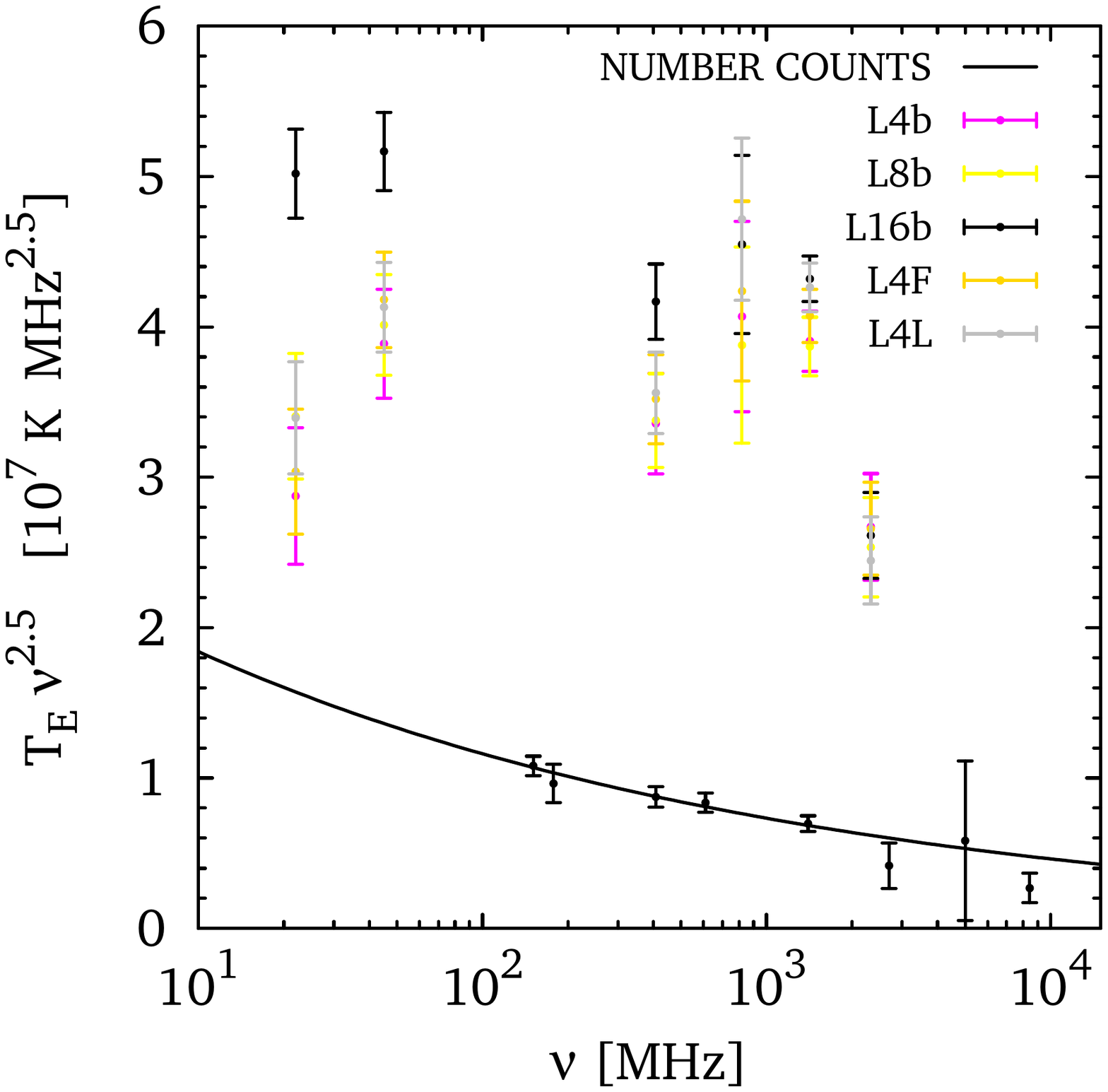}\hspace{-0.4cm}
 \includegraphics[width=0.34\textwidth]{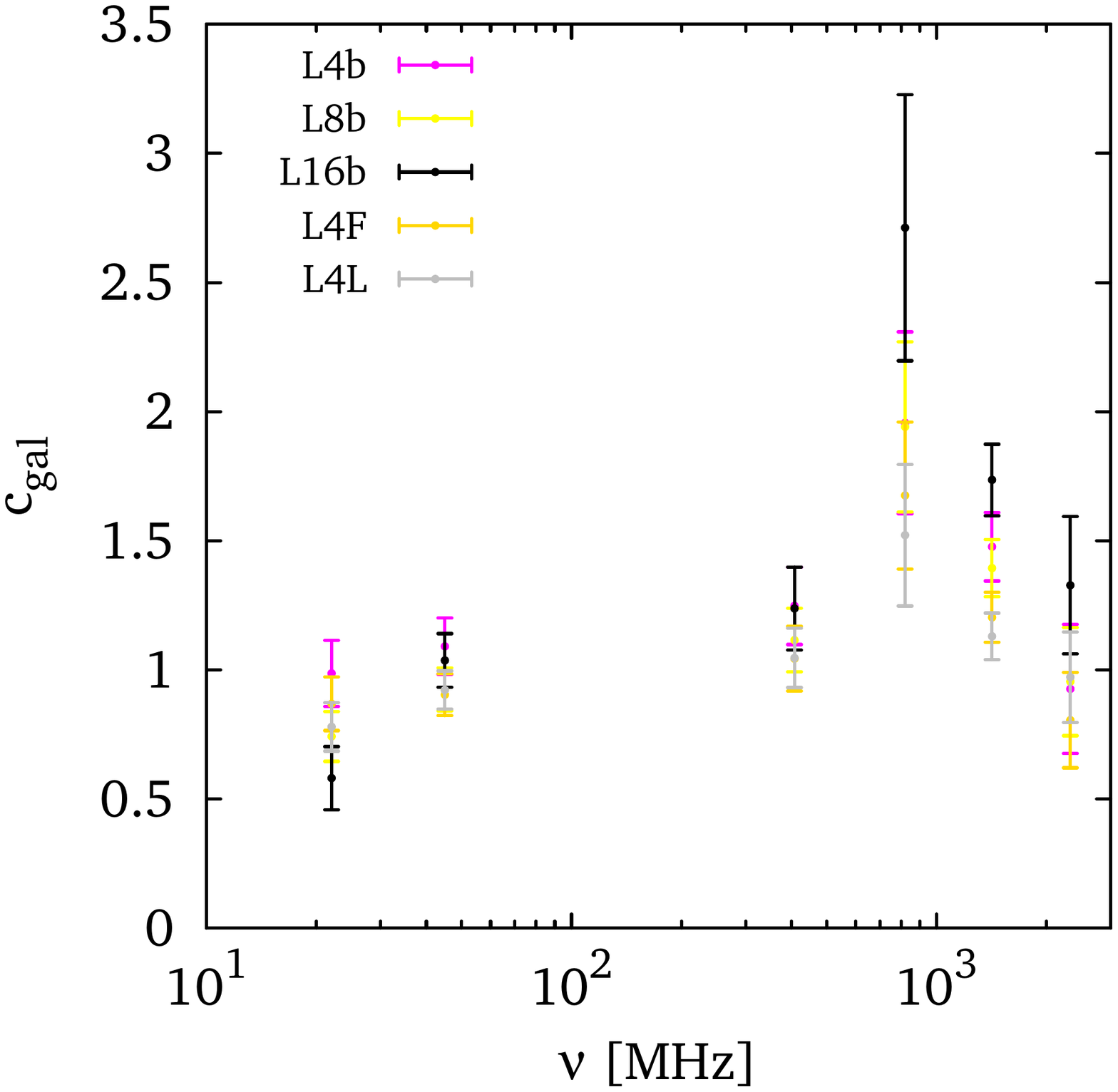}\hspace{-0.4cm}
 \includegraphics[width=0.34\textwidth]{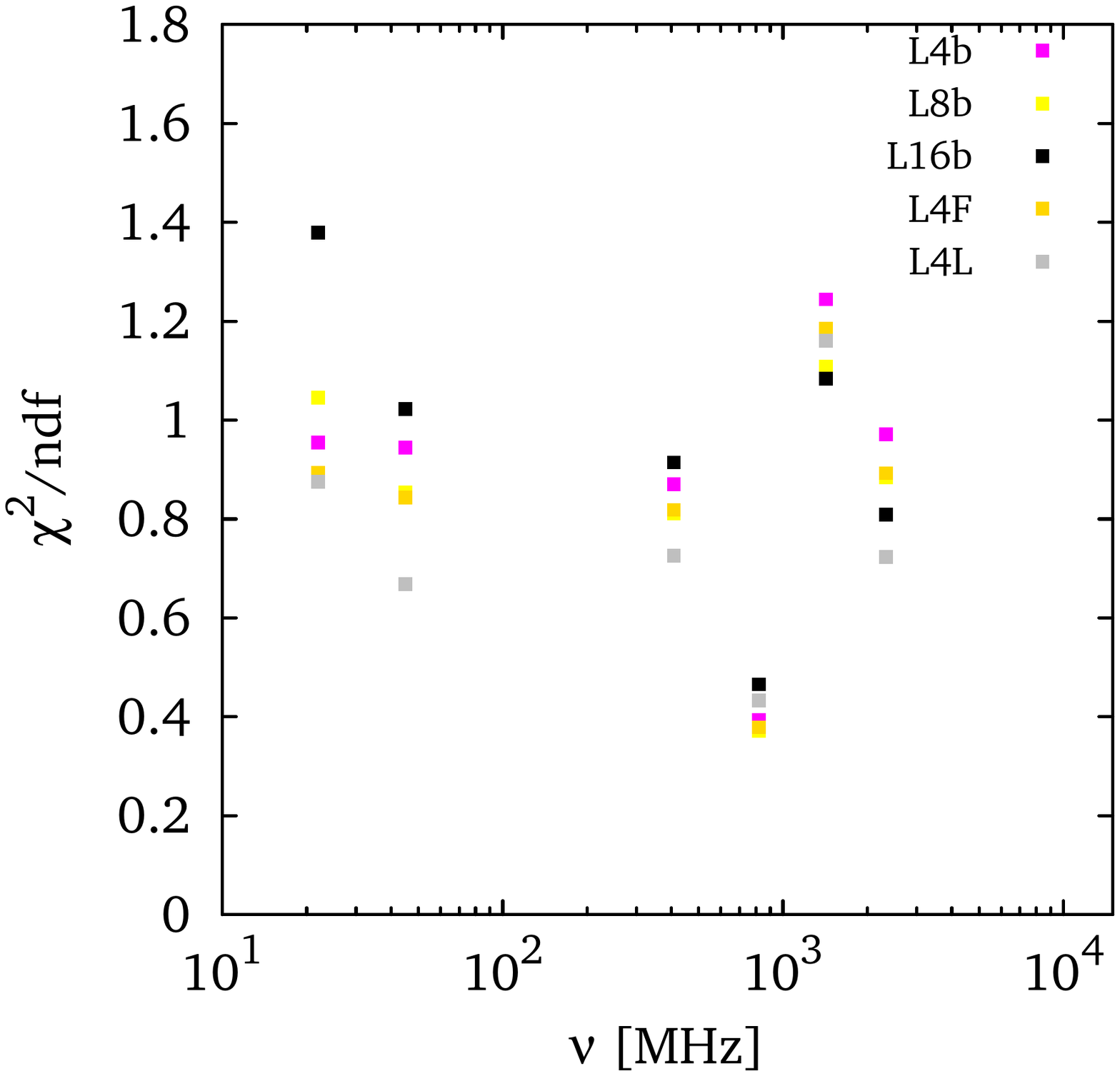}\\ \vspace{-1cm}
 \includegraphics[width=0.34\textwidth]{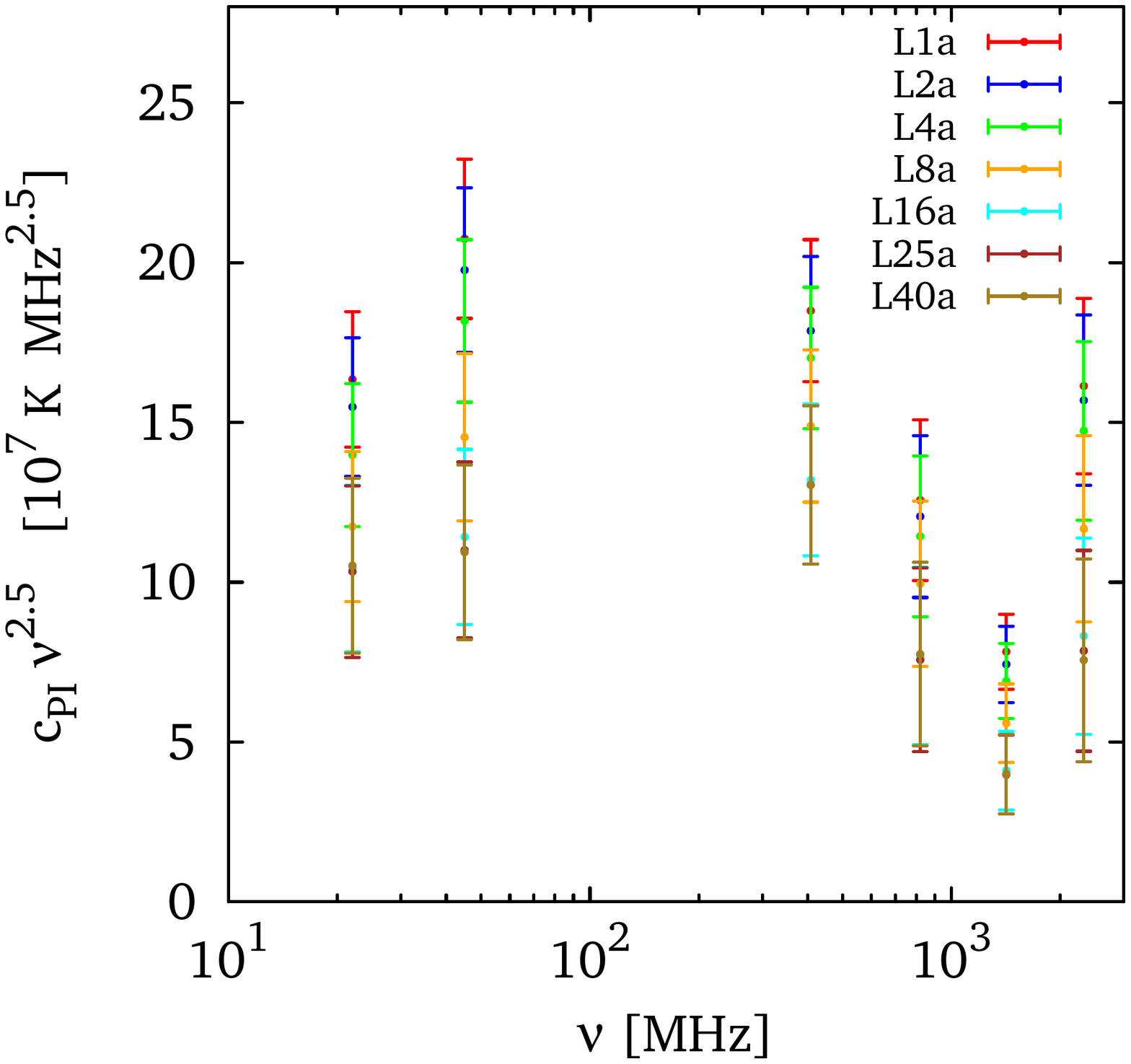}\hspace{-0.4cm}
 \includegraphics[width=0.34\textwidth]{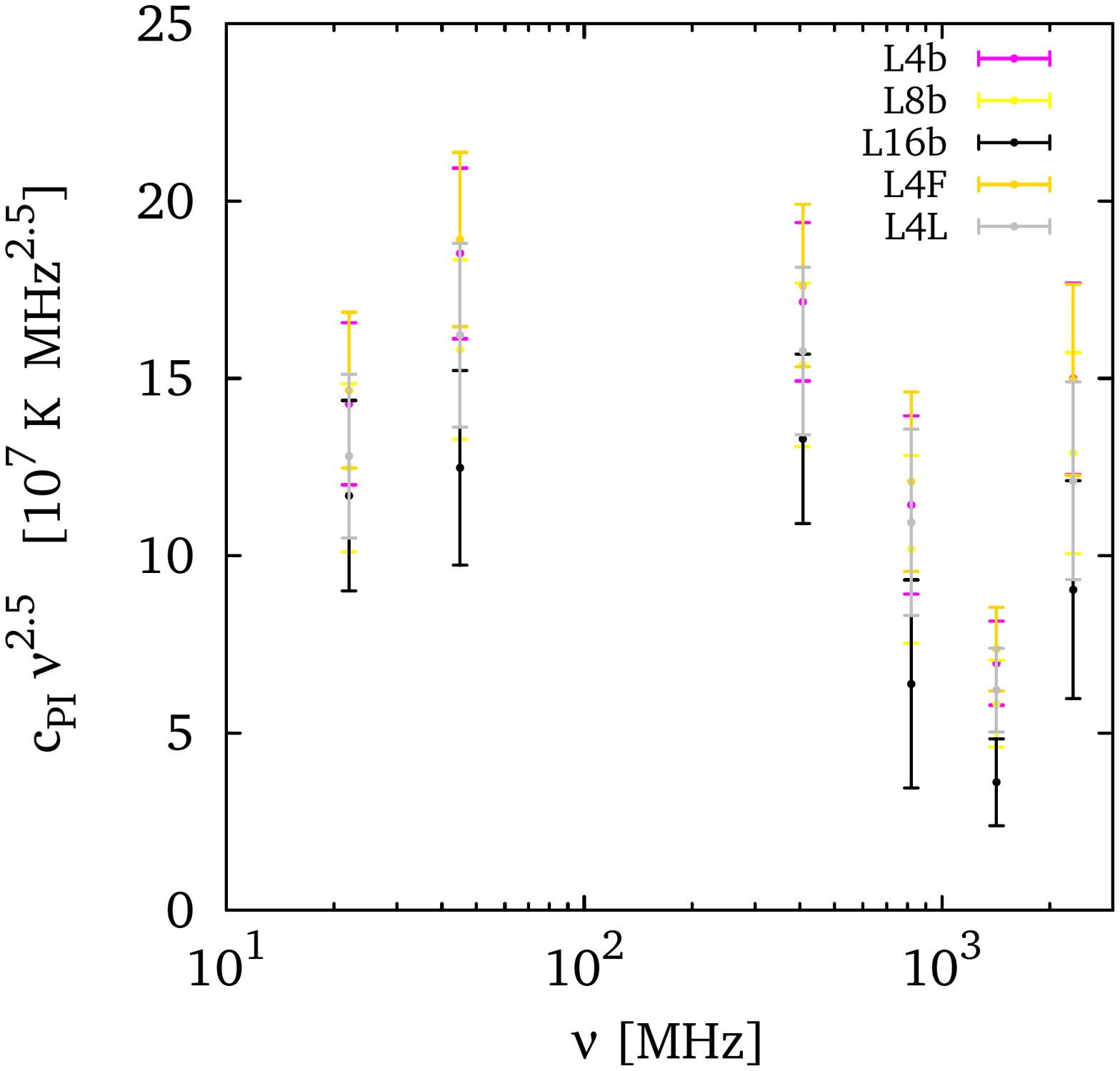}\hspace{-0.4cm}
\caption{Results of the fits obtained following the method of Sect. \ref{sect:template}, i.e. when templates are adopted. The panels show the same type of information of Fig. \ref{Fit3a}, with the inclusion of a third raw, where the values of the additional normalization coefficients $c_{\rm PI}$ (multiplied by $\nu^{2.5}$) of the polarization template are reported.
}
\label{Fit4a}
\end{center}
\end{figure*}

\begin{figure*}[t]
\begin{center}
 \includegraphics[width=0.34\textwidth]{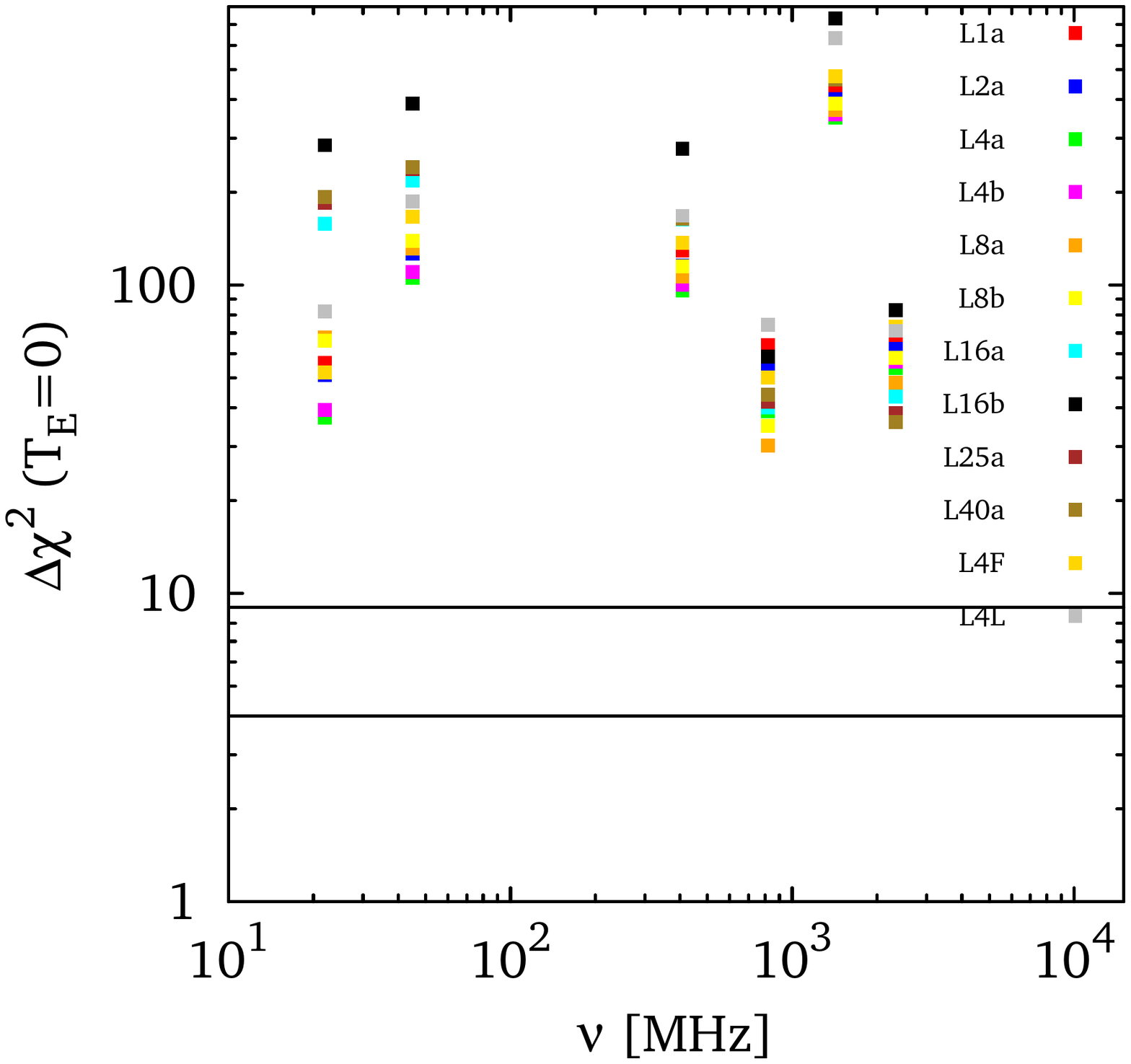}\hspace{-0.4cm}
 \includegraphics[width=0.34\textwidth]{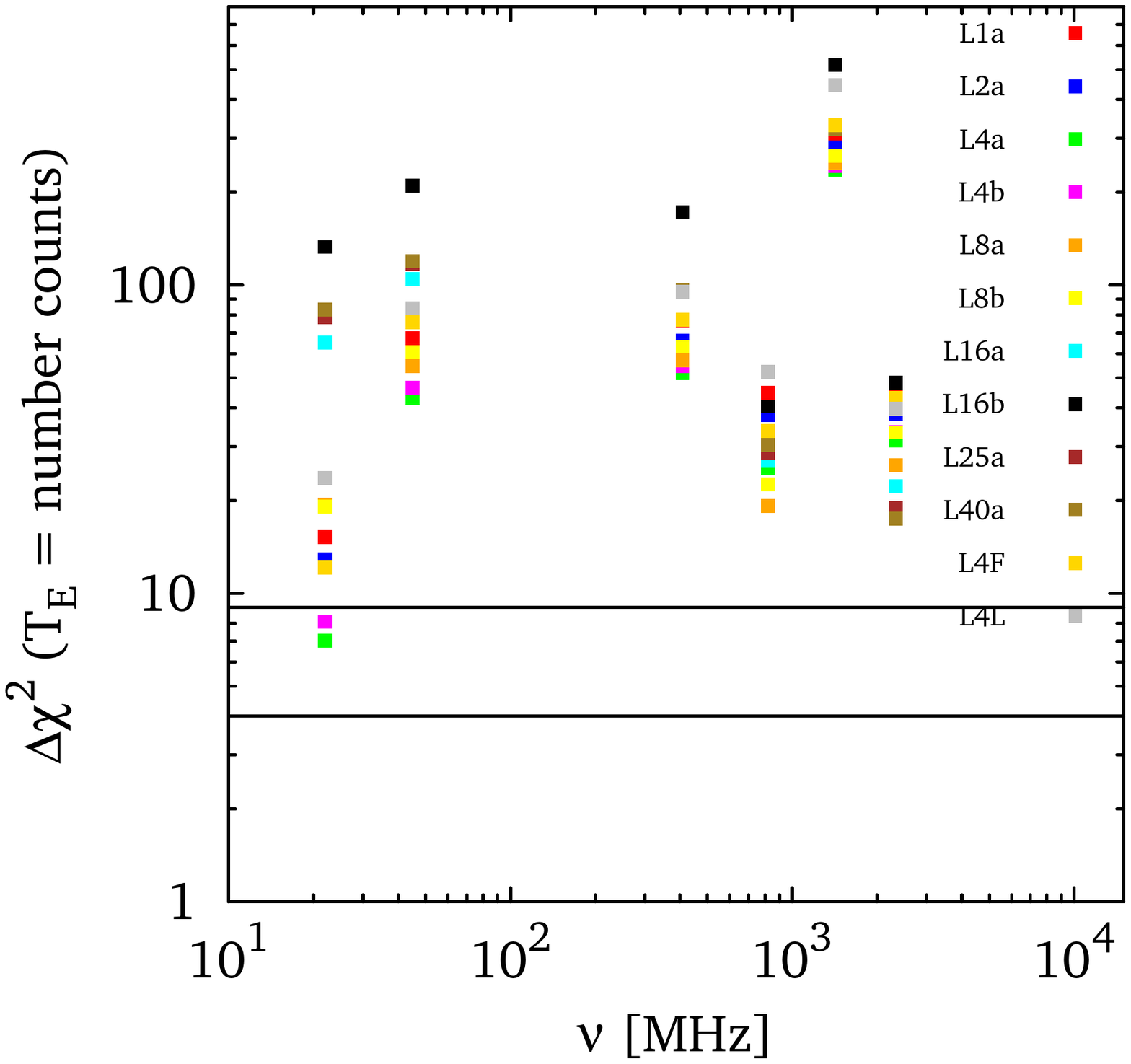}\hspace{-0.4cm}
\caption{The same as in Fig. \ref{Fit3b}, but for the method of Sect.\ref{sect:template},
i.e. when source templates are adopted.
}
\label{Fit4b}
\end{center}
\end{figure*}

\begin{figure*}[t]
\begin{center}
 \includegraphics[width=0.32\textwidth]{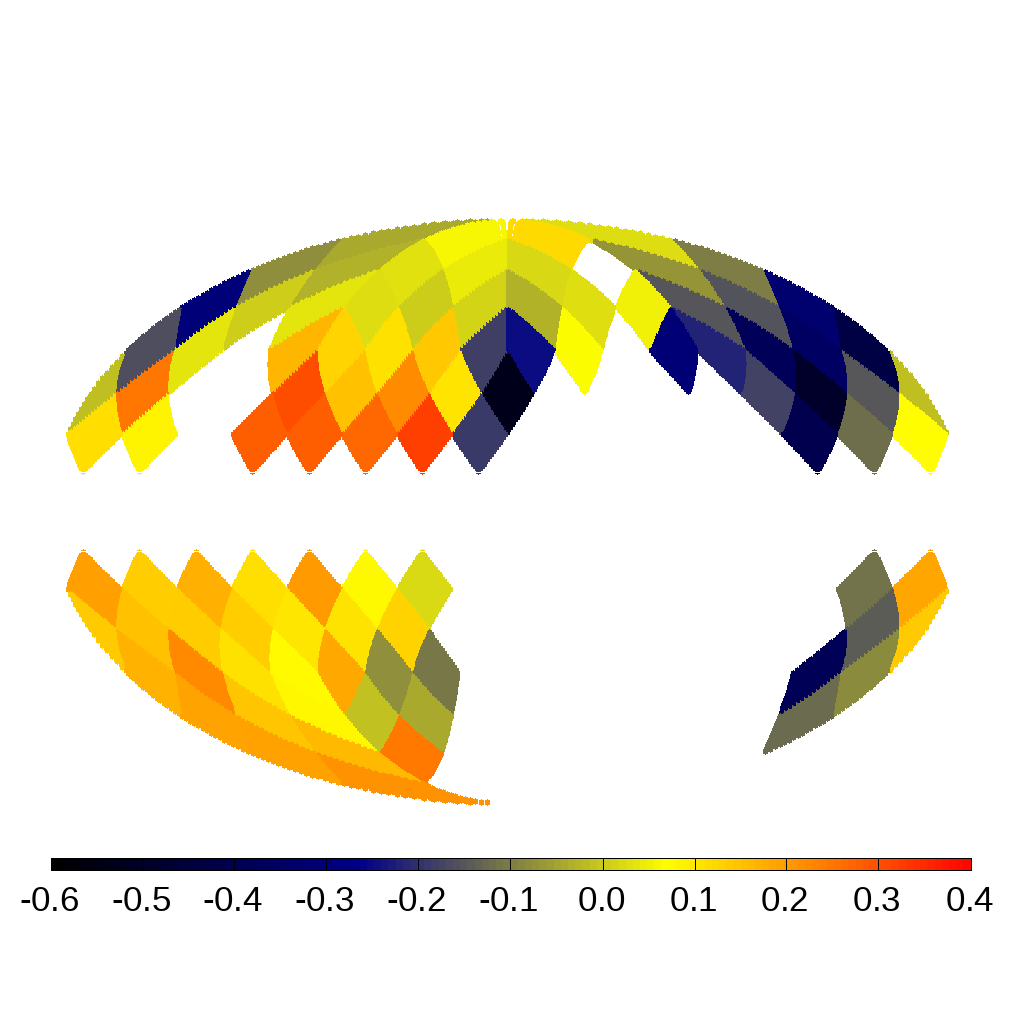}
 \includegraphics[width=0.32\textwidth]{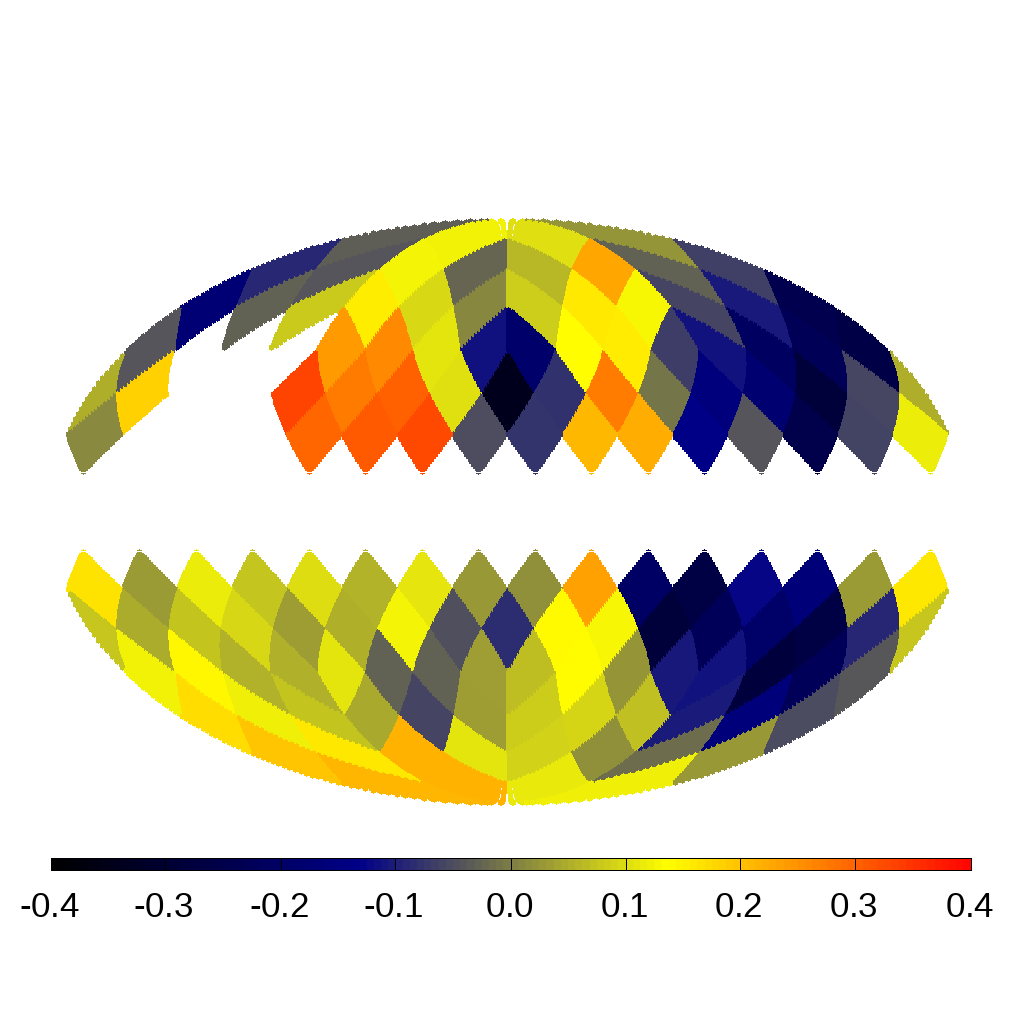}
 \includegraphics[width=0.32\textwidth]{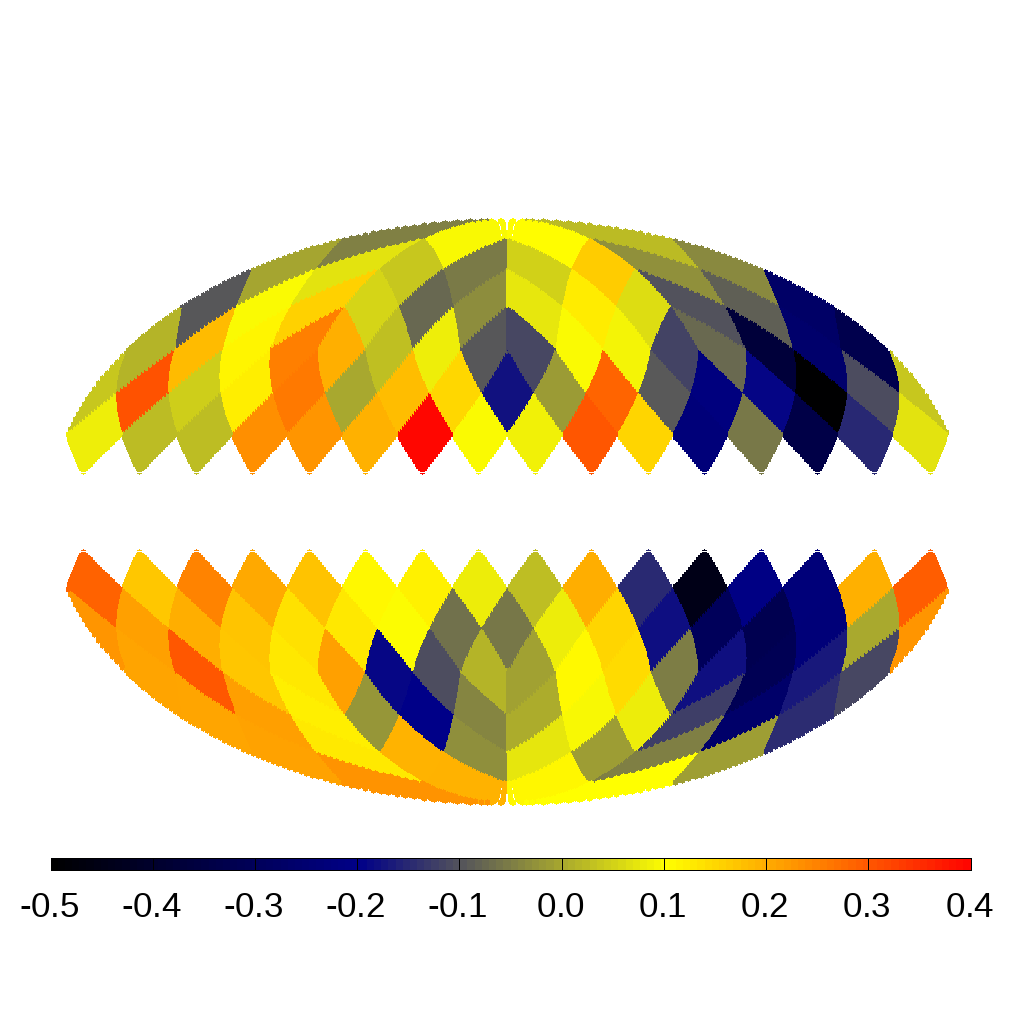}\\
 \includegraphics[width=0.32\textwidth]{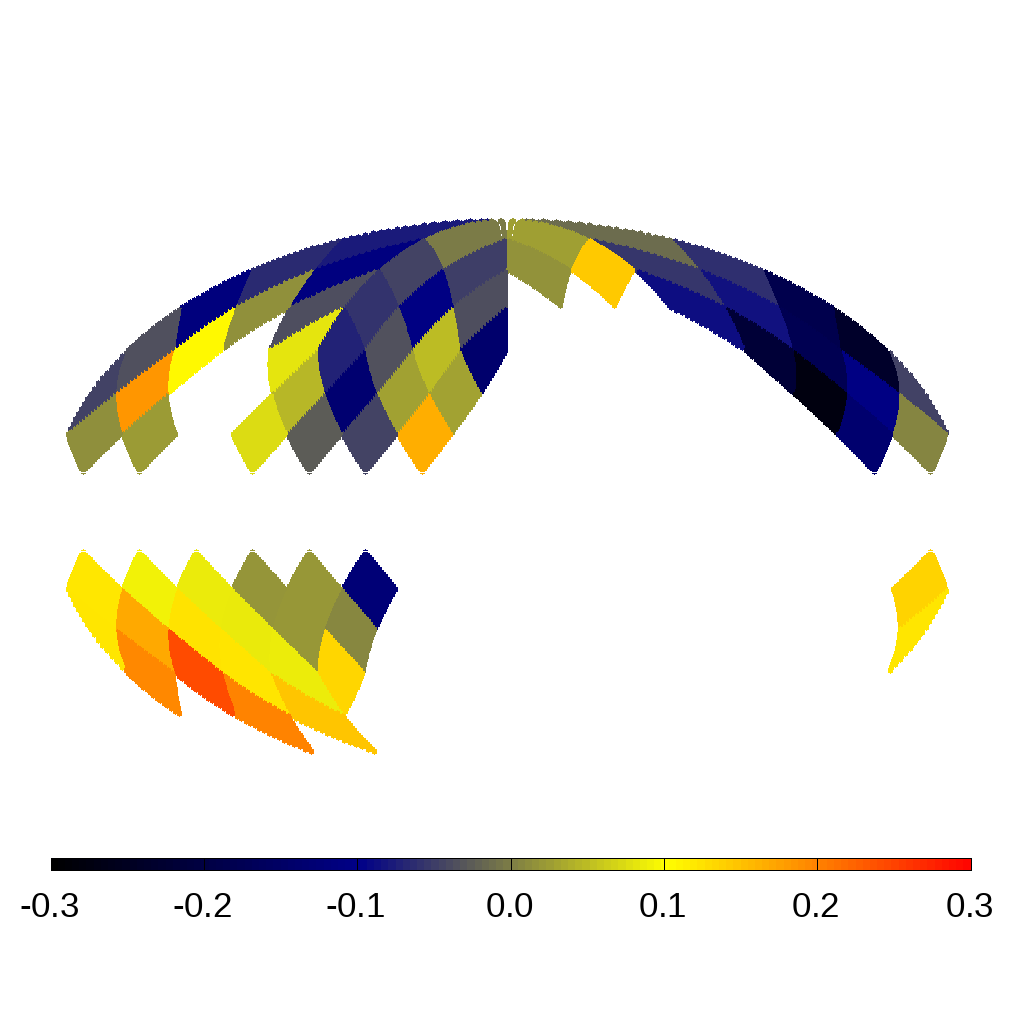}
 \includegraphics[width=0.32\textwidth]{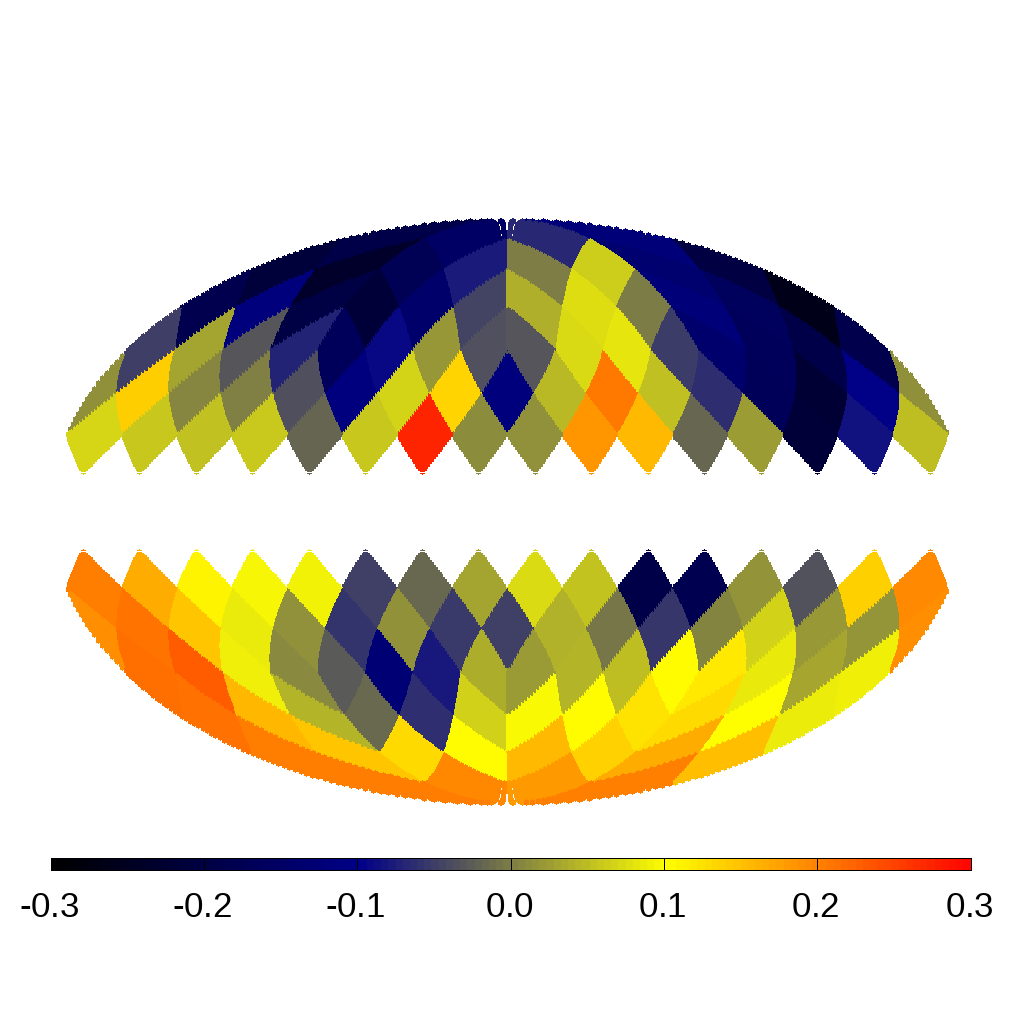}
 \includegraphics[width=0.32\textwidth]{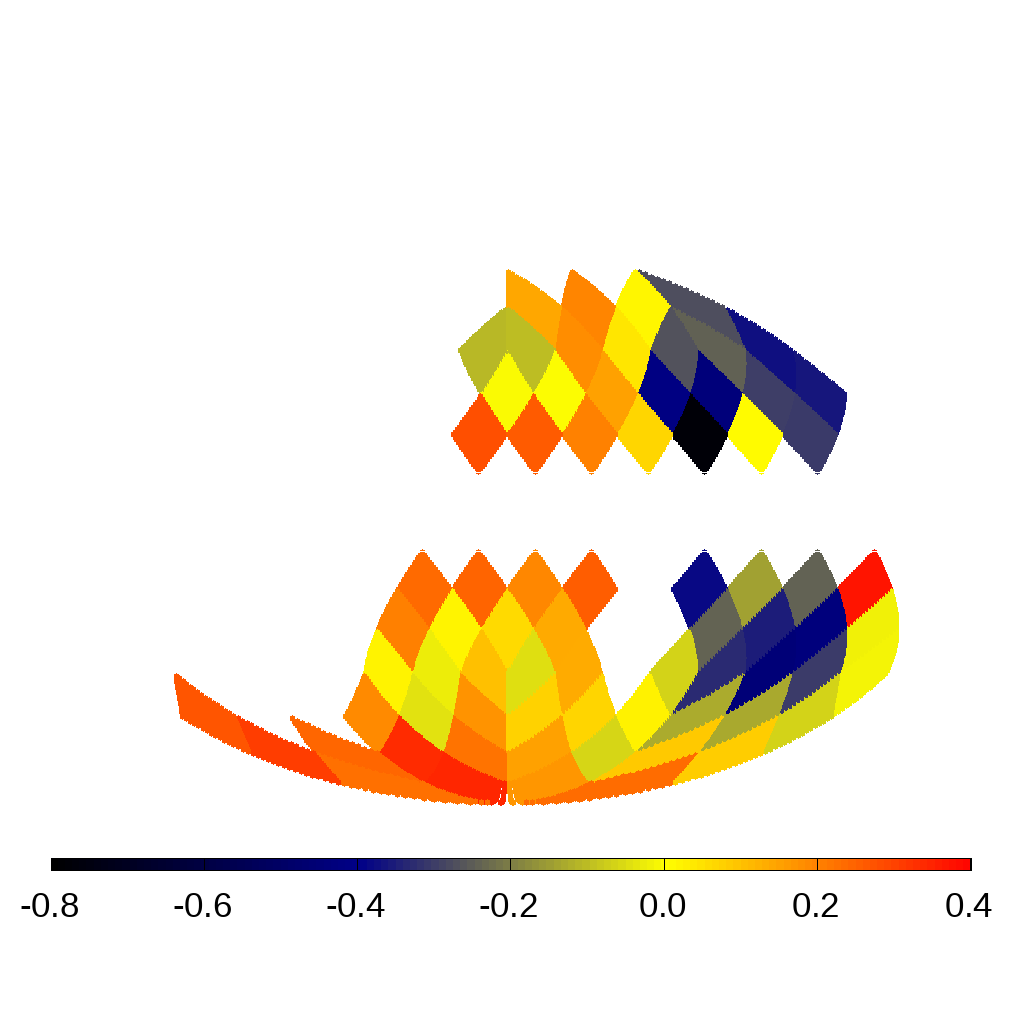}
\caption{Fractional residuals of the model L8a, defined as $(T_i^{\rm data} -T_i^{\rm model})/T_i^{\rm data}$, for the source template method of Sect. \ref{sect:template}.}
\label{Fit4res}
\end{center}
\end{figure*}

When we adopt the template method for accounting for single sources, as described in Sect.~\ref{sect:template} (instead of the mask technique, as done in the discussion so far) the results are very similar. They are shown in Fig.~\ref{Fit4a}, Fig.~\ref{Fit4b} and \ref{Fit4res}.
However, the significance of the excess with respect to the isotropic background inferred from number counts, is now larger and becomes quite relevant at all frequencies (basically above $3\sigma$ in all cases). This is due to the larger sky coverage considered in this analysis (since now there is no masked region, except for $|b|<10$ degrees). The average values of the
fractional residuals $|R_i^{\%}|$ for the model L8a are now: 0.15, 0.13, 0.14, 0.09, 0.10 and 0.19 at 22 MHz, 45 MHz, 408 MHz, 820 MHz, 1420 MHz, and 2326 MHz, respectively.

\begin{figure*}[t]
\begin{center}
 \includegraphics[width=0.48\textwidth]{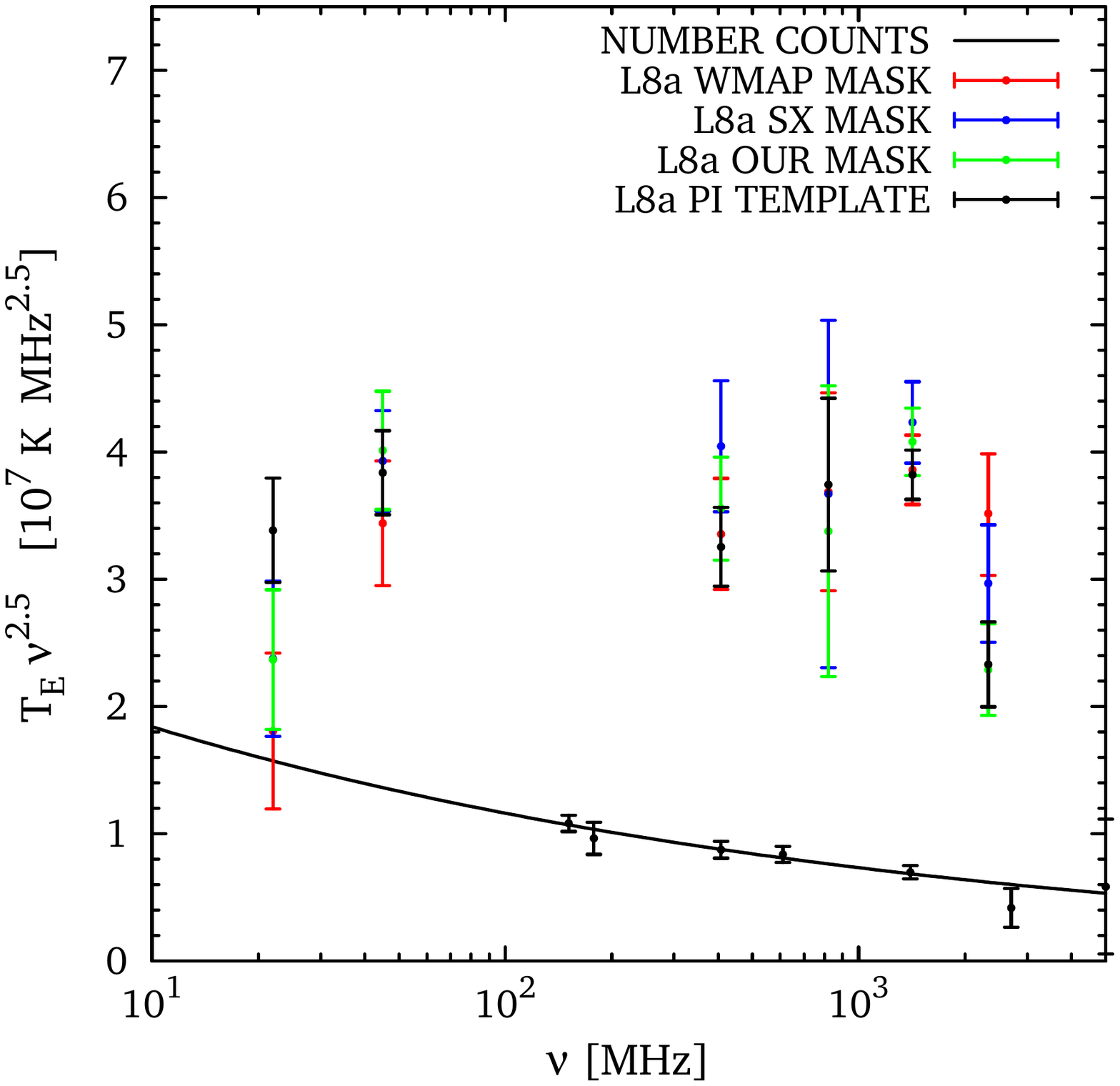}
 \includegraphics[width=0.48\textwidth]{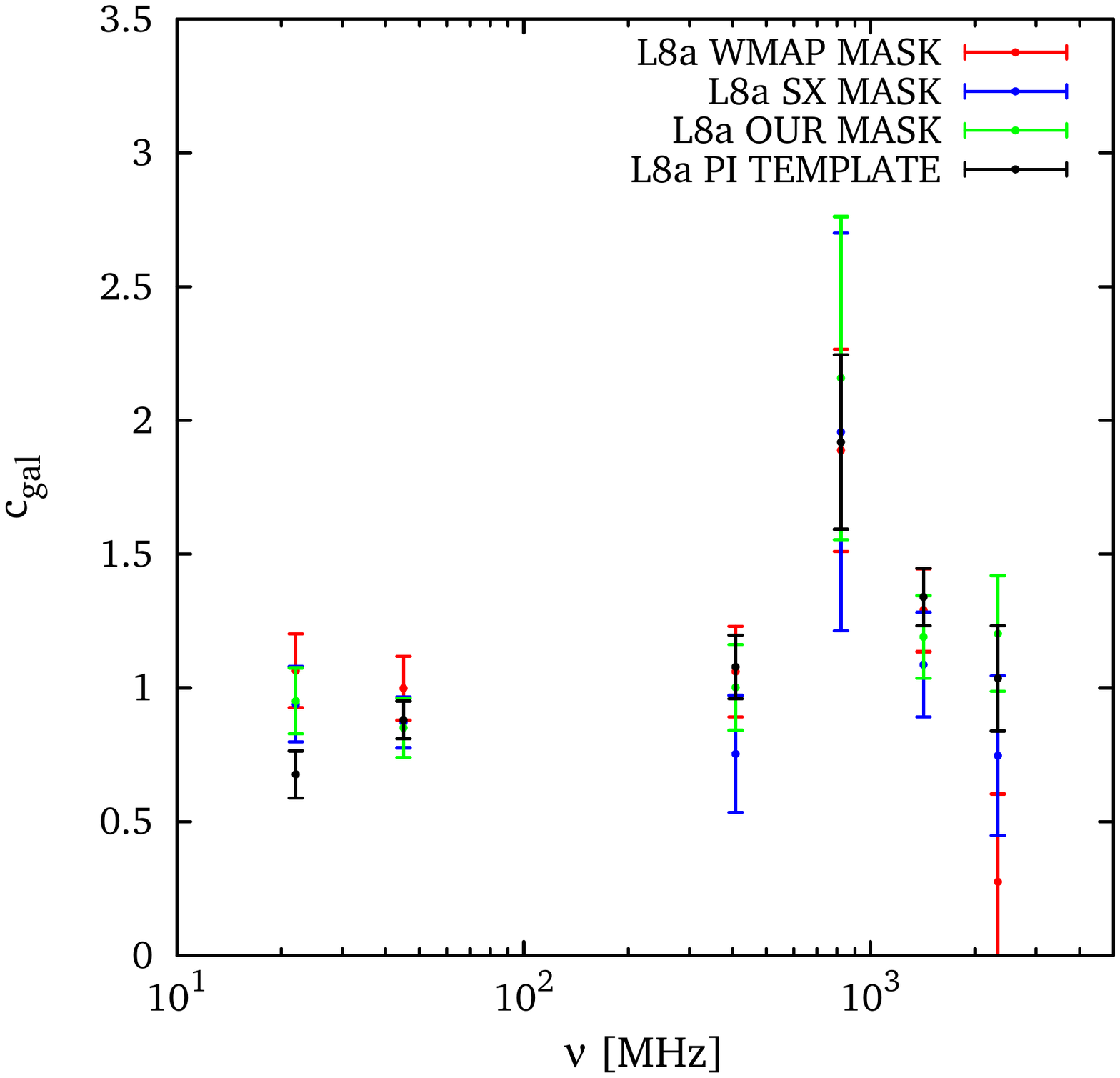}
\caption{{\sl Left:} Comparison of the best-fit values of the isotropic temperature $T_E$ at the different frequencies $\nu$, obtained with the various methods adopted in our analysis. The lower black points \cite{Gervasi:2008rr} and the solid line, which fits them, show the extragalactic temperature expected from number counts. {\sl Right:} Derived values of the normalization
parameter $c_{\rm gal}$ of the Galactic synchrotron emission.
Both plots refer to the L8a Galactic model. See 
Sect. \ref{sec:Results} for more details.
}
\label{Fit_comparison1}
\end{center}
\end{figure*}

As discussed in Sect.~\ref{sect:mask}, we have further considered additional masks, one from the WMAP collaboration~\cite{Bennett:2012zja} and 6 masks (one for each frequency) computed by means of the package SExtractor~\cite{Bertin:1996fj}.
For the sake of conciseness, we do not show again all the detailed outcomes, which are again similar to the ones already discusses.  Rather, we focus on  model L8a and, in Fig.~\ref{Fit_comparison1}, we compare the estimates of the isotropic temperature $T_E$ and of the Galactic coefficients considering the different treatments of sources we employed (three different masks and a template).
The results are quite consistent among different techniques, and we found similar plots when taking models other than L8a for the Galactic synchrotron emission. 
As said before, in maps with limited sky coverage, the Galactic and extragalactic components become highly degenerate. This explains why at 2326 MHz and with a large mask (WMAP case), the coefficient normalizing the Galactic emission is very small (even consistent with zero).

\subsection{Other sources of uncertainty}
\label{comparison}
In the previous Section, we discussed how the estimate of the isotropic background can be affected by the variation of $L$, the size of the diffusion box, which is the most crucial parameter in the modeling of the Galactic emission.
In this Section we aim at studying the role of other sources of uncertainties.

We start with the Galactic magnetic field, which is also an important input for the synchrotron emission. As already mentioned, the feature which can in principle provide the biggest uncertainty is its $z$-extension. We discussed in Sect. \ref{sec:magnetic}, that for the vertical exponential dependence, two different cases for its scale $z_B$ are investigated: $z_B=L$ (model $a$) and $z_B=2$ kpc (model $b$). As shown  in Figs. \ref{Fit3a} and \ref{Fit4a}, the impact on the results is quite modest for the models with $L=4$ kpc and $L=8$ kpc.
For larger sizes of the halo (model $L=16$ kpc), model $b$ gives a slightly larger estimate of the isotropic background. This is because small values of $z_B$ suppress the emission at large latitudes and thus a larger isotropic component is needed in order to fit the data. 

\begin{figure*}[t]
\begin{center}
 \includegraphics[width=0.34\textwidth]{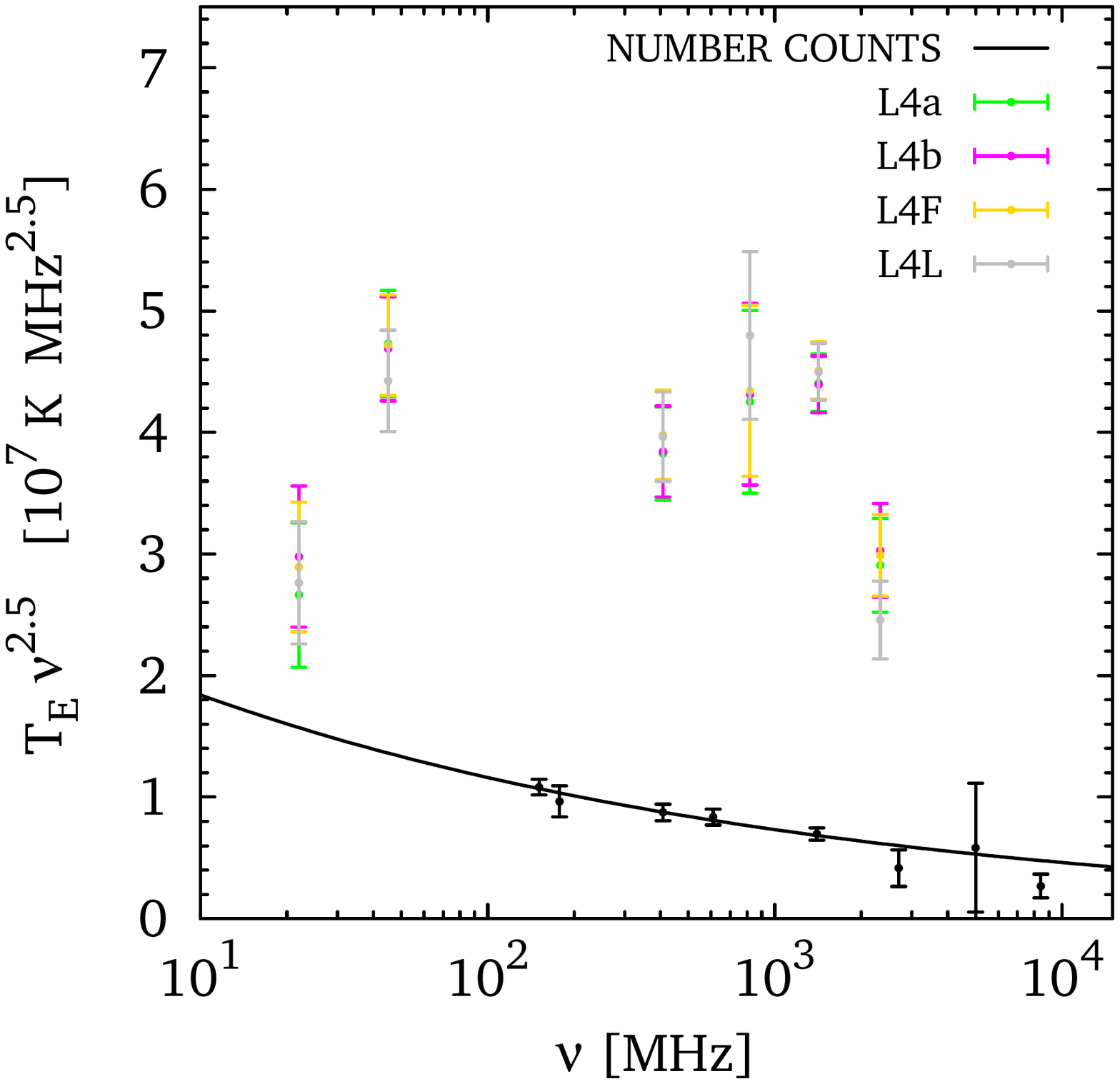}\hspace{-0.4cm}
 \includegraphics[width=0.34\textwidth]{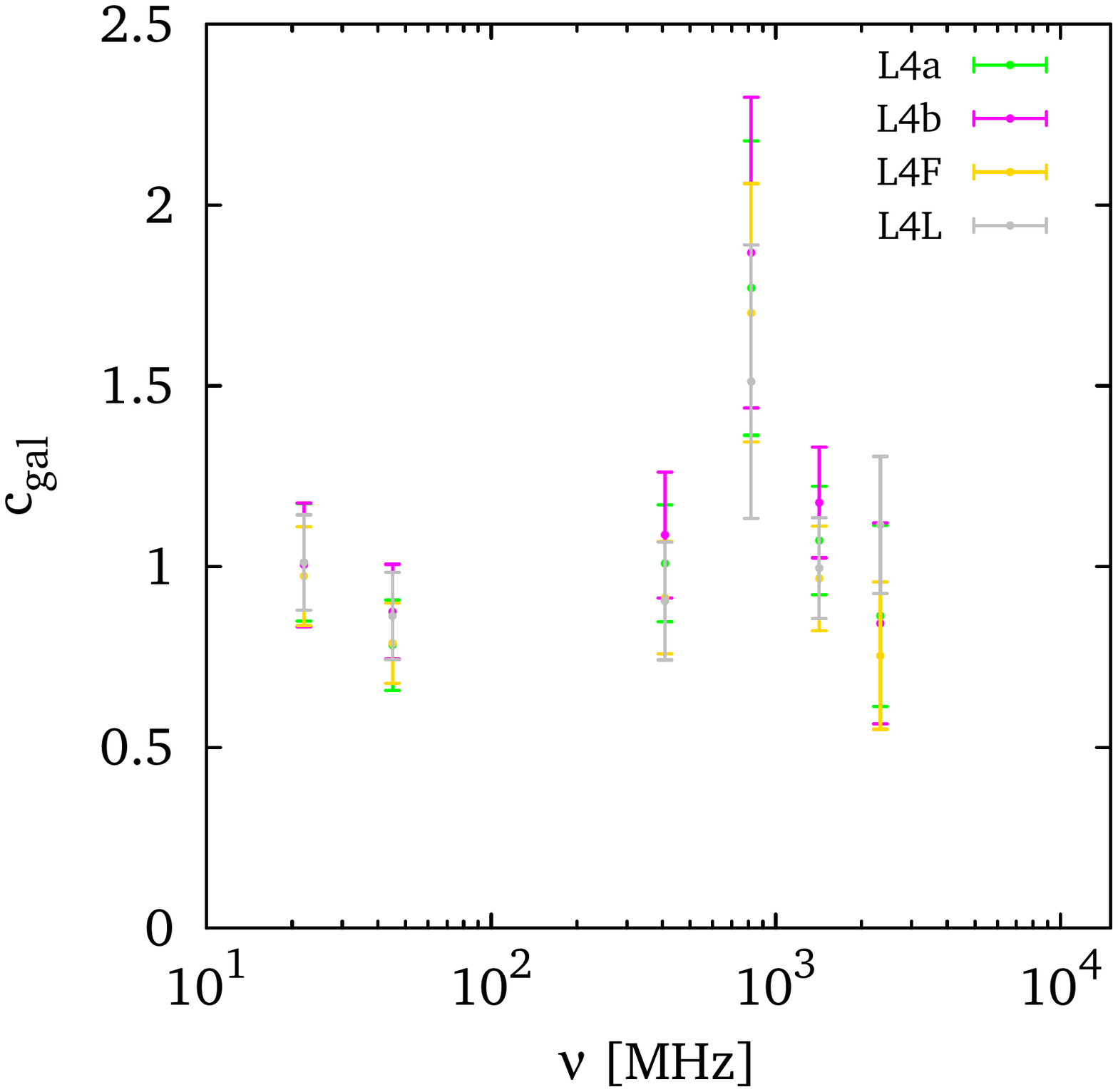}\hspace{-0.4cm}
 \includegraphics[width=0.34\textwidth]{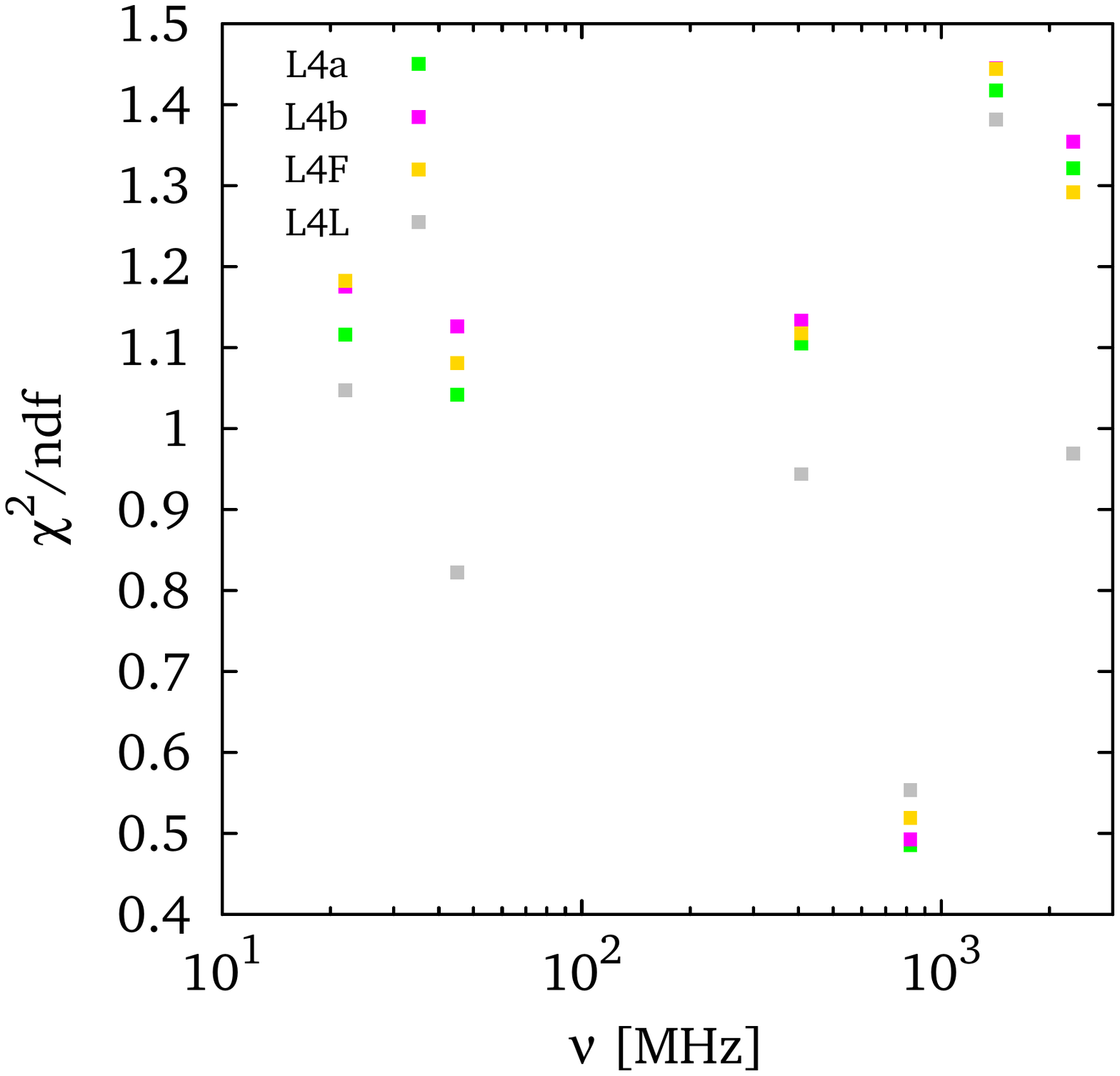}\\ \vspace{-2cm}
 \includegraphics[width=0.34\textwidth]{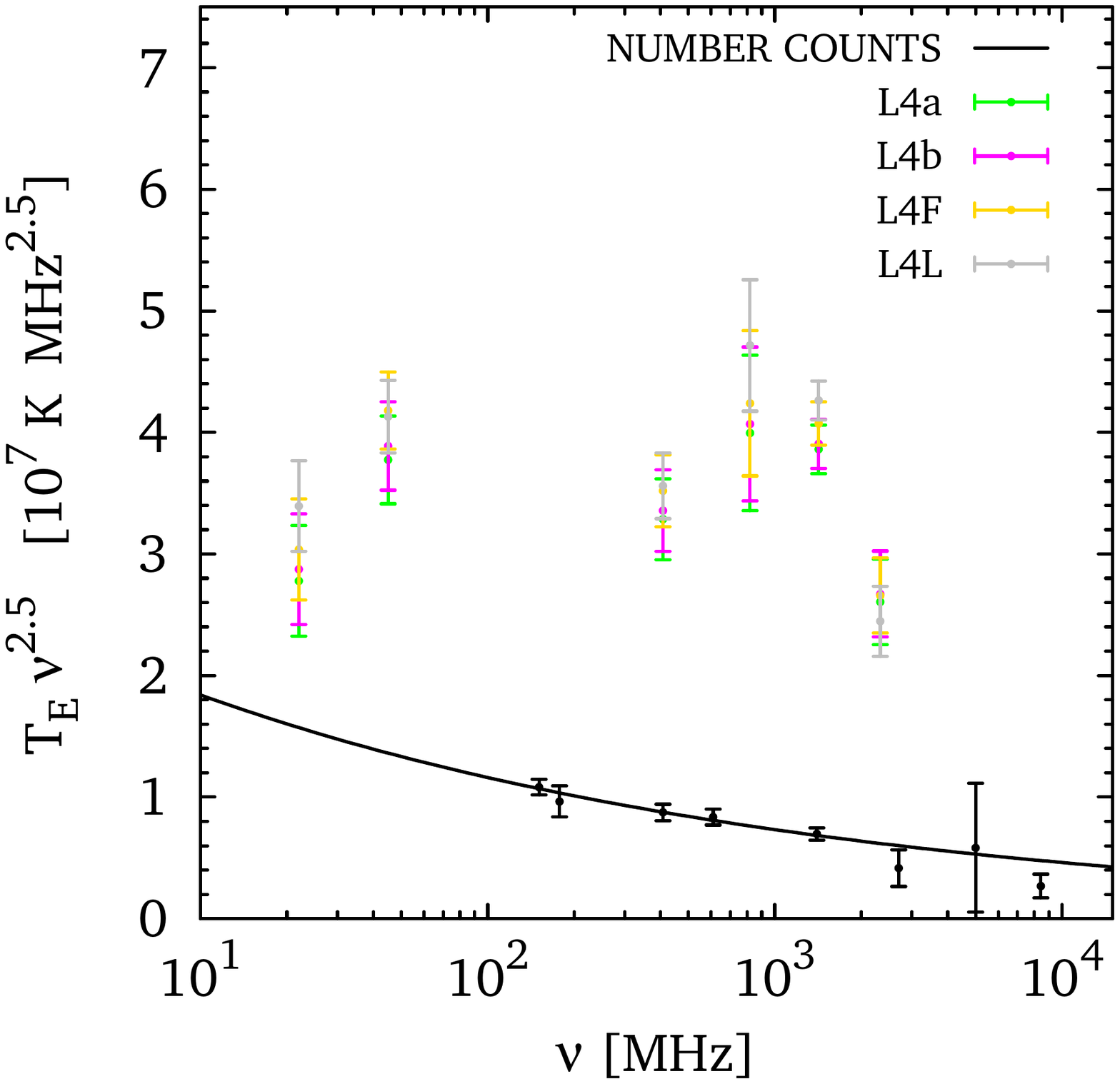}\hspace{-0.4cm}
 \includegraphics[width=0.34\textwidth]{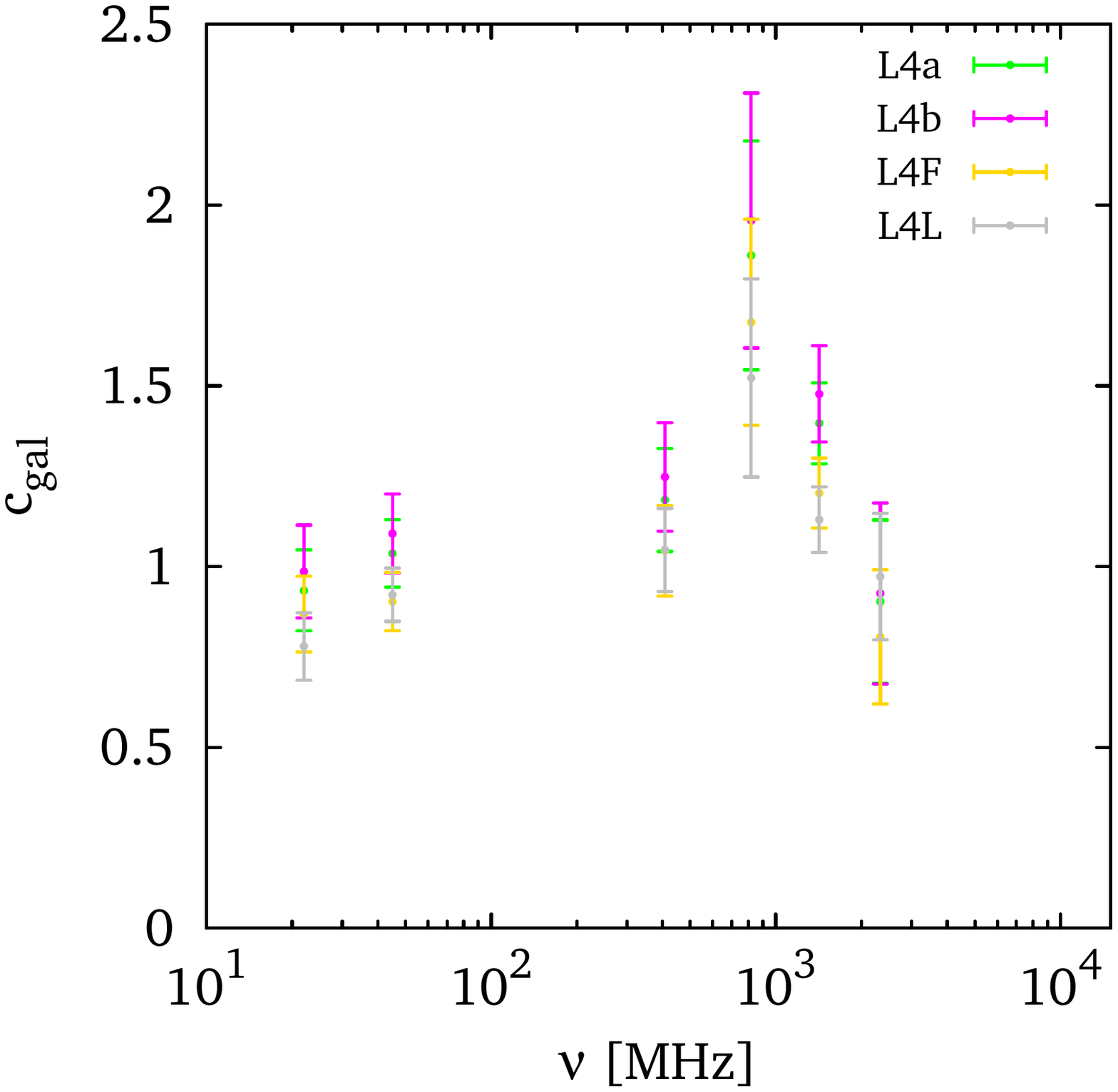}\hspace{-0.4cm}
 \includegraphics[width=0.34\textwidth]{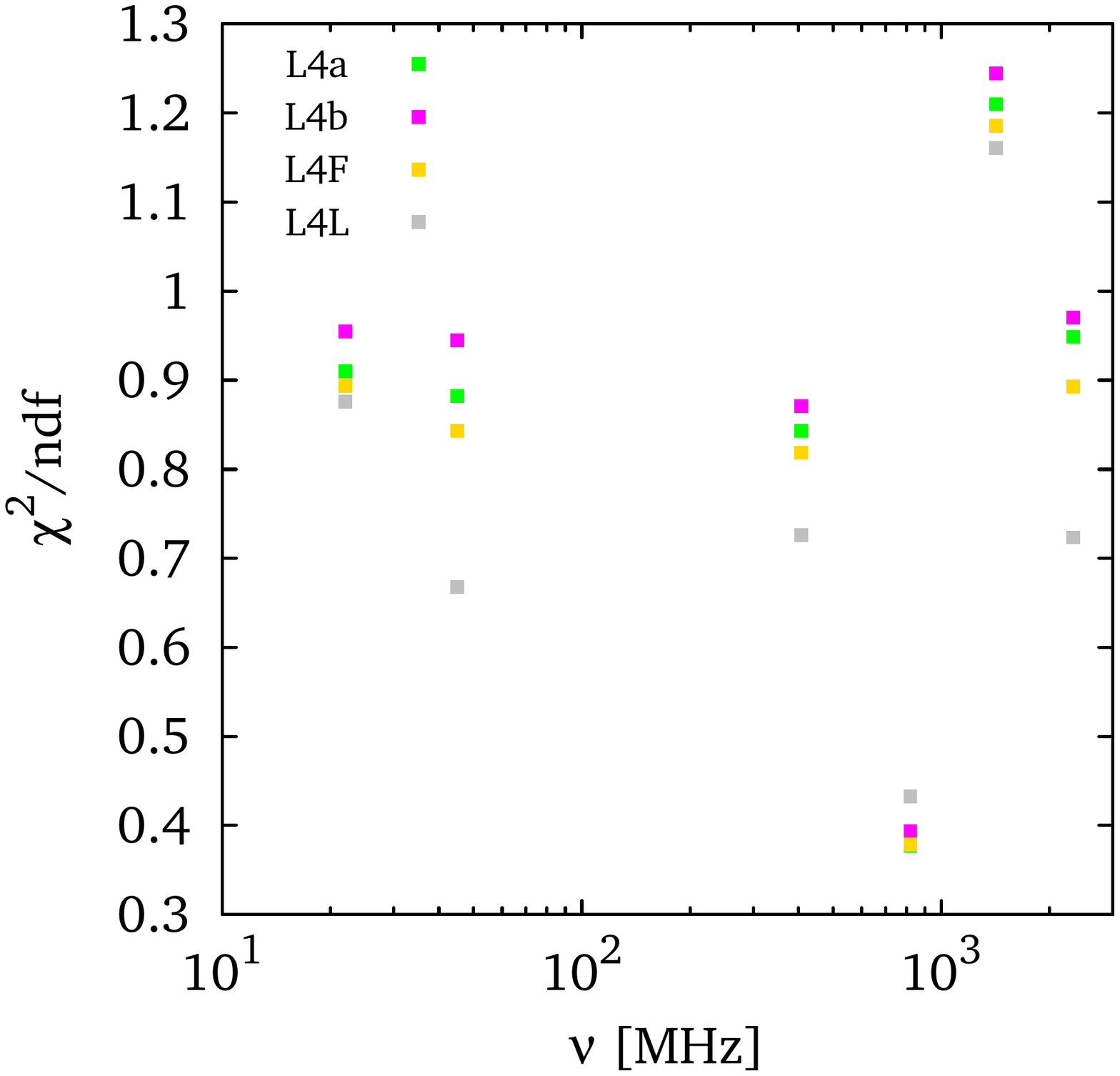}
\caption{Comparison of the results obtained with variations on the models of Galactic emission in the case of $L=4$ kpc. Starting from the benchmark model L4a, we change the scale height of the magnetic field (model L4b), the model of random magnetic field (model L4F) and the spatial distribution of cosmic-rays sources (model L4L). More details are in Sect. \ref{comparison}.
The left column shows the isotropic temperature $T_E$ at different frequencies, the central
column the normalizations parameters $c_{\rm gal}$ of the Galactic models, and the right column the reduced $\chi^2$ for the various
models at different frequencies.
The top rows refers to the mask method of Sect. \ref{sect:mask}, while the bottom row stands for the template method of Sect. \ref{sect:template}.}
\label{Fit_comparison2}
\end{center}
\end{figure*}

On top of the double exponential law, we have also considered the more complex model of random magnetic field suggested in Ref.~\cite{Jansson:2012rt}. It contains a disk component, with a central region and eight spiral arms, and a halo component which has a cylindrical symmetry, with a Gaussian dependence in $z$ and an exponential decrease in the radial coordinate.
We have implemented this parametrization for a propagation model with $L=4$ kpc (labeled as model L4F). As shown in Fig.\ref{Fit_comparison2}, this does not change the estimate of
the isotropic background, for both source treatments (mask and template) that we have employed. 

\begin{figure*}[t]
\begin{center}
 \includegraphics[width=0.32\textwidth]{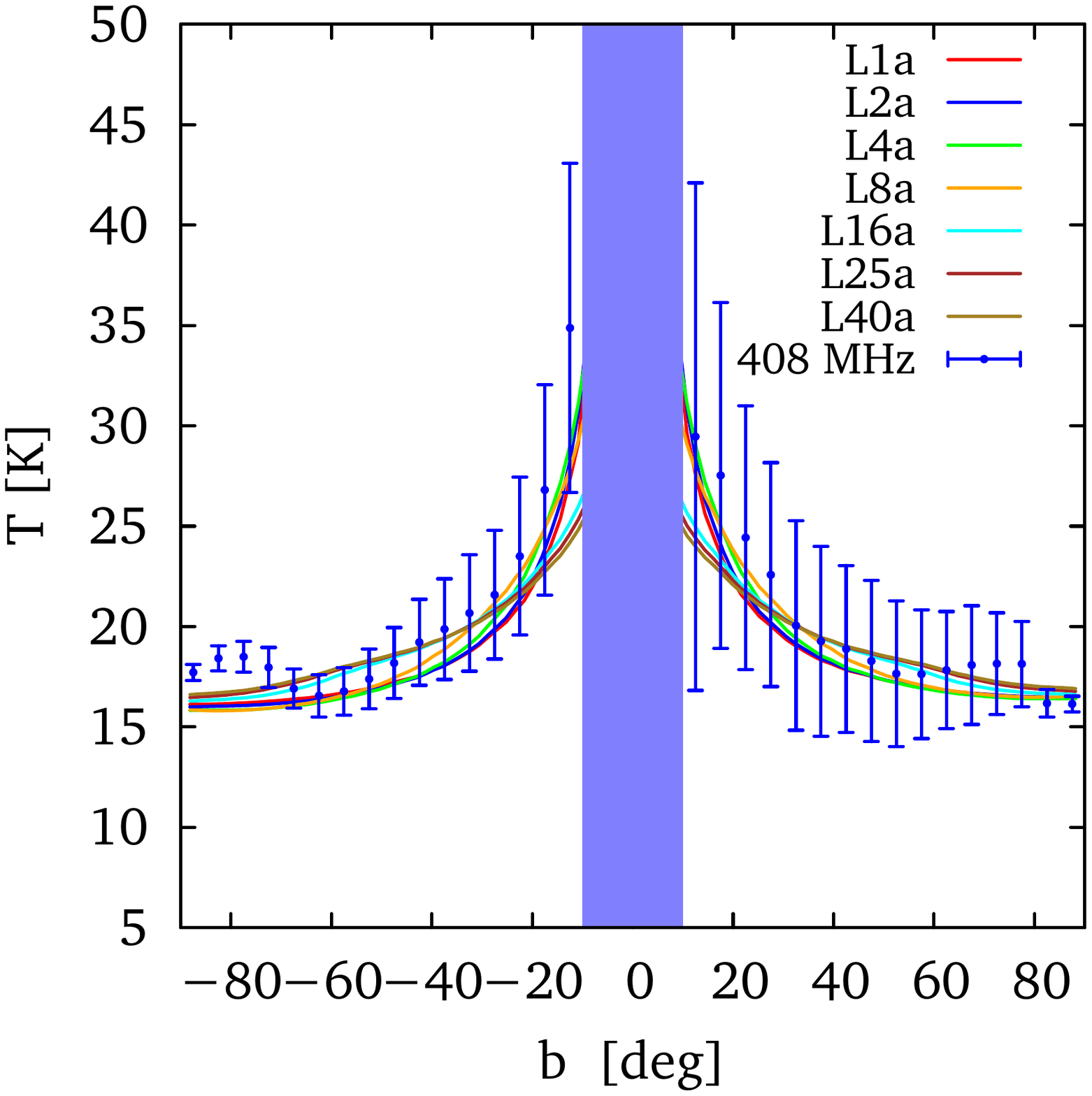}
 \includegraphics[width=0.32\textwidth]{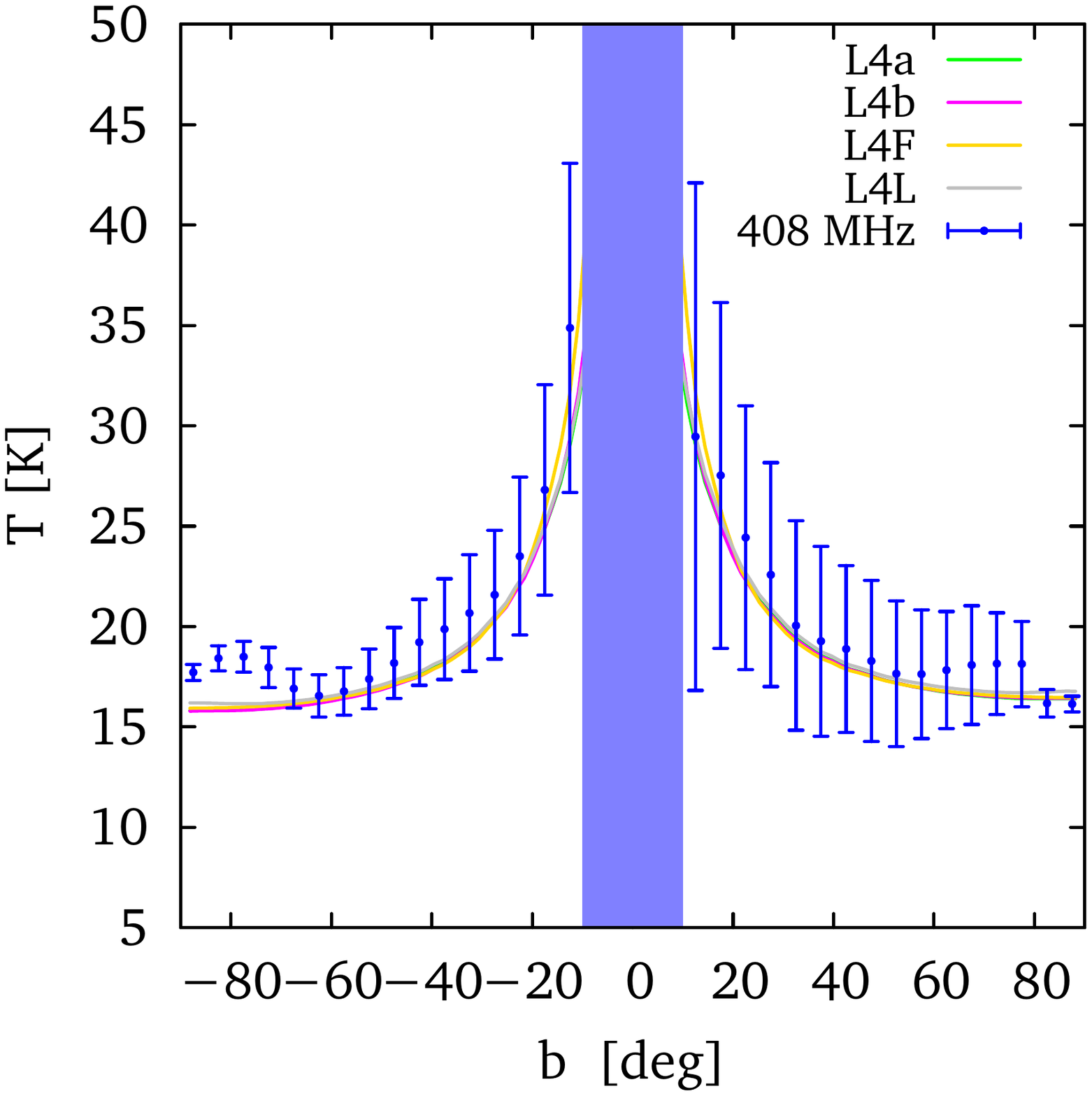}
 \includegraphics[width=0.32\textwidth]{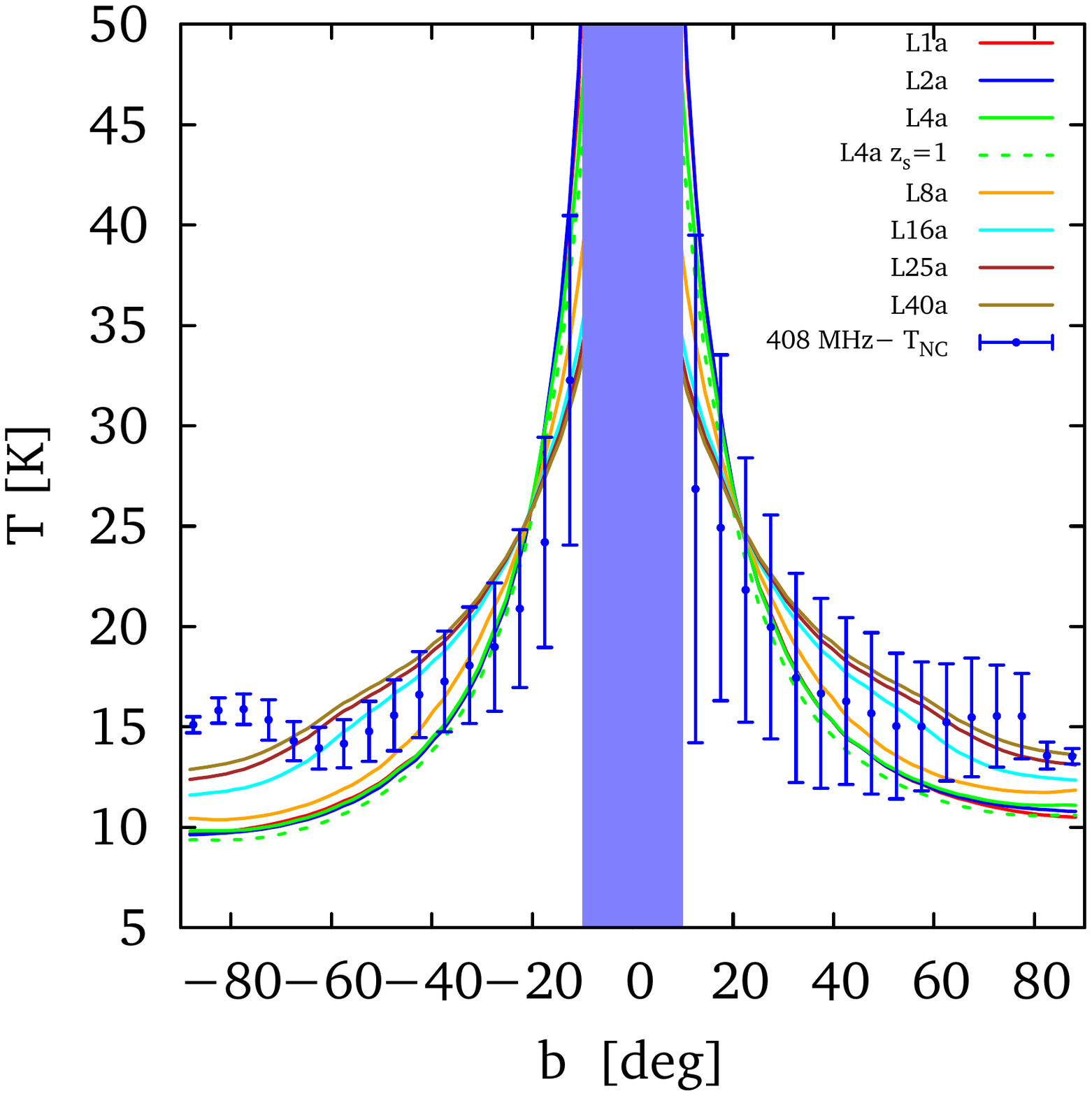}
\caption{{\sl Left}: Latitude profiles of the best-fit models (Galactic + isotropic emission) at 408 MHz. Data points correspond to the average temperature of the 408 MHz survey (CMB subtracted and after applying the WMAP7 mask) in strips of constant latitude $b$ and width of 5 degrees. The errors are the semi-dispersion of the data in each strip. {\sl Center:} The same as in the left plot, but for different Galactic models. {\sl Right}: Same type of latitude profiles, but the estimate of the extragalactic background derived from number counts in Ref. \cite{Gervasi:2008rr} has been subtracted from data; the solid lines are Galactic models (without any extra isotropic contribution) with arbitrary normalization.}

\label{profiles}
\end{center}
\end{figure*}

The spatial distribution of CRs sources inside the galaxy affects the morphology of the synchrotron emission as well. So far, we have considered the model of SNRs described in Ref. \cite{Strong:2010pr}. A steeper radial distribution has been previously proposed in Ref. \cite{Lorimer:2006qs}. Since the two models differ only for the radial dependence, the most relevant change is in the longitude profile of the synchrotron emission, while the latitude profile is unaffected (see Fig.~\ref{profiles}, central panel). For this reason, the estimate of the isotropic emission is almost unchanged, as shown for the benchmark model $L=4$ kpc in Fig.\ref{Fit_comparison2} (the model is labelled with L4L).
We note that the quality of the fits improve using the distribution of Ref. \cite{Lorimer:2006qs}. This is the opposite of what has been found in Ref. \cite{Orlando:2013ysa} for the 408 MHz survey. However, the analysis performed in Ref. \cite{Orlando:2013ysa} is different from ours. They have considered a resolution $N_{\rm side}=64$ and they have included all the sky in the fit. Performing their same calculation (although we have a different definition for the $\sigma$'s) we confirmed their results. However, imposing a cut on the Galactic plane ($|b|<10$ degrees), we find again a preference for the model in Ref.~\cite{Lorimer:2006qs}. On the other hand, including small latitudes, the $\chi^2$ quickly increases, since the picture around the disk is quite complex and the large scale Galactic model considered here should probably be made more complex. A detailed analysis on the disk emission and on the modeling of the SNRs distribution is beyond the scope of this paper.

The distribution of CRs sources along the $z$-direction is typically assumed to follow from the thickness of the disk. We take an exponential law with $z_s=0.2$ kpc.
As far as we reasonably assume $z_s\ll L$, we don't see any appreciable change in the Galactic synchrotron profile, as shown in Fig.~\ref{profiles} for the benchmark model $L=4$ kpc and for $z_s=1$ kpc. 
This is because low-energy electrons have a large confinement time, so diffusive processes have time to reshape the initial distribution.  Obviously, if $z_s\gtrsim L$, the picture would be different but we consider this hypothesis quite extreme.

As discussed in Sect.~\ref{sourceterm}, the spectral index of injection for Galactic primary electrons has been tuned to reproduce the synchrotron frequency scaling.
However, this again has a mild impact on the estimate of the extragalactic emission.
It can be understood looking the case tuned on AMS-02~\cite{AMS} data of $e^++e^-$. It has a different low-energy spectral break, but this does not affect the extragalactic estimate.

\begin{figure*}[t]
\begin{center}
 \includegraphics[width=0.48\textwidth]{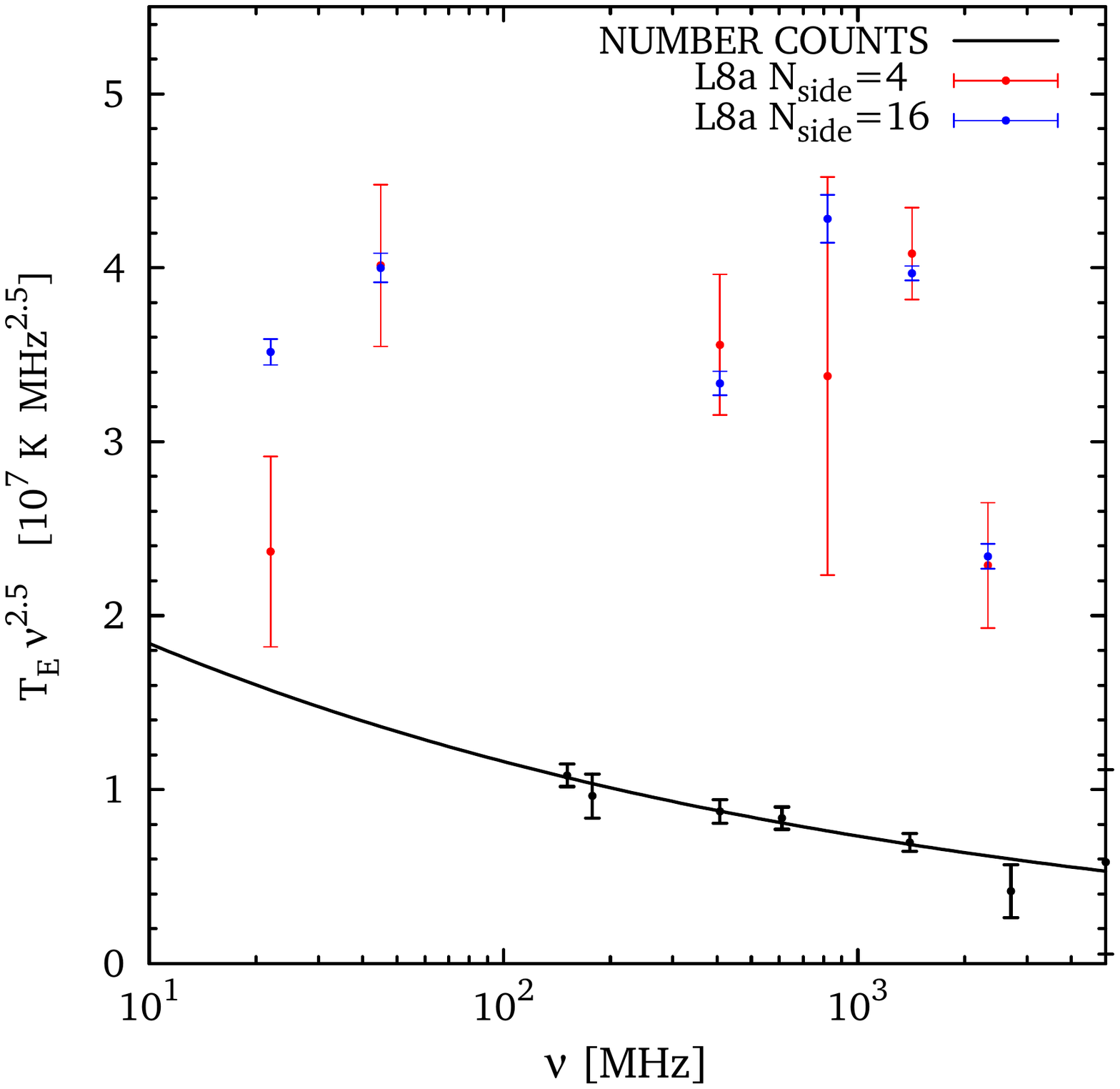}
 \includegraphics[width=0.48\textwidth]{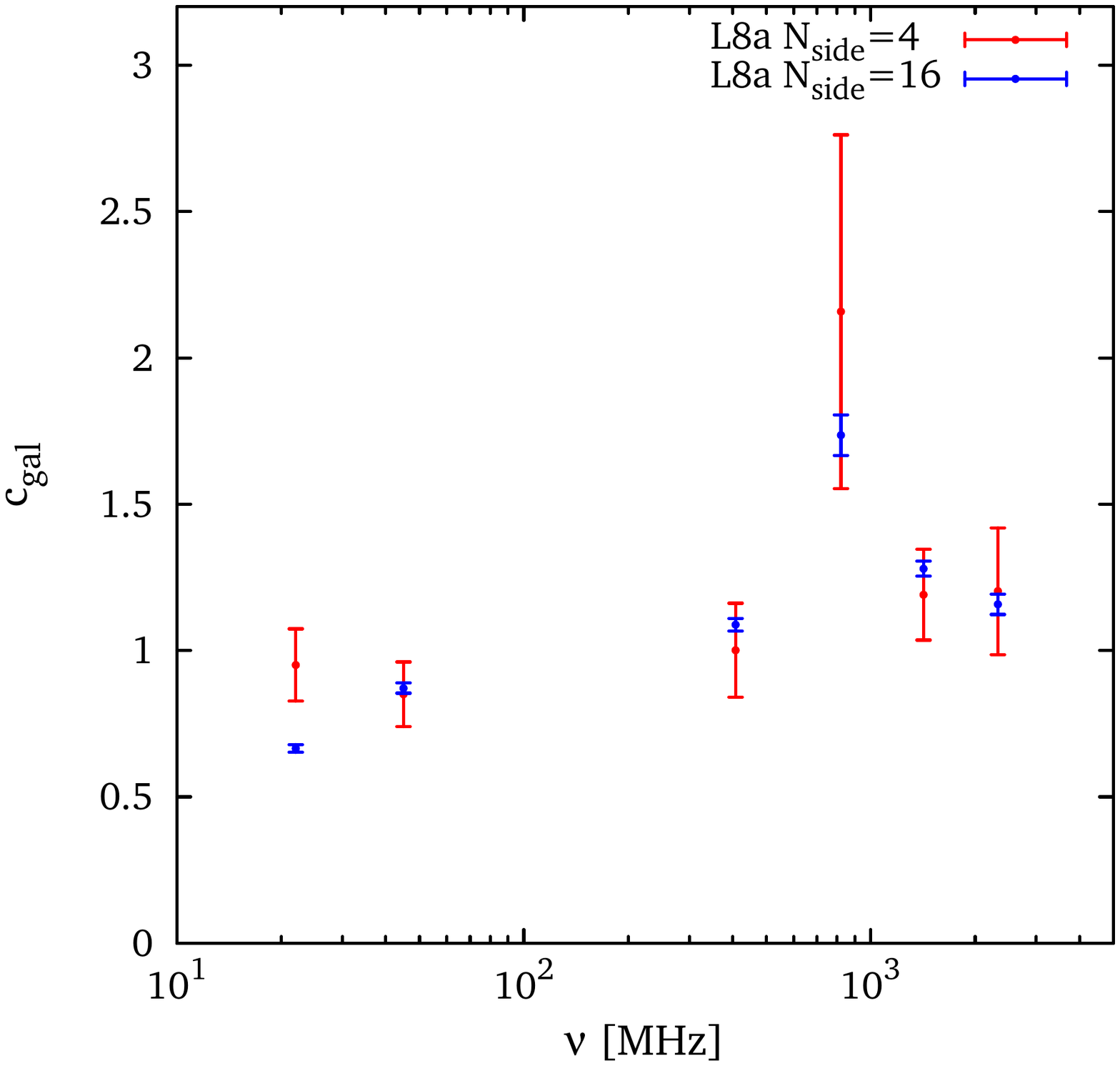}
\caption{Comparison of the results for the isotropic temperature $T_E$ (left) and the normalization coefficients $c_{\rm gal}$ (right) obtained with different angular resolutions. The figures refer to model L8a and to the mask method of Sect.\ref{sect:mask}.}
\label{Fit_Nside}
\end{center}
\end{figure*}

Finally, we have tested the robustness of our results with respect to a different choice of the resolution in the maps used in the analysis. As discussed in Sect.~\ref{sec:Fit}, this is an important aspect in order to account for the turbulent nature of the Galactic magnetic field. We have chosen to average our maps on a scale around 15 degrees, more specifically adopting the resolution $N_{\rm side}=4.$ We have repeated the analysis using a smaller angular scale, $N_{\rm side}=16,$ corresponding to 3.7 degrees. For consistency, we have recomputed the maps of $\sigma_i$ described in Sect.~\ref{sec:Fit}, averaging over the same scale. The significant increase of the statistics reduces the error bars associated to the fitting parameters. The results are fully consistent with those obtained with the previous resolution, except for 22 MHz (in some Galactic synchrotron models). In the latter case, the fraction of available sky is small and can be quite different at different resolutions (remember that, when we downgrade the maps, the masked region become significantly larger in $N_{\rm side}=4$).
In Fig.~\ref{Fit_Nside}, we show our findings for the model L8a and for the two angular resolutions considered.
The $\chi^2/{\rm ndf}$ increases moving from $N_{\rm side}=4$ to $N_{\rm side}=16,$ probably suggesting that for the case of $N_{\rm side}=16$ the models that we use are not appropriate to describe the Galactic emission at these small angular scales (i.e., fluctuations are underestimated).
$N_{\rm side}=4$ corresponds to a more conservative choice and leads to more conservative results for what concerns the error bars associated to the fitting parameters, and in turn to the isotropic component.

\section{Discussion}
\label{sec:Discussion}

\begin{figure*}[t]
\begin{center}
 \begin{minipage}[htb]{5.7cm}
   \centering
   \includegraphics[width=\textwidth]{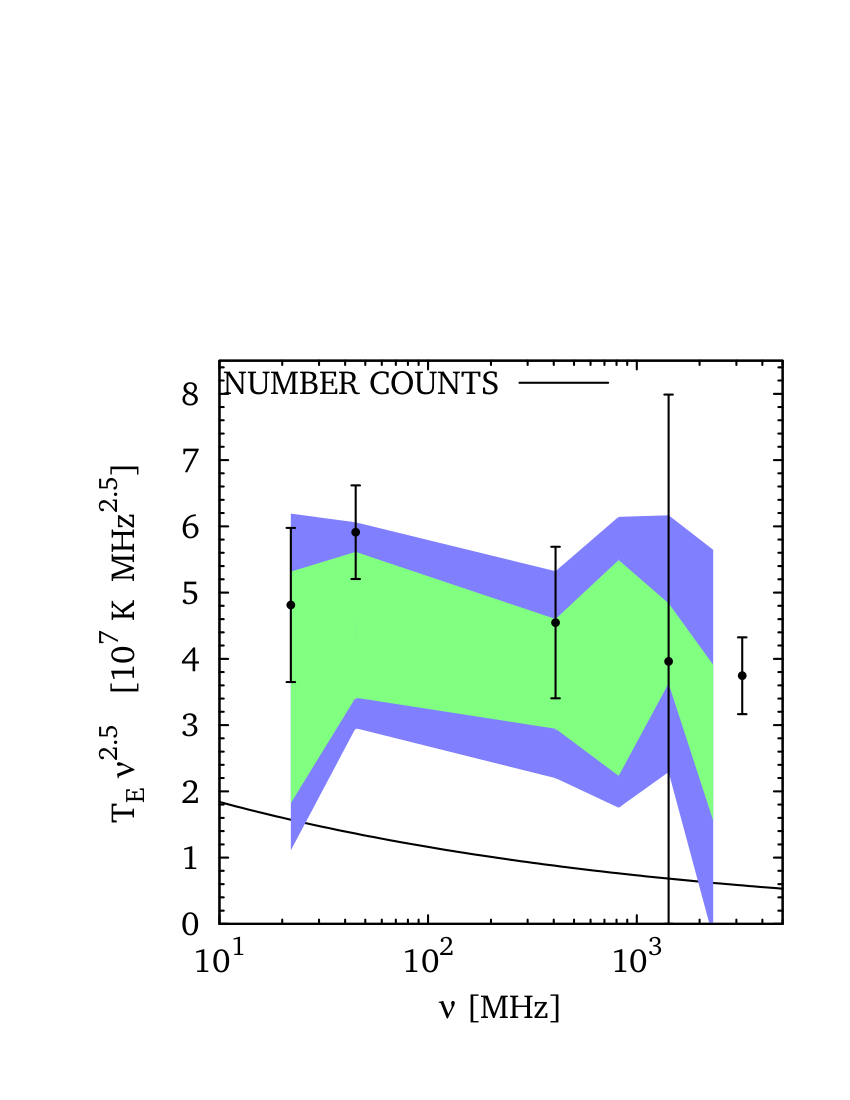}
 \end{minipage}
\hspace{-10mm}
 \begin{minipage}[htb]{5.7cm}
   \centering
   \includegraphics[width=\textwidth]{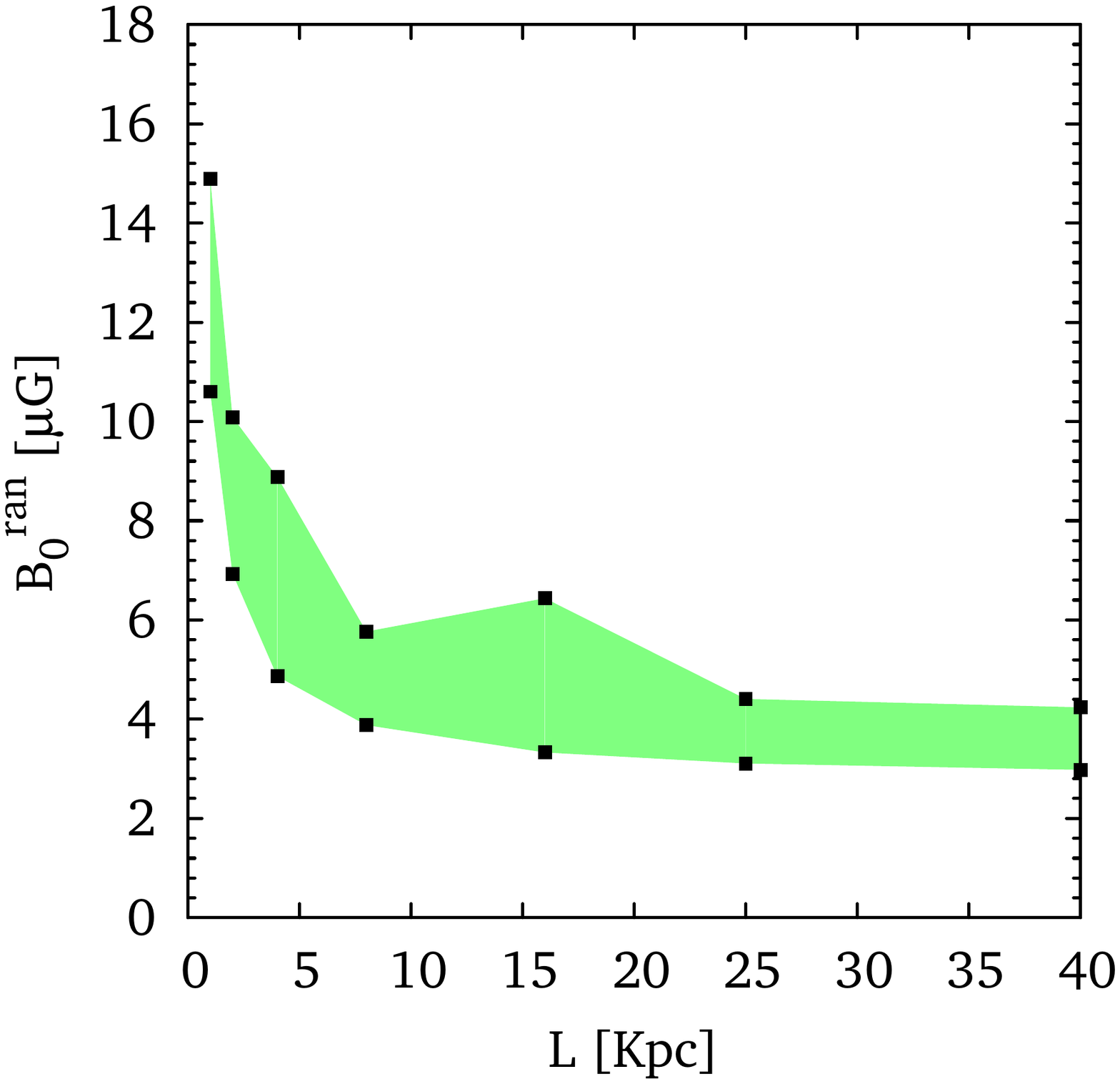}
 \end{minipage}
\hspace{-2mm}
 \begin{minipage}[htb]{4.6cm}
   \centering
\vspace{17.5mm}
   \includegraphics[width=\textwidth]{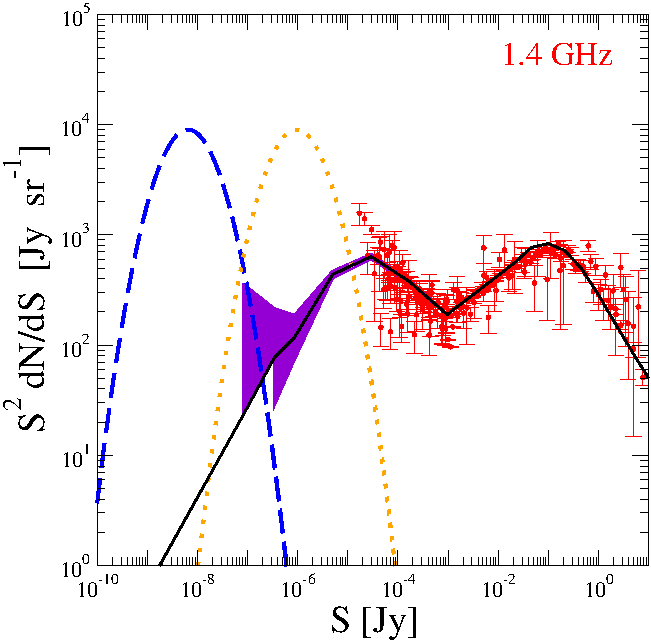}
 \end{minipage}

\caption{{\sl Left}: Uncertainty band of our estimate of the isotropic background temperature $T_E$, compared to the ARCADE-2 determination (points with error bars) and to the number count estimate (black solid line). The inner (green) area is obtained by
convolving all the results obtained by considering the variation of the Galactic modeling
and the different analysis techniques. The outer (blue) region further takes into account the experimental
zero-level error.
{\sl Center}: Strength of the random component $B_0^{\rm ran}$ of the magnetic field which best-fits the data, within the models we considered.
{\sl Right}: Source counts at 1.4 GHz. We show data (red points) of observed sources, the best-fit (black solid line, extrapolated following the evolutionary model of Ref. \cite{Condon:1984}) and the 1-$\sigma$ region (violet) of the P(D) analysis of Ref.\cite{Vernstrom:2013vva}. We additionally show two different examples (Gaussian bumps) of source counts distributions which could account for the excess in the isotropic emission obtained in our analysis: one just below the current observational threshold (dotted orange line), one for a population of fainter radio emitters (blue dashed line) satisfying the bound from \cite{Vernstrom:2013vva}.}
\label{TEB}
\end{center}
\end{figure*}

Our estimates of the isotropic radio background are summarized in Fig.~\ref{TEB}. The (inner) green area in the left panel shows the convolution of the estimates of $T_E$ obtained with the different methods adopted for source descriptions and for all the models of Galactic synchrotron emission discussed in the previous Sections. This band corresponds to a quite conservative assessment of the uncertainties, since some Galactic models are clearly favored by the data and, excluding the cases which provide a poor fit, the band would actually shrink. 

To associate a statistical confidence level to the green band, we could perform a statistical marginalization over the synchrotron models. However, to perform such a full scan of
the parameter space of Galactic modeling is extremely computer-time consuming, since is would require to perform a full numerical modeling for each set of Galactic models parameters, taken on a sufficient fine grid to allow us to perform a posteriori statistical analysis and marginalization, and to repeat the masking or template adaptation techniques discussed in Sect. \ref{sec:Fit} for each of these models. On the other hand, since our benchmark choices of propagation setups and Galactic magnetic fields explore the most relevant sources of uncertainties in the modeling of the Galactic emission, we do not expect an increase of the scatter in the estimates of the isotropic component $T_E$ by varying other parameters (as discussed in the previous Sections). By including in our results all the synchrotron models described in the previous Sections, we believe to have a rather conservative estimate of the uncertainty band associated to the isotropic emission. 

The blue (outer) area in the left panel of Fig.~\ref{TEB} shows the uncertainty band once we further add (in quadrature) the experimental zero-level error, reported in Table~\ref{tab:data} for the various radio surveys.
The estimate of the isotropic background obtained by the ARCADE-2 collaboration (the black points with error bars) \cite{Fixsen:2009xn} is all contained in this area, and therefore is fully compatible with our estimate. 
However, we wish to recall that if we focus on the models of Galactic emission with better $\chi^2$, the isotropic contribution tends to be somehow smaller than the estimate of ARCADE-2, as shown in Figs.~\ref{Fit3a} and \ref{Fit4a}.

In the previous Sections we have quantified the excess with respect to the level of isotropic emission expected from number counts of extragalactic sources. A simple way to visualize this excess is shown in Fig.~\ref{profiles}. In the left and center panels, we show the latitude profiles of the best-fit models and of the data, focusing at a the frequency of 408 MHz. For the experimental points, we have considered the average temperature in strips of constant latitude (with a width of 5 degrees). We masked bright sources using the WMAP-7 mask (see Sect.~\ref{sect:mask}).
The error bars are the semi-dispersions of the data in each strip. 
Then we subtract from the data the intensity of the isotropic background suggested by number counts: this is shown in the right panel. The resulting profile is compared to the latitudes profiles of the various Galactic models (with arbitrary normalizations). It is clear that the synchrotron emission from Galactic CRs tends to be too steep 
and some additional emission is needed to increase the compatibility with the data at high latitudes.
These plots make the physical picture clearer, but we warn that they are shown for illustrative purposes only. For the estimates of the isotropic background and confidence intervals, including the level of disagreement of the model with the isotropic contribution set by number counts estimates, see the previous Sections. 

An excess at high latitudes has been previously pointed out in Ref.~\cite{Orlando:2013ysa}, see in particular the figures of the latitude profile. 
Ref.~\cite{Orlando:2013ysa} fitted radio maps with various models of Galactic emissions and an offset term, analogous to the isotropic background that we have considered here, and founds that the offset is particularly large for Galactic models with moderate size of the halo (Ref.~\cite{Orlando:2013ysa} considered $L=4$ kpc), exceeding the expectations from number counts of extragalactic sources. The excess reduces for larger halos, although the model considered in Ref. ~\cite{Orlando:2013ysa} does not
provide a good fit to the data. The results of Ref.~\cite{Orlando:2013ysa} are similar to our findings, although
differences in the two analyses are present:  the statistical method employed is different, also because the analysis of Ref.~\cite{Orlando:2013ysa} is not specifically focussed to determine the isotropic radio background; Ref. ~\cite{Orlando:2013ysa} analyzes the whole sky, without masking or modeling discrete radio sources; in our analysis, we downgrade the maps resolution, for the reasons discussed above, while Ref.~\cite{Orlando:2013ysa} considers the original maps resolution. For these reasons, Ref.~\cite{Orlando:2013ysa} obtaines a different estimate of the isotropic background from our findings.

The inclusion of the zero-level uncertainty of the maps partially reduces the discrepancy between our estimates of the isotropic emission and the one from number counts integration. For instance, at 22 MHz and 2326 MHz, the latter falls into the uncertainty band we draw. However, for the most reliable frequencies (45 MHz, 408 MHz, and 1.4 GHz) there is still a factor of $\sim 3$ between the two estimates.
A conservative assessment of the presence of the excess can be done by selecting at each frequency the Galactic model for which the lowest isotropic temperature is obtained. The best-fit values of $T_E$ and the corresponding confidence intervals (including the zero level uncertainty) associated to this model are then taken to compute how many standard deviations $\sigma$ away the number count value is: this quantifies the discrepancy. We find, for the different radio maps from 22 MHz to 2326 MHz: $0.6\sigma,$  $2.9\sigma,$   $2.2\sigma,$  $1.6\sigma,$ $2.0 \sigma$ and $0.6\sigma$. They are all systematically positive deviations, with an overall significance of the excess of $4.5\sigma$.

In Table~\ref{tab:results}, we show our estimates (with the associated statistical errors) of the isotropic emission $T_E$ in our best-fit model at different frequencies.
They are compared to the number count estimates and to the zero level uncertainty.
As a sort of independent upper bound for the isotropic level, we also report, at each frequency, the mean temperature of the coldest spot in the map, which has been defined as a circle of 15 degrees of radius.
Obviously, with a too small region one could run into a statistical fluctuation. On the other hand, a too large region won't be representative of coldest temperatures.
We checked that although the size of 15 degrees is sufficiently large, there is no significant amount of pixels which have a much smaller temperature than the reported mean.
Table~\ref{tab:results} shows that the isotropic estimate is consistent with the the coldest patch at all frequencies. However, the two temperatures are not far from each other, leaving little room for a Galactic contribution.
In other words, as it is clear from the residuals in Figs.~\ref{Fit3res} and \ref{Fit4res}, our model overshoots the data in the coldest spots (which is the top-right region in Fig.~\ref{surveys} for basically all the maps). The residual amounts to about 30\% of the isotropic contribution and is due to the fact that the radio maps show some remarkable (visible even by eye) east-west and north-south asymmetries, while our Galactic model is nearly symmetric (with moderate asymmetries due to the magnetic field shape).
The first possibility to account for this mismatch is that the correct Galactic model will have a deficit in the eastern region.
The other case is instead that the isotropic component would have to be damped to reduce the overall emission and then the Galactic contribution should increase in the western part.

Adding up source count estimates and zero-level offsets one typically reaches about 50\% of the estimated isotropic level, so it is clear that if we would take the coldest patch as the region to be fitted (i.e., if we would take an isotropic estimate reduced by 30\% with respect to our determinations), the evidence for an excess would be significantly weakened.
On the other hand, this would lead to an underestimation of the temperature at the poles. Since the profile of the emission is flat at high latitudes, as already mentioned, it would be very difficult to be fitted by a Galactic component. So, in order to have a consistent picture, a completely different Galactic model would be needed. We also remind the reader that the employed Galactic models fit local CR spectra and gamma-ray data, and it is not simple to accommodate a significant departure from them without violating such bounds.
The other possibility (that we consider more likely) is instead that either the Galactic contribution has a deficit in the cold-spot direction (e.g., because of the magnetic field shape) or the maps need some small zero-level adjustment. In this case, the estimate of the isotropic component would not change with respect to what reported here.

The source number counts allow to infer the contribution of detected extragalactic radio sources to the isotropic background. Then, one should add the populations of faint undetected sources. Obviously, their contribution is unknown, so the final estimate of the extragalactic background depends on the extrapolation of number counts below the experimental threshold. 
However, it has been recently shown in Ref. \cite{Condon:2012ug,Vernstrom:2013vva} that from a P(D) analysis it is possible to set constraints on the counts at flux levels almost two orders of magnitude below instrumental and confusion noise. In particular, at 1.4 GHz, the current threshold is about few tens of $\mu$Jy, and Ref.~\cite{Vernstrom:2013vva} found that no significant contribution above the standard extrapolation is allowed above few hundreds of nJy.
Therefore, modifications to the evolutionary models of standard sources appear unlikely to be able to explain the excess.

In the context of the ARCADE excess \cite{Singal:2009xq}, different astrophysical scenarios have been carefully scrutinized as possible origin of the extragalactic background. The contribution from  radio supernovae, radio quiet quasars and diffuse emission from intergalactic medium and clusters have been shown to be quite modest. Star forming galaxies with a non-standard evolutionary model are constrained by bounds from the far IR-radio correlation and from the P(D) analysis.
If the origin of the excess is from a population of extragalactic sources, they have to be very faint (significantly contributing to number counts well below the $\mu$Jy level) and very numerous \cite{Singal:2009dv,Vernstrom:2011xt,Condon:2012ug,Vernstrom:2013vva}.

A possibility in this direction is represented by synchrotron radiation in DM halos induced by annihilations of DM particles, as proposed in Ref. \cite{paper1}.
It satisfies the P(D) bound, see, e.g., Fig. 3 of Ref. \cite{paper1}.

In the right panel of Fig.~\ref{TEB}, we show two examples (dashed and dotted lines) of how the number counts could be modified in order to account for the excess at 1.4 GHz.
On top of data-points referring to counts of observed sources, we show the best-fit (black-line, extrapolated in the unsampled part following the evolutionary model of Ref. \cite{Condon:1984}) and its $1\sigma$ region (violet area) obtained through the P(D) analysis of Ref. \cite{Vernstrom:2013vva} which reaches brightnesses well below the detection threshold.
For simplicity, the examples are chosen to be Gaussian bumps. The first example peaks at $\mu$Jy level (dotted line) and affects counts only below the observed fluxes, but it is in conflict with the mentioned P(D) analysis.
The second example, with a peak at about 10 nJy (dashed line), is instead in principle viable. Similar results, but with weaker constraints (because of fewer data and analyses) can be found at the other frequencies considered in this work.

A different possibility to account for the excess might be given by a Galactic explanation. 
In an attempt to find a Galactic explanation, we considered three different Galactic diffuse synchrotron sources: one still connected with SN or pulsar, but with a broader distribution along the vertical direction ($z_s=1$ kpc, thus no longer strictly confined to the disc); two cases related to possible particle DM signals with a profile following the DM density $\rho_{DM}$ (decaying) or the density squared $\rho_{DM}^2$ (annihilating), with $\rho_{DM}$ described by a cored isothermal distribution.
We found that none of these models can actually account for the excess, because they exhibit a significant variation at high latitudes, while a flat component is needed to fit the data. This can be seen in the right panel of Fig.~\ref{profiles} for the $z_s=1$ kpc case.

Another possibility could be that the offset in the surveys has been underestimated. On the other hand, we find a positive excess in all maps and of a similar size. 
This would imply that all the surveys share a common calibration issue, a possibility that appears unlikely.

Finally, let us note that in our analysis we have constrained the Galactic synchrotron emission, which in turns depends on the interstellar electron field and on the Galactic magnetic field. 
In the central panel of Fig.~\ref{TEB}, we show our estimate of $B_0,$ the normalization of the random magnetic field at Earth, for different models of CRs propagation (the coherent part of the magnetic field is fixed from Ref. \cite{Jansson:2012pc} and its strength at the Earth position is 2 $\mu$G). In particular, we show $B_0$ as a function of $L$, the size of the diffusion box. The uncertainty band has been estimated from the confidence intervals of $c_{\rm gal},$ which is the fitting coefficient normalizing the synchrotron contributions, and assuming that the intensity of the synchrotron emission scales as $T\propto B_{\rm ran}^2$ (which is approximately true in the range of interest). The plot is shown for illustrative purposes, since (as already stated and commented) we do not perform a full scan over all the parameters involved in Galactic modeling, including the magnetic field, not being this full scan the focus of this work. However, it should be noted that low-$L$ models seem to prefer magnetic strength with extremely large and rather unlikely values, and therefore they might be disfavored by radio data.

\section{Conclusions}
\label{sec:Conclusions}

\begin{table}[t]
\begin{center}
\begin{tabular}{|c|c|c|c|c|}
\hline
Frequency [MHz]& $T_E$ [K] & $T_{NC}$ [K] & zero-level [K] & $T_{\rm cold\,spot}$ [K]\tabularnewline
\hline
22 & $(1.04 \pm 0.24) \times 10^4$ & $6.92 \times 10^3$  & 5000 & $1.80\times 10^4$
\tabularnewline
\hline
45 & $(2.95 \pm 0.34) \times 10^3$ & $1.0 \times 10^3$  & 544 & $3.84\times 10^3$
\tabularnewline
\hline
408 & 11.8 $\pm 1.1$  & 2.61  & 3 & 12.14
\tabularnewline
\hline
820 & 2.21 $\pm 0.39 $ & 0.39  & 0.6 & 2.91
\tabularnewline
\hline
1420 & 0.580 $\pm 0.025$  & 0.09  & 0.2 & 0.589
\tabularnewline
\hline
2326 & 0.073 $\pm 0.013 $  & 0.024  & 0.08 & 0.098
\tabularnewline
\hline
\end{tabular}
\end{center}
\caption{Estimates of the isotropic component $T_E$ at different frequencies from the best-fit model in our analysis, compared to the estimates $T_{NC}$ obtained from number counts, the zero-level error of the surveys, and the temperature $T_{\rm cold\,spot}$ of the coldest spot ($15^\circ$ of radius) in the maps.}
\label{tab:results}
\end{table}

We collected the most complete observational sky maps at radio frequencies ranging from 22 MHz to 2326 MHz.
The observed temperature is theoretically modeled by means of four components: Galactic diffuse synchrotron radiation, Galactic thermal bremsstrahlung, single sources, and isotropic emission.
We employed different methods to assess the main uncertainties associated to Galactic modeling and to the treatment of source through masks or templates.
The synchrotron emission is the main Galactic component and has been computed adopting a purely diffusive model for CRs (solving the transport equation with the GALPROP code~\cite{Strong:1998pw}) and a magnetic field based on Refs. \cite{Jansson:2012rt,Jansson:2012pc}.
We considered different cases for the parameters defining its latitude profile (such as the vertical scale of the diffusion box $L$ and the $z$-behaviour of the magnetic field).
Our main focus has been the determination of the isotropic background temperature $T_E$, and
we found fully compatible results among the different methods, with a moderate scatter, which makes us to believe that the estimate is robust.

Our findings on $T_E$ lie above expectations from what can be estimated by extrapolation to low intensities of number counts of observed extragalactic sources: the
reconstructed values of $T_E$ exceed them by a factor $\gtrsim 3$ in the maps with the largest fraction of sky available (45 MHz, 408 MHz, and 1.4 GHz), which is where our estimates are more solid.
Therefore, our results confirm the so called ``ARCADE'' excess, although our estimates are somewhat smaller than the results in Ref. \cite{Fixsen:2009xn}. 

In the radio maps, the isotropic component dominates the emission at high latitudes making the total profile quite flat at $b\gtrsim 50^\circ$.
A Galactic component with a spatial distribution related either to the disc, or to a DM halo (as in the case of annihilating or decaying DM), or being very local would introduce some latitude dependence which is not seen in the data.
For this reason, a Galactic origin for the mismatch between our estimate and the integral of number counts appears unlikely, unless some new model, with quite different properties with respect to what is usually considered, will be developed. 

On the other hand, from statistical estimates of the source count at fluxes below the faintest sources that can be counted individually, no significant deviation from the standard extrapolation of extragalactic source count is expected above few hundreds of nJy~\cite{Vernstrom:2013vva}.

We therefore conclude that observational radio maps point towards a puzzling excess.
It could be due to to a systematic offset in the surveys (although in the same positive direction for all the maps), to the need of a profound modification of our current understanding of the Galactic synchrotron emission, or to a novel population of very faint extragalactic sources. The extragalactic interpretation is quite interesting, since its origin could be linked to a dark matter emission \cite{paper1}.

Next future telescopes, and in particular LOFAR~\cite{LOFAR} for the Galactic diffuse emission and SKA~\cite{SKA} (with its precursors) for the source counts, will bring important tiles to solve this intriguing puzzle.

{\acknowledgments
We thank Patricia Reich for kindly providing the data of the total intensity all-sky map at 1.4 GHz used in this work.
We also acknowledge the MPIfR's Survey Sampler from which some of the datasets were downloaded. M.T. thanks Marco Cirelli and Gaelle Giesen for discussions.
This
work is supported by the research grant {\sl Theoretical Astroparticle Physics} number 2012CPPYP7 under the program PRIN 2012  funded by the Ministero dell'Istruzione, Universit\`a e della Ricerca (MIUR), by the research grant {\sl TAsP (Theoretical Astroparticle Physics)}
funded by the Istituto Nazionale di Fisica Nucleare (INFN), by the  {\sl Strategic Research Grant: Origin and Detection of Galactic and Extragalactic Cosmic Rays} funded by Torino University and Compagnia di San Paolo, by the Spanish MINECO under grants FPA2011-22975 and MULTIDARK CSD2009-00064 (Consolider-Ingenio 2010 Programme), by Prometeo/2009/091 (Generalitat Valenciana), and by the EU ITN UNILHC PITN-GA-2009-237920. M.T. is supported  by the European Research Council ({\sc Erc}) under the EU Seventh Framework Programme (FP7/2007-2013) / {\sc Erc} Starting Grant (agreement n. 278234 - `{\sc NewDark}' project).
R.L. is supported by a Juan de la Cierva contract (MINECO). 
}

\bibliography{./bib/references}

\end{document}